\documentclass[longauth]{aa} % for the long lists of affiliations 

\usepackage[varg]{txfonts}
\usepackage{placeins}
\usepackage{verbatim}
\usepackage{xcolor}
\newcommand{\HL}[1]{\textcolor{black}{#1}}

\newcommand{\HLr}[1]{\textcolor{black}{#1}}
\usepackage{amsmath}
\usepackage[version=4]{mhchem}
\usepackage{longtable}
\usepackage{float}
\usepackage{lscape}
\usepackage{appendix}

\begin{document}

\title{CARMENES input catalog of M dwarfs}
\subtitle{\HLr{VII.} New rotation periods for the survey stars and their correlations with stellar activity}

\titlerunning{CARMENES GTO Rotation Periods}

\authorrunning{Y.~Shan, D.~Revilla, S.~Skrzypinski, et al.}

\author{Y.~Shan \inst{\ref{inst:ceed},\ref{inst:iag}}
\and D.~Revilla \inst{\ref{inst:ucm},\ref{inst:iaa}} 
\and S.\,L.~Skrzypinski \inst{\ref{inst:ucm}} 
\and S.~Dreizler \inst{\ref{inst:iag}}
\and V.\,J.\,S.~B\'ejar \inst{\ref{inst:iac},\ref{inst:ull}}
\and J.\,A.~Caballero \inst{\ref{inst:cabesac}}
\and C.~Cardona Guill\'en \inst{\ref{inst:iac},\ref{inst:ull}}
\and C.~Cifuentes \inst{\ref{inst:cabesac}}
\and B.~Fuhrmeister \inst{\ref{inst:hs}}
\and A.~Reiners \inst{\ref{inst:iag}} 
\and S.~Vanaverbeke \inst{\ref{inst:kul}}
\and I.~Ribas \inst{\ref{inst:ice},\ref{inst:ieec}}
\and A.~Quirrenbach \inst{\ref{inst:lsw}}
\and P.\,J.~Amado \inst{\ref{inst:iaa}}
\and F.\,J.~Aceituno \inst{\ref{inst:iaa}}
\and V.~Casanova \inst{\ref{inst:iaa}}
\and M.~Cort\'es-Contreras \inst{\ref{inst:ucm}}
\and F.~Dubois \inst{\ref{inst:astrolab},\ref{inst:vvs}}
\and P.~Gorrini \inst{\ref{inst:iag}}
\and Th.~Henning \inst{\ref{inst:mpia}}
\and E.~Herrero \inst{\ref{inst:ice},\ref{inst:ieec}}
\and S.\,V.~Jeffers \inst{\ref{inst:mps}}
\and J.~Kemmer \inst{\ref{inst:lsw}}
\and S.~Lalitha \inst{\ref{inst:birm}}
\and N.~Lodieu \inst{\ref{inst:iac}}
\and L.~Logie \inst{\ref{inst:astrolab},\ref{inst:vvs}}
\and M.~J.~L\'opez Gonz\'alez \inst{\ref{inst:iaa}}
\and S.~Mart\'\i n-Ruiz \inst{\ref{inst:iaa}}
\and D.~Montes \inst{\ref{inst:ucm}}
\and J.\,C.~Morales \inst{\ref{inst:ice},\ref{inst:ieec}}
\and E.~Nagel \inst{\ref{inst:iag},\ref{inst:tls}}
\and E.~Pall\'e \inst{\ref{inst:iac},\ref{inst:ull}}
\and V.~Perdelwitz \inst{\ref{inst:ariel},\ref{inst:hs}}
\and M.~P\'erez-Torres\inst{\ref{inst:iaa},\HLr{\ref{inst:capa},\ref{inst:cyp}}}
\and D.~Pollacco \inst{\ref{inst:war}}
\and S.~Rau \inst{\ref{inst:astrolab},\ref{inst:vvs}}
\and C.~Rodr\'iguez-L\'opez \inst{\ref{inst:iaa}}
\and E.~Rodr\'iguez\inst{\ref{inst:iaa}}
\and P.~Sch{\"o}fer\inst{\ref{inst:iaa}}
\and W.~Seifert\inst{\ref{inst:lsw}}
\and A.~Sota\inst{\ref{inst:iaa}}
\and M.\,R.~Zapatero Osorio\inst{\ref{inst:cabesac}}
\and M.~Zechmeister \inst{\ref{inst:iag}}
}

\institute{
\label{inst:ceed}Centre for Planetary Habitability, Department of Geosciences, University of Oslo, Sem Saelands vei 2b 0315 Oslo, Norway %01
\and
\label{inst:iag}Institut f\"ur Astrophysik, Georg-August-Universit\"at, Friedrich-Hund-Platz 1, 37077 G\"ottingen, Germany \\ %02
\email{yutong.shan@uni-goettingen.de} 
\and
\label{inst:ucm}Departamento de F\'{i}sica de la Tierra y Astrof\'{i}sica and IPARCOS-UCM (Instituto de F\'{i}sica de Part\'{i}culas y del Cosmos de la UCM), Facultad de Ciencias F\'{i}sicas, Universidad Complutense de Madrid, 28040, Madrid, Spain %03
\and
\label{inst:iaa}Instituto de Astrof\'isica de Andaluc\'ia (CSIC), Glorieta de la Astronom\'ia s/n, 18008 Granada, Spain %04
\and
\label{inst:iac}Instituto de Astrof\'isica de Canarias, Via L\'{a}ctea s/n, 38205 La Laguna, Tenerife, Spain %05
\and 
\label{inst:ull}Departamento de Astrof\'isica, Universidad de La Laguna, 38206 La Laguna, Tenerife, Spain %06
\and 
\label{inst:cabesac}Centro de Astrobiolog\'ia (CSIC-INTA), ESAC campus, Camino bajo del castillo s/n, 28692 Villanueva de la Ca\~nada, Madrid, Spain %07
\and
\label{inst:hs}Hamburger Sternwarte, Universit\"at Hamburg, Gojenbergsweg 112, 21029 Hamburg, Germany %08
\and
\label{inst:kul}Centre for Mathematical Plasma Astrophysics, Katholieke Universiteit Leuven, Celestijnenlaan 200B, bus 2400, 3001 Leuven, Belgium %09
\and
\label{inst:ice}Institut de Ci\`encies de l’Espai (ICE, CSIC), Campus UAB, Can Magrans s/n, 08193 Bellaterra, Spain %10
\and 
\label{inst:ieec}Institut d’Estudis Espacials de Catalunya (IEEC), 08034 Barcelona, Spain %11
\and
\label{inst:lsw}Landessternwarte, Zentrum f\"ur Astronomie der Universit\"at Heidelberg, K\"onigstuhl 12, 69117 Heidelberg, Germany %12
\and
\label{inst:astrolab}AstroLAB IRIS, Provinciaal Domein ``De Palingbeek'', Verbrande-molenstraat 5, 8902 Zillebeke, Ieper, Belgium %13
\and
\label{inst:vvs}Vereniging Voor Sterrenkunde, Oude Bleken 12, 2400 Mol, Belgium %14
\and
\label{inst:mpia}Max-Planck-Institut f\"ur Astronomie, K\"onigstuhl 17, 69117 Heidelberg, Germany %15
\and
\label{inst:mps}Max-Planck-Institut f\"ur Sonnensystemforschung, Justus-von-Liebig-Weg 3, 37077 G\"ottingen, Germany %16
\and
\label{inst:birm}School of Physics \& Astronomy, University of Birmingham, Edgbaston, Birmingham B15 2TT, UK %17
\and
\label{inst:tls}Th\"uringer Landessternwarte Tautenburg, Sternwarte 5, 07778 Tautenburg, Germany %18
\and
\label{inst:ariel}Department of Physics, Ariel University, Ariel 40700, Israel %19 
\and
\label{inst:capa}Center for Astroparticles and High Energy Physics (CAPA), Universidad de
Zaragoza, E-50009 Zaragoza, Spain %20
\and
\label{inst:cyp}School of Sciences, European University Cyprus, Diogenes street, Engomi,
1516 Nicosia, Cyprus %21
\and
\label{inst:war}Department of Physics, University of Warwick, Gibbet Hill Road, Coventry CV4 7AL, United Kingdom %22
}

\date{Received 02 May 2023 / Accepted 07 Dec 2023}

\abstract
{} 
{Knowledge of rotation periods ($P_{\rm rot}$) is important for understanding the magnetic activity and angular momentum evolution of late-type stars, as well as for evaluating radial velocity signals of potential exoplanets and identifying false positives. We measured photometric and spectroscopic $P_{\rm rot}$ for a large sample of nearby bright M dwarfs with spectral types from M0 to M9, as part of our continual effort to fully characterize the Guaranteed Time Observation programme stars of the CARMENES survey. }
{We analyse light curves chiefly from the SuperWASP survey and TESS data. We supplemented these with our own follow-up photometric monitoring programme from ground-based facilities, as well as spectroscopic indicator time series derived directly from the CARMENES spectra. }
{From our own analysis, we determined $P_{\rm rot}$ for 129 stars. Combined with the literature, we tabulated $P_{\rm rot}$ for 261 stars, or 75\% of our sample. We developed a framework to evaluate the plausibility of all periods available for this sample by comparing them with activity signatures and checking for consistency between multiple measurements. We find that 166 of these stars have independent evidence that confirmed their $P_{\rm rot}$. There are inconsistencies in 27 periods, which we classify as debated. A further 68 periods are identified as provisional detections that could benefit from independent verification. We provide an empirical relation for the $P_{\rm rot}$ uncertainty as a function of the $P_{\rm rot}$ value, based on the dispersion of the measurements. We show that published formal errors seem to be often underestimated for periods longwards of $\sim 10$\,d.
We examined rotation-activity relations with emission in X-rays, H$\alpha$, Ca~{\sc ii} H\&K, and surface magnetic field strengths for this sample of M dwarfs. We find overall agreement with previous works, as well as tentative differences in the partially versus fully convective subsamples. We show $P_{\rm rot}$ as a function of stellar mass, age, and galactic kinematics. With the notable exception of three transiting planet systems and TZ Ari, all known planet hosts in this sample have $P_{\rm rot} \gtrsim 15$\,d.}   
{Inherent challenges in determining accurate and precise stellar $P_{\rm rot}$ means independent verification is important, especially for inactive M dwarfs. Evidence of potential mass dependence in activity-rotation relations would suggest physical changes in the magnetic dynamo that warrants further investigation using larger samples of M dwarfs on both sides of the fully convective boundary. Important limitations need to be overcome before the radial velocity technique can be routinely used to detect and study planets around young and active stars. 
}


\maketitle

%===========================================
\section{Introduction}\label{s:intro} 

Owing in large part to a rapidly growing interest in the search for potentially habitable planets in the Galaxy, working towards a full understanding of the most prevalent type of star has gained a sense of urgency. M dwarfs are prolific hosts of rocky, temperate planets \citep[e.g.][]{DC15,Sabotta21, Burn21,Ribas23}, and are the focus of several dedicated planet search programmes, including transiting planet surveys such as MEarth \citep{Nutzman08}, SPECULOOS \citep{Delrez18}, and EDEN \citep{Gibbs20}, and the radial velocity (RV) survey CARMENES \citep{Quirrenbach18}. 

When evaluating the veracity of a planet candidate detection as well as assessing the formation and evolution environment of confirmed planets, the properties and behaviour of the host star are paramount considerations. 
%Relevant stellar parameters include mass, radius, temperature, metallicity, as well as age, Galactic kinematics, multiplicity, and activity. 
A particularly important property is the rotation period ($P_{\rm rot}$). While a star's rotation may not by itself directly influence the planet's fate, it connects many of the other factors that do matter. For example, rotation controls the magnetic activity and field strength of the star, which in turn govern its chromospheric emission and the dosage of high-energy radiation received by orbiting planets, thus shaping their atmospheres \citep[e.g.][]{Lammer03, Ribas05, SanzForcada11, Owen12, Lampon21}. The rotational evolution of the host star therefore influences the history of a planet's evolution. Relatedly, the length of this history, that is, the ages of planet-hosting stars, can be inferred from $P_{\rm rot}$ through gyrochronological relations \citep[e.g.][]{Gaidos23,Engle23}. Moreover, the visibility of stellar surface inhomogeneities in the form of spots and faculae repeats with the periodicity of rotation and can be mistaken for, or distract from, planetary signals in both photometry and spectroscopy \citep[e.g.][]{Knutson11,Oshagh14,Perryman18,Rackham18,Cale21}.  

Quantifying stellar rotation is also crucial for understanding the nature of stars themselves. Measuring the spin evolution of stars can shed light on their interior structure and the principles of the magnetic dynamo and its coupling to angular momentum loss mechanisms. Rotation rates are linked to the age and activity level of stars through various gyrochronology and age-rotation-activity relationships \citep[e.g.][]{Barnes07,Wright11,Reiners14, Gaidos23,Engle23}. For example, rapid rotation is usually associated with youth and high levels of activity. There have been many recent efforts to characterize these relationships for late-type stars \citep[e.g.][]{Reiners08,McQuillan13,Newton16,Stelzer16,AD17a,Newton17,Newton18,SM18,Wright18,GA19,Raetz20,Magaudda20,Magaudda22,Medina20,Ramsay20,Popinchalk21,Boudreaux22,Pass22}. 
For instance, \citet{McQuillan13} showed a well-defined upper limit in $P_{\rm rot}$ for early-M dwarfs that depends monotonically and continuously on stellar mass. \citet{Newton16} pointed out that the period distribution of fully convective field M dwarfs hints at a rapid spin-down phase from $< 10$ d to $> 50$ d. 
Several authors noticed a very tight correlation between Ca~{\sc ii} H\&K emission and $P_{\rm rot}$ for slow rotators \citep{AD17a, SM18}, which is useful for predicting $P_{\rm rot}$ whenever $R'_{\rm HK}$ ($\equiv L_{\rm Ca~{\sc ii}~ H\&K}/L_{\rm bol}$) could be readily measured. According to \citet{Magaudda22}, the X-ray saturation regime shows hints of mass-dependent behaviour. However, few of these aforementioned studies involved by themselves a large sample that spans the entire range of M dwarf spectral subtypes, ages, and rotation periods, or had at their disposal stellar properties that have been comprehensively and uniformly characterized. 

As part of its planet search programme, the CARMENES\footnote{CARMENES: Calar Alto high-Resolution search for M dwarfs with Exoearths with Near-infrared and optical \'Echelle Spectrographs, \url{https://carmenes.caha.es}} survey has intensively monitored a sample of more than 300 nearby field M dwarf stars (M0--9\,V) using high-resolution spectroscopy during its guaranteed time observations \citep[GTO --][]{Quirrenbach14, Quirrenbach20, Alonso-Floriano15, Reiners18, Ribas23}. One important goal of the survey is to build a large set of benchmarks for studies of late-type stars and the planet population around them. This requires a comprehensive characterization of these stars. In order to accomplish this, we have constructed the Carmencita\footnote{Carmencita: CARMENES Cool dwarf Information and daTa
Archive.} catalogue \citep{Caballero16}, where we have assembled extensive information on every survey star, from the literature, archival data, as well as from our own observations. This includes astrometry, imaging, photometry, and spectroscopy, from which properties such as kinematics, fundamental stellar parameters, multiplicity, planet-hosting status, activity, age, and rotation are derived \citep[e.g.][]{CC17,Reiners18,Jeffers18,Passegger19,Schoefer19,Schweitzer19,Cifuentes20,Baroch21,Marfil21,Lafarga21,Perdelwitz21}. 

Stellar $P_{\rm rot}$ is  most readily measurable via modulations in light curves, but also increasingly commonly through time series of activity-sensitive spectroscopic signatures and in spectropolarimetry data directly sensitive to surface magnetic fields. For M dwarfs, substantial catalogues of $P_{\rm rot}$'s have been made available based on, for example, large transiting planet surveys from both space \citep[e.g.][]{Reinhold13,McQuillan14,Raetz20,Ramsay20,Gordon21,Magaudda21} and the ground \citep[e.g.][]{KS07,Hartman11,Newton16,SM16}, as well as from RV \citep[e.g.][]{SM15,SM17a} and spectropolarimetry surveys \citep[e.g.][]{Fouque23,Donati23}. Currently, for the CARMENES M dwarfs, the Carmencita catalogue compiles $P_{\rm rot}$ from a variety of literature sources. 
A significant contributor is \citet{DA19}, hereafter DA19, who measured $P_{\rm rot}$ for a 142 CARMENES survey stars from a homogeneous analysis of archival data taken by several photometric monitoring surveys. Complementary searches for periodicity in the CARMENES~GTO sample have also been undertaken using spectroscopic activity indicators \citep{Lafarga21,Schoefer22,Fuhrmeister23}. In addition, planet hosts often have their $P_{\rm rot}$ constrained in the discovery papers \citep[e.g.][and references therein]{Ribas23}. However, many stars in this prime sample are still missing $P_{\rm rot}$ determinations, and some stars have conflicting or implausible periods based on comparison with multiple references or with activity signatures. 

Since DA19, more photometric observations have become available for our sample. Here we continue DA19's endeavour using light curves from survey data as well as our own ground-based follow-up efforts to populate the $P_{\rm rot}$ catalogue for CARMENES survey stars. We also add periods from recent literature references, and perform a critical analysis of the reliability of the $P_{\rm rot}$ determinations for each star. Our new catalogue, therefore, builds on and supersedes DA19, and should be considered the reference $P_{\rm rot}$ catalogue for the CARMENES survey stars. Combined with newly available stellar parameters for this sample, we put $P_{\rm rot}$ for nearby M dwarfs in the context of activity, mass, age, and planet-hosting status. We examine the spread in published $P_{\rm rot}$ values and discuss realistic uncertainties in rotation period measurements. 

In this manuscript, we describe the CARMENES survey sample in Sect.~\ref{ss:sample}. In the rest of Sect.~\ref{s:data}, we present the data used to measure $P_{\rm rot}$, including light curves from SuperWASP, TESS, and ground-based follow-up observations, as well as time series of CARMENES spectroscopic indicators. We describe the methods for analysing the light curves from each data set in Sect.~\ref{s:analysis}. Our measurements are presented in Sect.~\ref{s:prots}. Informed by comparisons with the literature (Sect.~\ref{ss:prot_lit}, ~\ref{ss:lit_compare}), $v\sin i$ (Sect.~\ref{ss:vsini}) and H$\alpha$ (Sect.~\ref{ss:pew_ha}) measurements, we assign the most credible value for each target using a framework outlined in Sect.~\ref{ss:prot_adopted}. Sect.~\ref{ss:lit_errors} characterizes the scatter in multiple $P_{\rm rot}$ measurements and discusses implications for realistic uncertainties. Rotation-activity relations, including X-ray, H$\alpha$, $R'_{\rm HK}$, and magnetic field strengths are examined in Sect.~\ref{ss:rot-act}. Correlations with stellar mass, kinematics, and age are shown in Sect.~\ref{ss:mass_kinematics}. %Section~\ref{ss:longterm} shows examples of evolution in the rotational modulation of light curves spanning several years. 
The distribution of $P_{\rm rot}$'s for known planet hosts in this sample is discussed in Sect. \ref{ss:planethosts}. We summarize in Sect.~\ref{s:sum}.   

%===========================================
\section{Data}\label{s:data}
\subsection{Sample}\label{ss:sample}
    
\begin{figure}
\vspace{-0.5cm}
\begin{minipage}{0.5\textwidth}
    \centering
    \includegraphics[width=\hsize]{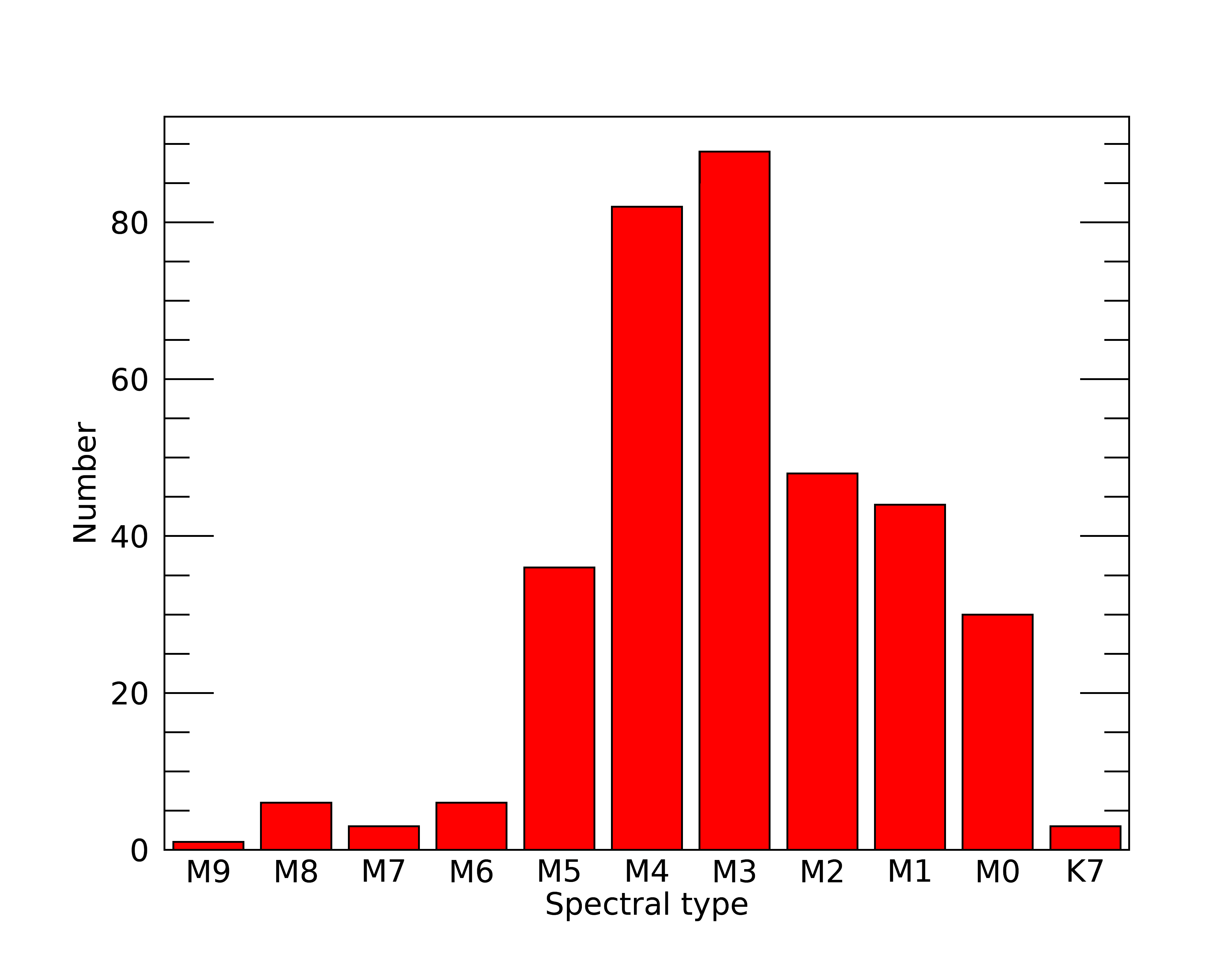}
\end{minipage}  
\vspace{-0.5cm}
\begin{minipage}{0.5\textwidth}
    \centering
   \includegraphics[width=\hsize]{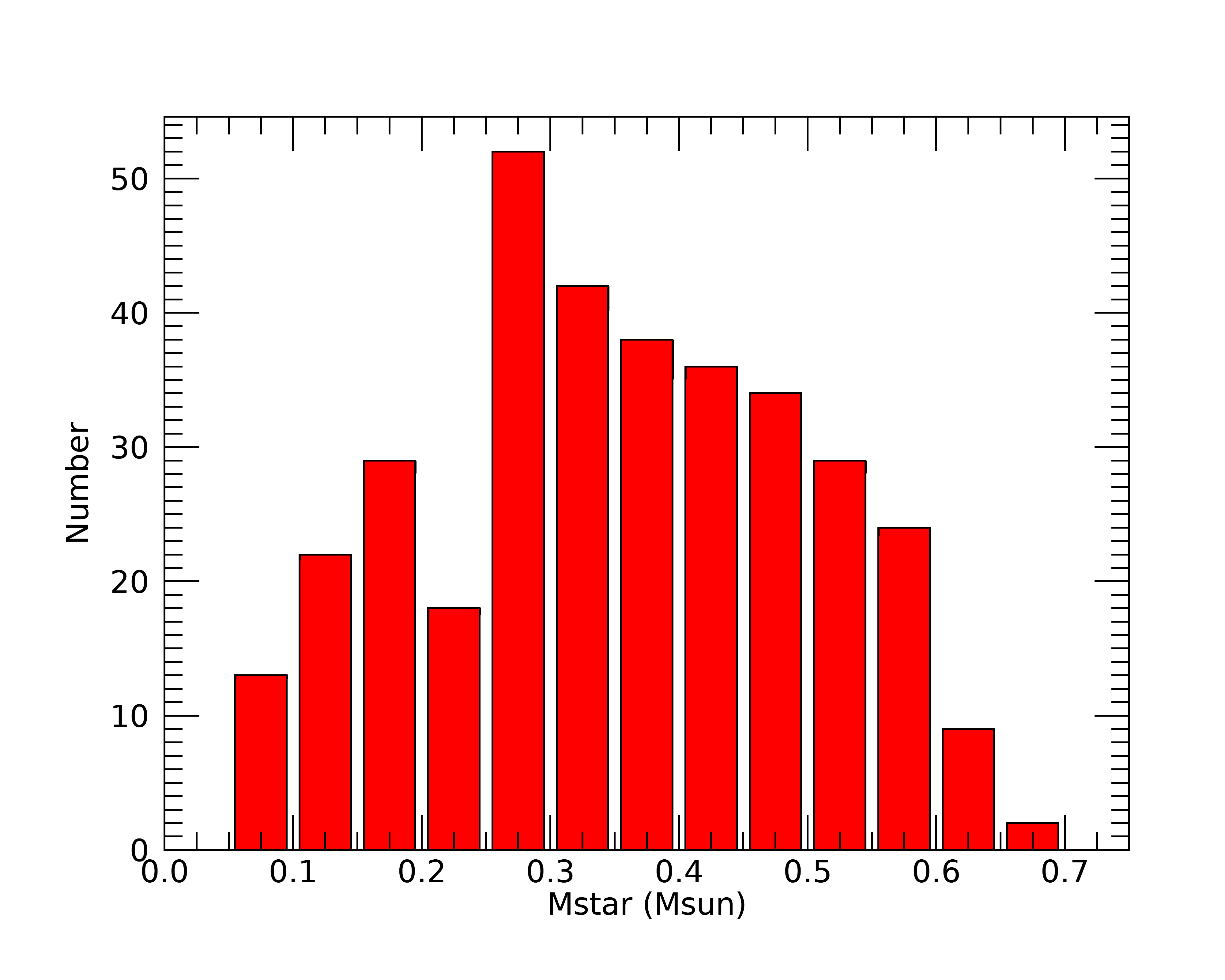}
\end{minipage}  
\caption{Distribution of spectral type (top) and stellar mass (bottom) of the CARMENES survey sample (total of 348 stars) studied in this work. }
\label{fig:spt}
\end{figure}

Our overall sample consists of 348 stars from the CARMENES RV survey, whose selection has been documented in previous works \citep[e.g.][]{Quirrenbach14,Alonso-Floriano15,Caballero16,Ribas23}. As shown in Fig.~\ref{fig:spt}, this sample spans the entire range of the M dwarf spectral type and masses. Most of these stars appear to be field stars, although 69 of them have been recently identified to belong to nearby stellar kinematic groups with ages $\lesssim 800$ Myr \citep{Cardona23}. 

Our goal is to compile as much information on the rotation periods of as many stars in this sample as possible. To this end, we first performed a literature search and found at least one rotation period estimate on record for 241 stars (see Sect. \ref{ss:prot_lit}). We also collected new photometric and spectroscopic data to help independently derive periods for a subset of the sample stars. Table \ref{t:surveys} presents an overview of the sources of our data and the numbers of light curves provided by each of them. More detailed descriptions are given below.\footnote{\HLr{Non-public light curves (i.e.
SuperWASP, AstroLAB, OSN, LCOGT, TJO) can be provided upon request to the first
author.}}  

\begin{table*}[htb]
\begin{center}
\caption{Overview of sources of light curve data for the targets in the CARMENES survey.}
\label{t:surveys}
\begin{tabular}{llllc}
\hline
\hline
\noalign{\smallskip}
Survey & Location & Instrument configuration & Band & No. of  \\
 &  &  &  & light curves\textsuperscript{a}  \\
\noalign{\smallskip}
\hline
\noalign{\smallskip}
  SuperWASP & Roque de los Muchachos, Spain & 8 $\times$ Andor DW436 & Clear, Broad (0.4--0.7\,$\mu$m) & 112 (99) \\
   & Sutherland, South Africa & & Broad (0.4-0.7\,$\mu$m) & \\
  AstroLAB IRIS & Zillebeke, Belgium & 0.68\,m Keller & $B$, $V$, $R$ & 6 (6) \\
   &  & + SBIG STL-6303E & & \\
  OSN & Observatorio de Sierra Nevada, Spain & 0.90\,m, 1.50\,m & $B$, $V$, $R$, $I$ & 22 (14) \\
  LCOGT & Las Cumbres Observatory, Chile & 0.40\,m & $B$, $V$, $i$ & 13 (9) \\ 
  TJO & Montsec Observatory, Spain & 0.80\,m + LAIA & $R$ & 5 (3) \\
  TESS & Space & 4 $\times$ 0.10\,m + CCID-80 & $T$ (0.6--1.0\,$\mu$m) & 240 \\
\noalign{\smallskip}
\hline
\end{tabular}
\end{center}
%Notes: \\
\textsuperscript{a}: The number of light curves with $>25$ nightly-binned epochs, i.e., deemed admissible for analysis, are given in parentheses. Applicable only to ground-based data sets.  \\
\end{table*}

\subsection{Ground-based photometry}\label{ss:ground_data}

\subsubsection{SuperWASP}\label{sss:swasp_data}
The Super-Wide Angle Search for Planets (SuperWASP) Survey is an all-sky photometric survey for transiting exoplanets \citep{Pollacco06}. Two robotic telescopes, each consisting of eight cameras, observe from La Palma (Spain) and Sutherland (South Africa), chiefly through broadband filters (400--700\,nm). Since 2004, they have monitored the entire sky at high cadence (weather-permitting) for about four months per year. The typical photometric precision is $\lesssim 1\%$ for stars with $V \lesssim 11.5$\,mag. The first and so far only public data release (DR1) made available data collected up to 2008 \citep{Butters10}. This was one of several public data sets used by DA19. Data post-2008 can be obtained via direct request to the SuperWASP team.   
We made such a request and received SuperWASP light curves for 112 CARMENES survey stars. 
The light curves were produced by the SuperWASP team after custom extraction and detrending according to the {\tt Sys-Rem} algorithm \citep{Tamuz05,Mazeh07}, which corrects for systematics arising from the instrument or atmospheric extinction, leveraging the manifestation of these effects in large sets of simultaneously monitored stars. This extraction method is expected to preserve astrophysical signals such as rotational modulation. 

The observations span up to nine seasons (though not necessarily contiguous), each of which typically lasts for $\sim 130$ days and occurs during the fall and winter months. We note that SuperWASP light curves with fewer than about 25 nights of data (i.e. $\lesssim 25$ nightly-binned data points) tend to be too sparsely populated to give credible signals. The number of light curves that exceeds this threshold (i.e. deemed potentially useful) is 99.

\subsubsection{AstroLAB IRIS}\label{sss:astrolab}

Optical observations were collected for six CARMENES survey targets using the 684\,mm aperture Keller $f$4.1 Newtonian New Multi-Purpose Telescope of the public observatory AstroLAB IRIS located at Zillebeke, Belgium\footnote{\url{http://www.astrolab.be}}. The telescope is equipped with a Santa Barbara Instrument Group (SBIG) STL-6303E CCD camera operating at $-20^\circ$C.  A 4-inch Wynne corrector feeds the CCD at a final focal ratio of 4.39 and provides a nominal field of view of $20\arcmin\times30\arcmin$.
The 9\,$\mu$m physical pixels project to $0\farcs62$ and are read out binned to $3\times3$ pixels, or $1\farcs86$ per
combined pixel. We used the $B$, $V$, and $R$ filters from Astrodon optics. These filters closely match the Johnson-Cousins system \citep{Goldman06}. Differential photometry relative to suitable comparison stars in the field of view were used to construct the light curves using the {\tt Lesve} photometry package. 
%\footnote{http://www.dppobservatory.net/AstroPrograms/Software4VSObservers.php\#LesvePhotometry)}

\subsubsection{Observatorio de Sierra Nevada (OSN)}

A number of targets were also photometrically monitored at Observatorio de Sierra Nevada (OSN)\footnote{\url{https://www.osn.iaa.csic.es/en}} in Granada, Spain, using the T90 and T150 telescopes and different set of filters ($B$, $V$, $R$, $I$) in the Johnson photometric system. T90 is a 90\,cm Ritchey-Chr\'etien telescope equipped with a CCD camera Vers\-Array 2k$\times$2k with a resulting field of view %(FOV) 
of $13\farcm2\times13\farcm2$. The camera is based on a high quantum efficiency back-illuminated CCD chip, type Marconi-EEV CCD42-4, with optimized response in the ultraviolet \citep{Amado21}. The T150 telescope is a 150\,cm Ritchey-Chr\'etien telescope equipped with a CCD camera Andor Ikon-L DZ936N-BEX2-DD 2k$\times$2k, with a resulting field of view
of $7\farcm92\times7\farcm92$. The camera is based on a back-illuminated CCD chip, with high quantum efficiency from ultraviolet to near infrared. This camera also includes thermo-electrical cooling down to --100$^\circ$C for negligible dark current \citep{Quirrenbach22}. 

The data were reduced in the same way for both telescopes. All CCD measurements were obtained via synthetic aperture photometry using typically a 1$\times$1 binning (no binning). Each CCD frame was bias subtracted and flat-fielded in a standard way. Different aperture sizes were tested in order to choose the best one for our observations. A number of nearby and relatively bright stars within the frames were selected as check stars, and the best set is chosen to be used as reference stars. The data in each filter are presented as magnitude differences normalized to zero. Outliers due to bad weather conditions were previously removed. In particular, we used the same methodology as in previous works involving photometric monitoring of nearby M dwarfs with exoplanets \citep[e.g.][]{Amado21,Quirrenbach22}. Of the 22 stars for which we have OSN light curves, 14 have $\ge$ 25 nightly-binned epochs and were retained for analysis.

\subsubsection{Las Cumbres Observatory (LCOGT)}

Part of our sample was observed in the $B$, $V$, and $i^{'}$ bands using the 40\,cm telescopes of the Las Cumbres Observatory Global Telescope (LCOGT) network \citep{Brown13}, during different campaigns between 2016 and 2021 under the Instituto de Astrof\'isica de Canarias programmes IAC2016A-004, IAC2017AB-001, IAC2018A-001, IAC2018B-005, IAC2019A-001, IAC2020A-001, IAC2020B-001, and IAC2021B-002 (PI: V.\,J.\,S.~B\'ejar). The 40\,cm telescopes are equipped with a 3k $\times$ 2k SBIG CCD camera with a pixel scale of $0\farcs571$ providing a field of view of $29\farcm2\times19\farcm5$. Sky transparency conditions at the observatories were mostly clear during our observations, and the average seeing varied from one to a few arcseconds.
Raw data were processed using the {\tt BANZAI} pipeline \citep{Mccully18}, which includes bad pixel, bias, dark, and flat field corrections for each individual night. We performed aperture photometry for our targets and several reference stars in the same field of view, and estimated the relative flux between the target and references in order to derive the photometric light curves. We selected the most appropriate aperture for each target (typically around 10 pixels or $5\arcsec$) that minimizes the dispersion of the differential photometry. Nine of the 13 stars for which LCOGT light curves were collected have more than 25 nightly-binned epochs and are used for period analysis.

\subsubsection{Montsec Observatory (TJO)}

Part of the photometric monitoring was done with the Joan Or\'{o} Telescope (TJO) at the Montsec Observatory in Lleida (Spain).\footnote{\url{https://montsec.ieec.cat/en/}} The TJO is a 0.8\,m robotic telescope equipped with the LAIA instrument, a 4k $\times$ 4k back illuminated CCD camera Andor iKon XL with a pixel scale of $0\farcs4$ and a square field of view of $30\arcmin$. The instrument is also equipped with a set of Johnson $UBVRI$ photometric filters.

The photometric measurements of the TJO were collected between 2019 and 2022. For all the targets, the observational strategy consisted of executing blocks of five images every two or three nights during the visibility period throughout the year. All the observations were done using Johnson $R$ filter and 1$\times$1 binning. The resulting images were calibrated with darks, bias, and flat fields with the {\tt ICAT} pipeline \citep{Colome06} using standard procedures. Differential photometry was extracted with {\tt AstroImageJ} \citep{Collins17} using the aperture size and the set of comparison stars that minimized the root-mean-square (rms) of the photometry. Out of five targets with TJO observations, we rejected two for having fewer than 25 nightly-binned epochs, leaving three for analysis.

\subsection{TESS}\label{ss:tess_data}

The Transiting Exoplanet Survey Satellite (TESS) mission is a space-based all-sky photometric survey to look for transiting planets around bright stars \citep{Ricker15}. This mission divides the entire sky into 26 sectors and observes each field continuously at high cadence (2 or 30 minutes) for 27.4 days at a time. The first 26 sectors constituted the primary mission and were completed between July 2018 to July 2020. Subsequently, most fields are being revisited. In this work, we used data from the primary mission, covering the southern and most of the northern hemispheres. For the stars that already have periods recovered from the first 26 sectors, we also checked data from later sectors (up to sector 47) to ensure consistency.\footnote{With the exception of J10360+051, which was not in the primary mission, we determined its period based on \HLr{SAP data from sectors 45 and 46.}}

We cross-matched the stars from the CARMENES sample with those from version 8 of the TESS Input Catalog (TICv8) using {\em Gaia} DR2 identifiers, when available, and otherwise via 2MASS identifiers. Of them, 240 of the survey stars were observed in the first 26 sectors. The remaining 109 stars lie in missing patches at the edges of adjacent sectors and in the ecliptic region, which was not observed until sector 42.

For our analysis, we used chiefly the 2-minute-cadence light curve products from TESS: Simple Aperture Photometry (SAP) and Pre-search Data Conditioning SAP (PDCSAP). These were available for 210 stars in our sample. For the 30 stars that do not fall into the pre-selected TESS target list but are captured in the 30-minute-cadence Full Frame Images (FFIs), we custom extracted their light curves. All data were downloaded via the Mikulski Archive for Space Telescopes\footnote{\url{https://mast.stsci.edu/portal/Mashup/Clients/Mast/Portal.html}}. The data were reduced using the Python packages \texttt{Lightkurve} \citep{LightkurveCollaboration} as well as \texttt{tesscut} \citep{tesscut}. Below we describe how each data product was used. 

\subsubsection{PDCSAP light curves}\label{sss:pdcsap}
Since the PDCSAP is optimized for planetary transit searches \citep{Kinemuchi2012}, it is most useful for detecting variability with timescales of less than a few days (see also discussion in Section \ref{sss:pdcsap_analysis}). 
%These light curves tend to be readily usable with only minimal additional processing. 
In the majority of cases, we used them as-is. 
For a few stars where the light curves show effects of possible contamination from nearby sources, we examined their raw Target Pixel Files (TPF) and refined the aperture masks to optimize the target's summed light curve. In these cases, the light curves were not detrended prior to analysis. 

\subsubsection{SAP light curves}\label{sss:sap}
\begin{figure}
\centering
\includegraphics[width=\hsize]{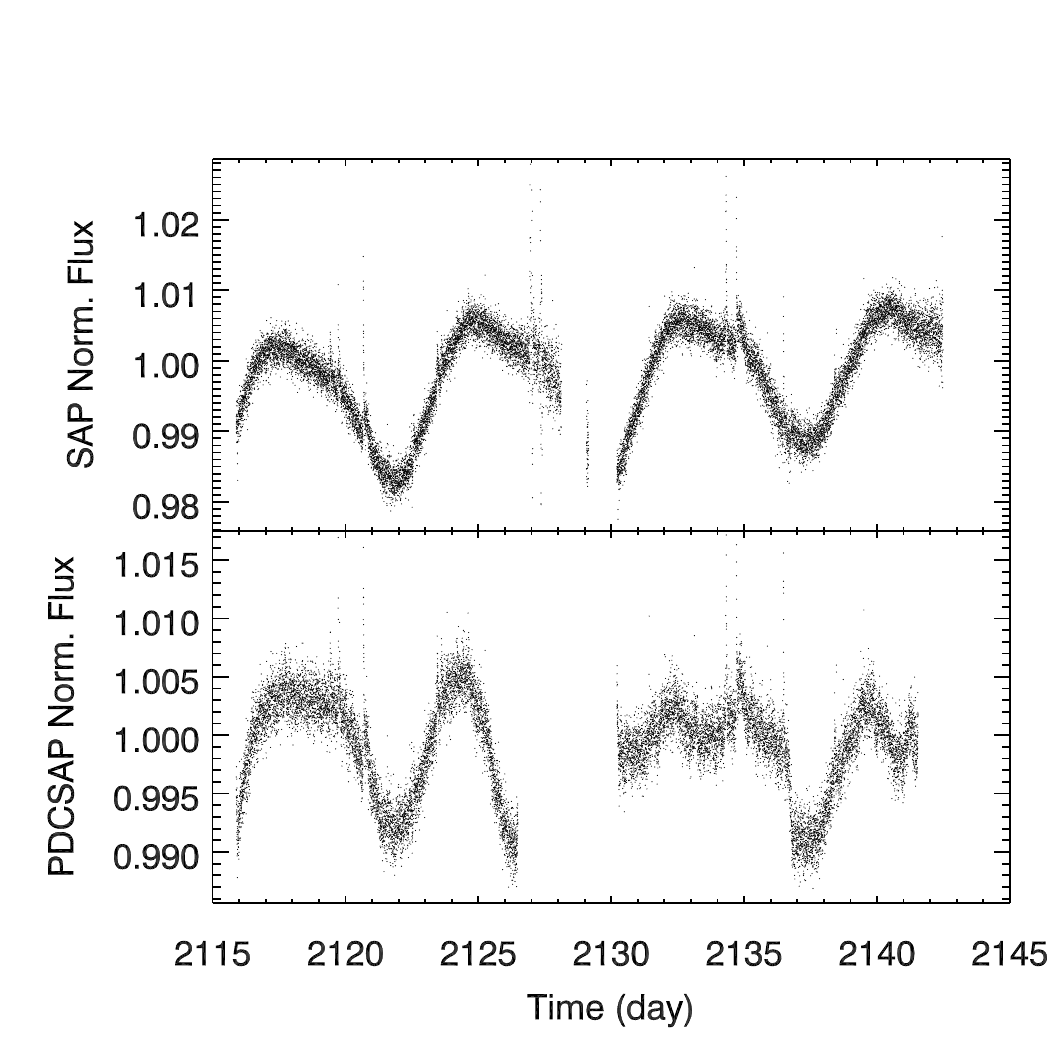}
\caption{SAP (top) and PDCSAP (bottom) light curves for J01339$-$176 from TESS sector 30. A periodic modulation at $\sim8$\,d is apparent in the SAP, but not in the PDCSAP. The time in the X axis is BJD -- 2\,457\,000 days.}
\label{fig:J01339-176}
\end{figure}

Despite being the `rawer' predecessor of the PDCSAP that can show large, instrument-related systematics and glitches, the SAP light curves are more suitable for measuring or indicating periods between $\sim 5$ and $\sim30$ days. Figure~\ref{fig:J01339-176} shows the example of J01339$-$176, whose SAP light curve exhibits an unambiguous modulating pattern repeating at $\sim 8$\,d, whereas its PDCSAP light curve appears to be more irregular. 
%another example: J05365+113, J07033+346 

\subsubsection{FFI light curves} \label{sss:ffi}
Using the \texttt{Python} tool \texttt{tesscut}, we recovered the 30-min cadence light curves for the 30 stars that did not have SAP. %\HL{(30 = 28 + 2 that saturated the detector)} for 28 of the 32 GTO stars that did not have SAP.
For each target star, we cut a 20$\times$20 pixel region of the FFI around its position to create a custom TPF. Since these TPFs do not come with a predefined optimal aperture mask, we generated them ourselves. We used \texttt{lightkurve}'s \texttt{create\_threshold\_mask} method, which draws an aperture according to a user-given flux threshold value. For each target, we visually inspected its TPF using {\tt{tpfplotter}} \citep{Aller20} to determine the optimal threshold value to use for creating the aperture mask. For some stars with bright close visual companions, we manually adjusted the aperture boundaries to best exclude the companions. 
Light curves were extracted from summing the pixels in the apertures.

\subsection{Spectroscopic indicators}\label{ss:spectro_ind}

Various spectroscopic indicators sensitive to stellar activity have been demonstrated to exhibit variability linked to the stellar $P_{\rm rot}$. \citet{Schoefer19,Schoefer22} and \citet{Lafarga21} have already analysed a subset of CARMENES stars in this way. Here, we collected updated spectroscopic indicator time series from the CARMENES survey (up to February 2022) to form a sample of 205 stars with more than 20 epochs of CARMENES spectra \citep[see][for more details]{Fuhrmeister23}. 
As chromospheric indicators we used the following features available in the CARMENES spectral range, which were demonstrated by \citet{Schoefer19} to most likely show detectable variability related to rotation: pseudo-equivalent width ($pEW$) of H$\alpha$ and the two bluer \ion{Ca}{ii} infrared triplet (IRT) lines (Ca-IRTab), following \citet[][see their Table 2 for the integration bands used]{Fuhrmeister19}, as well as the TiO bandhead index at 7050 \AA, defined as the quotient of the integrated flux density in two wavelength bands on both sides of the bandhead \citep[see Table 3 in][]{Schoefer19}. In addition, we used \ion{Ca}{ii} H\&K data as measured by $R'_{\rm HK}$, whose variability has also been associated to rotation \citep[e.g.][]{SM15}. $R'_{\rm HK}$ cannot be measured from the CARMENES spectra, which does not span the blue wavelength range. Rather, we used archival data from a collection of different telescopes, mainly HARPS \HLr{(High Accuracy Radial velocity Planet Searcher)} and HIRES \HLr{(High Resolution Echelle Spectrometer)}, as compiled by \citet{Perdelwitz21}. The subsample with $R'_{\rm HK}$ time series containing > 20\,epochs consists of 56 stars. 

%===========================================
\section{Data analysis}\label{s:analysis}

This section describes the procedures for processing and time series analysis of each dataset listed in Section \ref{s:data}. All retrieved $P_{\rm rot}$'s and other applicable information (e.g. uncertainty flags for the SuperWASP analysis) are given in Table \ref{t:allprots}. 

\subsection{Ground-based photometry}\label{ss:ground_analysis}

\subsubsection{SuperWASP}\label{sss:swasp_analysis}

The SuperWASP light curves were first filtered for outliers on a season-by-season basis. The standard deviation ($\sigma$) threshold used for iterative outlier rejection was set to the number of $\sigma$'s away from which we expect only one outlier for a given number of data points present in the data set assuming Gaussianity, as determined from an inverse error function. This requirement translates into a typical threshold of $3\sigma$ to  $4\sigma$. The light curves were subsequently nightly-binned, where each bin was assigned a time and flux value corresponding to the error-weighted mean of the points that fell in the bin. The flux error of the bin was computed as the standard error of the weighted mean. By default, analysis was carried out on nightly-binned data. For 17 H$\alpha$-active stars whose periods were suspected to be short, we also considered their unbinned light curves. Here we adopted the definition of H$\alpha$-active from \citet{Schoefer19}: $pEW'_{\rm H\alpha} < -0.3 \AA$, where $pEW'_{\rm H\alpha}$ is the H$\alpha$ pseudo-equivalent width measured with spectral subtraction from that work (see also Sect. \ref{ss:pew_ha}). Eight of the 17 stars have compelling evidence of a period $\gtrsim$ 10\,d, usually visible in the light curve itself and supported by multiple literature measurements. For those stars, the final analyses were performed on nightly-binned data.    

We constructed error-weighted Generalized Lomb-Scargle (GLS, \citealt{Zechmeister09}) periodograms from the light curve data, using the IDL implementation of GLS \footnote{\url{https://github.com/mzechmeister/GLS/blob/master/idl/gls.pro}}. The peaks of the GLS occur at the most salient sinusoidal frequencies in the data, whose significances are assessed from their false alarm probability (FAP). The FAP of a given measured peak power $P_n$ was calculated as 
\begin{equation}
{\rm FAP} (P_n) = 1 - [1-{\rm prob} (P > P_n)]^M, \label{eqn:fap}
\end{equation}
\noindent where ${\rm prob} (P > P_n)$ is the probability that a signal with power $P$ is greater than the threshold power $P_n$, and $M$ is the number of independent frequencies being tested. We used the following prescription:
\begin{equation}
{\rm prob} (P > P_n) = (1-P_n)^{(N-3)/2}
\label{eqn:prob_p}
\end{equation}
\noindent and $M = \Delta f \Delta t$, where $\Delta f$ is the range of frequencies searched and $\Delta t$ the total time baseline. $N$ is the number of data points in the time series. 
By default, we searched a range of frequencies from 1\,d$^{-1}$ to $10^{-3}$\,d$^{-1}$, concordant with the inverse of the typical \HLr{time} span of SuperWASP data ($\sim$ 2--4 seasons), such that $\Delta f \approx 1$\,d$^{-1}$. In a few cases where a very short period was expected, we searched from 10\,d$^{-1}$ in the unbinned light curve data. 

Since sampling from ground-based instruments can be opportunistic and the photometry noisy, the periodograms are often difficult to interpret and require analysis based at least in part on qualitative principles. We visually examined each GLS periodogram for significant peaks with FAP $<$ 0.1 \%. For objects with more than one season of coverage, we looked for periodicities in the full global data set as well as on a season-by-season basis. When a period is very apparent in one well-sampled season, we accepted it even if it is not found with as much significance in other seasons. Many periodograms have multiple peaks exceeding our FAP threshold. In these cases, we evaluated the plausibility of each period signal by a visual inspection of the corresponding phase-folded light curves, aided by comparison to any reported periods in the literature and activity signatures (see Sect.~\ref{s:prots}). Daily aliases are very common and generally manifest themselves as a peak close to 1\,d mirroring a strong peak at a longer period. Therefore, periods below 1.5\,d were usually dismissed in favour of any identifiable counterpart peak structure symmetric across the 0.5\,d$^{-1}$ axis, unless there were strong reasons to suspect fast rotation, as evidenced by $pEW'_{\rm H\alpha}$.    
 
In general, we considered a detection to be valid when it manifests itself as a single dominant peak, usually with FAP $\ll$ 0.1\%, either in global or in seasonal data (at least one season with more than 25 nightly-binned epochs), with the periodicity clearly visible by eye in phase-folded data. The period should not exceed 130 d (i.e. the average length of a season) in a global analysis consisting of more than one season, or half the seasonal baseline in a seasonal analysis. 

For several signals for which we are less confident, for example where multiple periods of comparable significance are present, or the formal significance is strong but the period is barely visible in the phase-folded data, we flagged them as `U' for uncertain, to denote a possible signal following the nomenclature of \citet{Newton16}. Among them are three periods that are uncomfortably close to the synodic lunar period (see below). We report these signals because they may help verify measurements in future works.

Four stars have periodicities of $\sim$ 28\,d as their most convincing signals. Since this coincides with the synodic lunar month, these periods received extra scrutiny. J13102+477 shows a 29.06\,d period, which agrees well with the measurement from MEarth (28.8\,d, \citealt{Newton16}). J23556$-$061, which is a single-lined spectroscopic binary (SB1), shows a strong 28.53\,d periodicity in one of SuperWASP seasons, driven by a repeating distinctly sharp trough pattern that is uncharacteristic of star spot modulation. However, with an orbital period > 5000 d \citep{Baroch21}, the binary nature of J23556$-$061 is unlikely to explain the 28-d variability. J10584$-$107 has a strong 27.87\,d signal in one season and a weak 28.47\,d signal in another. However, this star has a resolved $v\sin i = 2.8$\,km\,s$^{-1}$ \citep{Reiners22} and, therefore, it is suspected to be a fast rotator. In the SuperWASP data, J04376$-$110 shows a highly significant 28.64\,d period in its best-covered season, but a 36.69\,d signal is present in another season (albeit at lower significance and with poorer coverage). We report the $\sim$ 28\,d periods of all four stars, but flagged the latter three with `U'. 

We retrieved periods for 44 objects, with 34 considered to be reliable and 10 being uncertain, three of which due to similarity with the synodic month. 
An example of the SuperWASP analysis is shown in Fig.~\ref{fig:J22330+093_SWASP} for J22330+093. This light curve spans three seasons, the latter two of which have superior coverage and exhibit significant modulations with a period of $\sim37$\,d. We took the most significant signal from the second season to be the $P_{\rm rot}$ for this star.       

\subsubsection{Other ground-based surveys}\label{sss·ground_analysis}

We repeated the analysis on the light curves collected from the other ground-based facilities. We treated each data set from each telescope and bandpass separately, but we did not split any light curve into seasons. In nightly-binned data, we looked for periodicities with formal significance FAP < 0.1 \% and below
half the total length of the corresponding data set. If a qualifying signal is found in multiple data sets for the same star, we recorded the one with the lowest FAP. In total, \HL{12} periods were determined in this manner.

%-------------------------------------------
\subsection{TESS}\label{ss:tess_analysis}

\subsubsection{PDCSAP}\label{sss:pdcsap_analysis}
To prepare the TESS PDCSAP light curves for analysis, points that deviated by more than $5\sigma$ from the mean magnitude, mostly comprising flares and transit features, were removed on a sector-by-sector basis. Then, we used the GLS periodogram as in Sect.~\ref{sss:swasp_analysis} to find significant periods in the unbinned data. We searched for periods on a sector-by-sector basis, limiting the period search to a maximum of $\sim 13$\,d (i.e. half of the total time baseline per sector) and a minimum corresponding to the Nyquist frequency. We located all peaks in the periodograms with FAP $<1\%$ and analyzed them separately. We assessed the veracity of each period by visually inspecting the TPF, the GLS periodogram, and the overall and phase-folded light curves. Figure \ref{fig:J05337+019} shows an example `summary page' containing the relevant information used for diagnostics. If the periodogram displayed several significant peaks, we compared the light curve phase-folded on each candidate period to pinpoint the one that showed the clearest modulation signal at minimal scatter. In addition, we checked whether the detected period was consistent across sectors, when applicable, and was compatible with the rotation velocity $v \sin i$ \citep{Reiners18} or activity indicators, such as the $pEW'_{\rm H\alpha}$ \citep{Schoefer19}. 

Through this vetting process, we recovered \HL{41} periods from the PDCSAP light curves, all of which are $< 6$\,d. 
%When multiple sectors yield similar periods, \HL{(explain error assessment)}. 
In some cases, significant peaks were found beyond 6\,d, but closer inspection revealed them to be spurious and related to corrections introduced by producing the PDCSAP.  
We conclude that $\sim 6$\,d is the limit to which rotation modulation can be accurately represented in the PDCSAP. A similar conclusion was reached by \citet{Medina20}.

\subsubsection{SAP}\label{sss:sap_analysis} 

Whereas the PDCSAP destroys astrophysical variability above a few days, the SAP is unfiltered and therefore better preserves longer-period signals (see e.g. Fig.~\ref{fig:J01339-176}). We stitched the SAP light curves across intra- and inter-sector data gaps (where applicable) using median flux normalization. We used a GLS periodogram to obtain the preliminary $P_{\rm rot}$ of the star, which we confirmed and refined by visual inspection of the phase-folded light curve. 
%The typical error on the SAP periods is $\sim 10$\%.

From the SAP we measured $P_{\rm rot}$ for an additional 17 stars. All of these periods lie between 6\,d and 25\,d and show high consistency with literature values and activity signatures (when available). Therefore, the SAP light curves can be valuable for constraining the periods of intermediate-rotators where the PDCSAP would fail. This could be especially useful for mapping the transition region between active and inactive M dwarfs.

\subsubsection{FFI}\label{sss:ffi_analysis}

Analogous to the PDCSAP, we analyzed the 30-minute cadence light curves custom-extracted from the FFIs, from which we recovered two periods: for J05366+112 and J06574+740. J06574+740 was later observed in sectors 40 and 47, where the PDCSAP data confirmed the period from the FFI analysis  (see further discussion in Sect. \ref{a:vsini_outliers} and Fig. \ref{fig:J06574+740}).

\subsection{Spectroscopic activity indicators}\label{ss:spectro_ind_analysis}

To each time series of the chromospheric indicators, a 3$\sigma$-clipping was first applied to omit outliers due to flaring or weather and instrumental issues. Here, we used the GLS periodogram as implemented in {\tt PyAstronomy}\footnote{\url{https://github.com/sczesla/PyAstronomy}} \citep{pya} to search for periods longer than 1.5\,d (to exclude daily aliases) and shorter than 150\,d. We defined periods with FAP $<$ 0.1\% as significant. For a given star, we required that the top significant period in a majority (minimum two) of indicators to agree within 15\% to be associated with the rotation of the star. In this case, we designated the period as the simple average of all the agreeing period values.

In all except one case, the data sets yielding acceptable periods contain at least $25$ epochs. For J09468+760, the signal in three of the indicators show high significance and consistency in 24 epochs of data, and therefore we adopted the corresponding period. For several stars, we detrended the time series with a degree-two polynomial, because they revealed more convincing and plausible signals. Figure~\ref{fig:J14524_spectro_inds} shows an example of the spectroscopic indicator time series and periodogram for J14524+123, for which the H$\alpha$ and Ca-IRT exhibit significant signals at $\sim 23$\,d. A total of 44 stars have an admissible period measured in this manner.

%========================================
\section{Results}\label{s:prots}

In this section, we combine results from our analysis with work by previous authors (Sect. \ref{ss:prot_lit}) to provide the most comprehensive and up-to-date catalogue of $P_{\rm rot}$'s for the CARMENES M dwarf sample. Where applicable, we compare our measurements with the literature, examine several types of disagreements and consider how to decide between multiple discrepant values (Sect. \ref{ss:lit_compare}). For stars with resolved $v\sin i$ and stellar radius ($R_\star$) measurements, an upper limit on $P_{\rm rot}$ can be calculated. We verify that most of the measured $P_{\rm rot}$'s are consistent with their implied upper limits, and argue that exceeding this limit should arouse suspicion (Sect. \ref{ss:vsini}). We also show $P_{\rm rot}$ correlates with $pEW'_{\rm H\alpha}$ in a way that could help identify some types of spurious periods (Sect. \ref{ss:pew_ha}). Informed by these lessons, we devise a framework to choose the most reasonable $P_{\rm rot}$ for each star and assign a confidence rating based on available information (Sect. \ref{ss:prot_adopted}). We show the dispersion among multiple measurements and discuss typical errors as a function of $P_{\rm rot}$ length (Sect. \ref{ss:lit_errors}).

\subsection{Rotation periods from the literature}\label{ss:prot_lit}

Since the CARMENES survey sample consists chiefly of nearby stars, many have been monitored as part of photometric and spectroscopic surveys in addition to the ones presented here. For 241 stars in our sample, we found at least one, and often multiple period measurements in the literature. Our main references include $P_{\rm rot}$ catalogs from photometry \citep{KS07,Hartman11,Irwin11,Kiraga12,Newton16,Newton18,SM16,Oelkers18,DA19,Medina20,Medina22b,Irving23,Pass23}\footnote{We considered A, B, and U-rotators in \citet{Newton16, Newton18}, where `A' and `B' denote reliable periods and `U' represents uncertain.}, spectroscopic indicators \citep{SM15,SM17a,SM18,Lafarga21}\footnote{For stars from \citet{Lafarga21}, we only considered those for which at least two activity indicators exhibit periodicity with FAP < 0.1\% consistent to within 15\%.}, spectropolarimetry \citep{Morin08, Morin10,Donati08, Donati23,Hebrard16,Moutou17,Fouque23}, as well as planet discovery papers, which sometimes take hybrid approaches via combined analysis of diverse types of data sets. 
All literature periods and their references are given in Table \ref{t:allprots}.

\subsection{Rotation periods derived in this work}\label{ss:lit_compare}

\begin{figure}
\centering
\vspace{-1cm}
\hspace{-1cm}
\includegraphics[width=1.1\hsize]{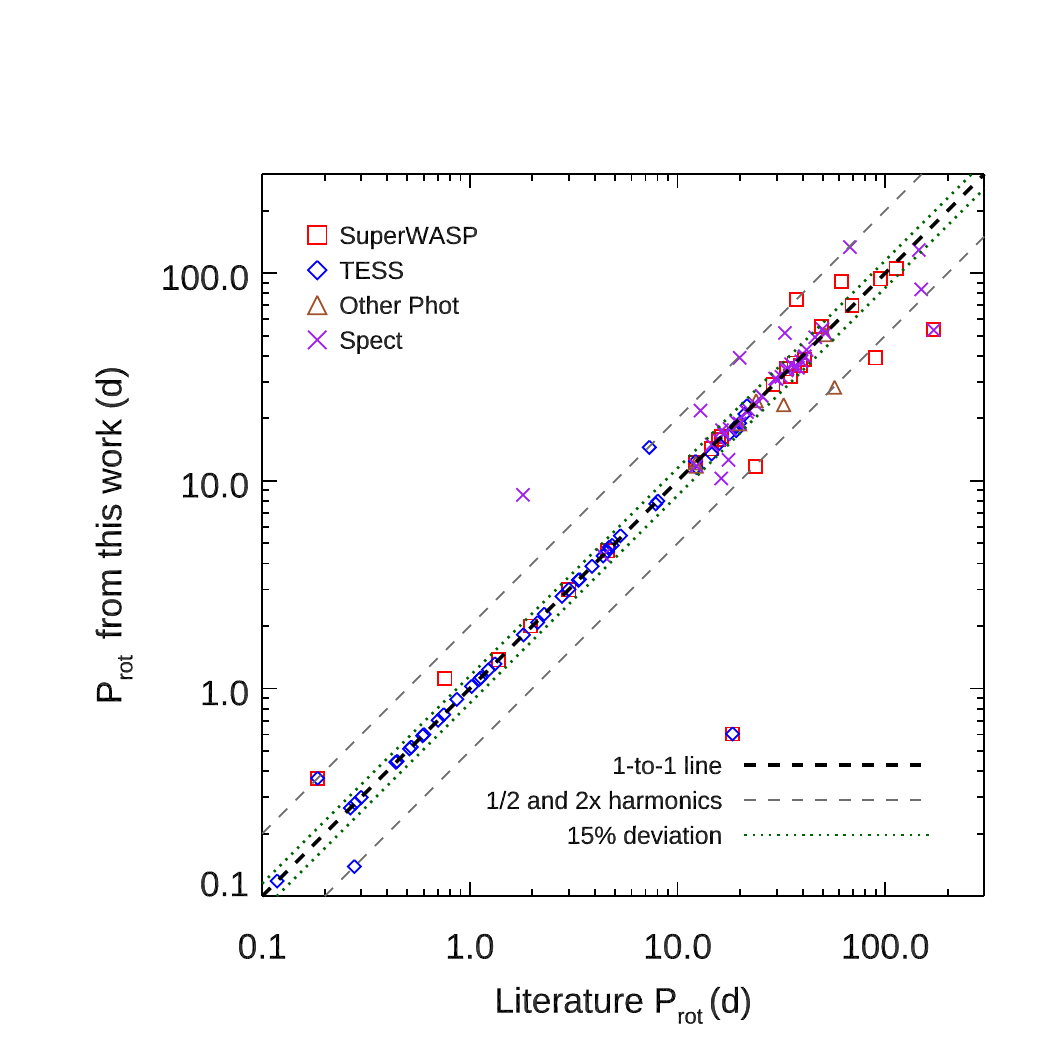}
\caption{Comparison between $P_{\rm rot}$'s determined in this work and from the literature for stars in common. The red squares, blue diamonds, brown triangles, and purple `X's represent periods from SuperWASP, TESS, other ground-based photometry, and spectroscopic indicators, respectively. The majority of periods are consistent with each other and are found directly on the one-to-one line (black dashed line) or within 15\% (green dotted line). The clear outliers from the 1-to-1 correspondence, some of which are obvious 1/2$\times$ and 2$\times$ harmonics (grey dashed lines), are discussed in Sect. \ref{a:lit_outliers}. Note that some stars have periods determined from independent data sets in this work. }
\label{fig:lit_compare}
\end{figure}

Following Sect.~\ref{s:analysis}, we recovered periodicities for 129 stars in our sample. Of these, 89 have at least one independent measurement from the literature. It is instructive to compare these periods, as in Fig. \ref{fig:lit_compare}. 
For simplicity, in the case where multiple periods have been published, we plot the one deemed most reasonable (i.e. consistent with other periods and activity signatures). 

While most $P_{\rm rot}$'s from this work are in excellent agreement with the literature, some of them deviate from the one-to-one relation by more than 15\%. Of these, 7 are likely harmonics of each other and \HL{12} are otherwise discrepant. The discrepant values are most often associated with the chromospheric activity indicators, suggesting that these periods may be less reliable than their photometric counterparts and should be used with more caution. We discuss these outliers individually in Sect. \ref{a:lit_outliers}. Therein we also document our reasoning for picking the most plausible period when presented with multiple measurements. 

The simplification of using one literature $P_{\rm rot}$ per star in this comparison entails that not every discrepancy with and between the literature values can be fully described in the text. This subsection and the outlier list in Sect. \ref{a:lit_outliers} are meant to give a representative view of the types of disagreements that occur. We direct the reader to Sect. \ref{ss:prot_adopted} to see how contesting periods are evaluated in this work to help determine the most recommended value, and to Sect. \ref{ss:lit_errors} for a discussion on the general level of consistency in independent period measurements and its implications.   

\subsection{Consistency with $v\sin 
 i$}\label{ss:vsini}
We compare the $P_{\rm rot}$'s for the CARMENES sample against the upper limits set by resolved $v\sin i$ measurements, where available (i.e. predominantly for fast rotators), together with the $R_\star$, determined from effective temperature ($T_{\rm eff}$) and bolometric luminosity ($L_{\rm bol}$) measurements according to the methods of \citet{Schweitzer19}. Such a comparison gives an independent indication of the plausibility of the documented $P_{\rm rot}$'s for the fastest rotators in the sample. The relation between $P_{\rm rot}$, $v \sin i$, and $R_\star$ is given by Equation \ref{eqn:prot_max}, by which we also define the quantity $P_{\rm rot, max}$:

\begin{equation}
P_{\rm rot} \le \frac{P_{\rm rot}}{\sin i}=\frac{2\pi R_\star}{v\sin i} \equiv P_{\rm rot, max}.
\label{eqn:prot_max}
\end{equation}

The majority of our $v\sin i$ values were uniformly measured from high-resolution CARMENES spectra ($\mathcal{R}\sim90\,000$) by \citet{Reiners22}, taking into account line broadening effects such as the Zeeman effect. Since $v\sin i$ values $\le 2$\,km/s cannot be reliably recovered in these spectra \citep{Reiners18}, we replaced all $v\sin i$ with nominal values below 2\,km/s with an upper limit of 2\,km/s and did not use them for computing $P_{\rm rot, max}$. We assigned an uncertainty of at least 2\,km/s or 10\% of the measured value, whichever is larger, to capture modelling errors in rotation profile, limb darkening, instrumental broadening and turbulence. For stars without a $v\sin i$ in \citet{Reiners22}, we used values from the literature. We gave preference to $v\sin i$ values measured by \citet{Reiners18}. We also discarded all $v\sin i$ measurements of known double-lined spectroscopic binaries (SB2's). 

From the same spectra, \citet{Marfil21} measured $T_{\rm eff}$ using spectral synthesis fitting. $L_{\rm bol}$ values were computed following the methodology of \citet{Cifuentes20} using the latest Gaia eDR3 data \citep{Gaia2021}. $R_\star$ was then derived via the Stefan-Boltzmann Law combining $T_{\rm eff}$ and $L_{\rm bol}$. In general, stellar mass ($M_\star$) was calculated using a mass-radius relation given by \citet{Schweitzer19}, which is anchored in eclipsing binaries. Since the fast rotators are predominantly young and active, inferring robust stellar parameters from spectra is more challenging and errors in $T_{\rm eff}$ could affect the accuracy of $R_\star$ for this subsample. We mitigated this effect for the set of stars which we could assign to nearby young moving groups (and therefore estimate an age) based on their Galactic kinematics \citep{Cardona23}. For those stars, we reevaluated their $M_\star$ and $T_{\rm eff}$ using their bolometric luminosities and PARSEC isochrones \citep{Bressan12} for the corresponding ages. We derived the radii from these parameters, again via the Stefan-Boltzmann Law. Since in this case $R_\star$ is model-dependent and only indirectly measured, a comparison against the measured $P_{\rm rot}$ values is in a sense also a consistency check for the stellar radius values listed for this sample. For known SBs in the sample as reported by \citet{Baroch18,Baroch21}, we adopted the estimated mass and radius of the primary component in computing $P_{\rm rot, max}$, since we assumed that the measured rotation period can be attributed to the brighter companion. Table \ref{t:bestprots} tabulates the $v\sin i$, $R_\star$, $M_\star$, and $P_{\rm rot, max}$ for this sample. The typical formal uncertainty on $R_\star/R_\sun$ and $M_\star/M_\sun$ are 0.01--0.03. 
 
\begin{figure}
\centering
\vspace{-1cm}
\hspace{-1cm}
\includegraphics[width=1.1\hsize]{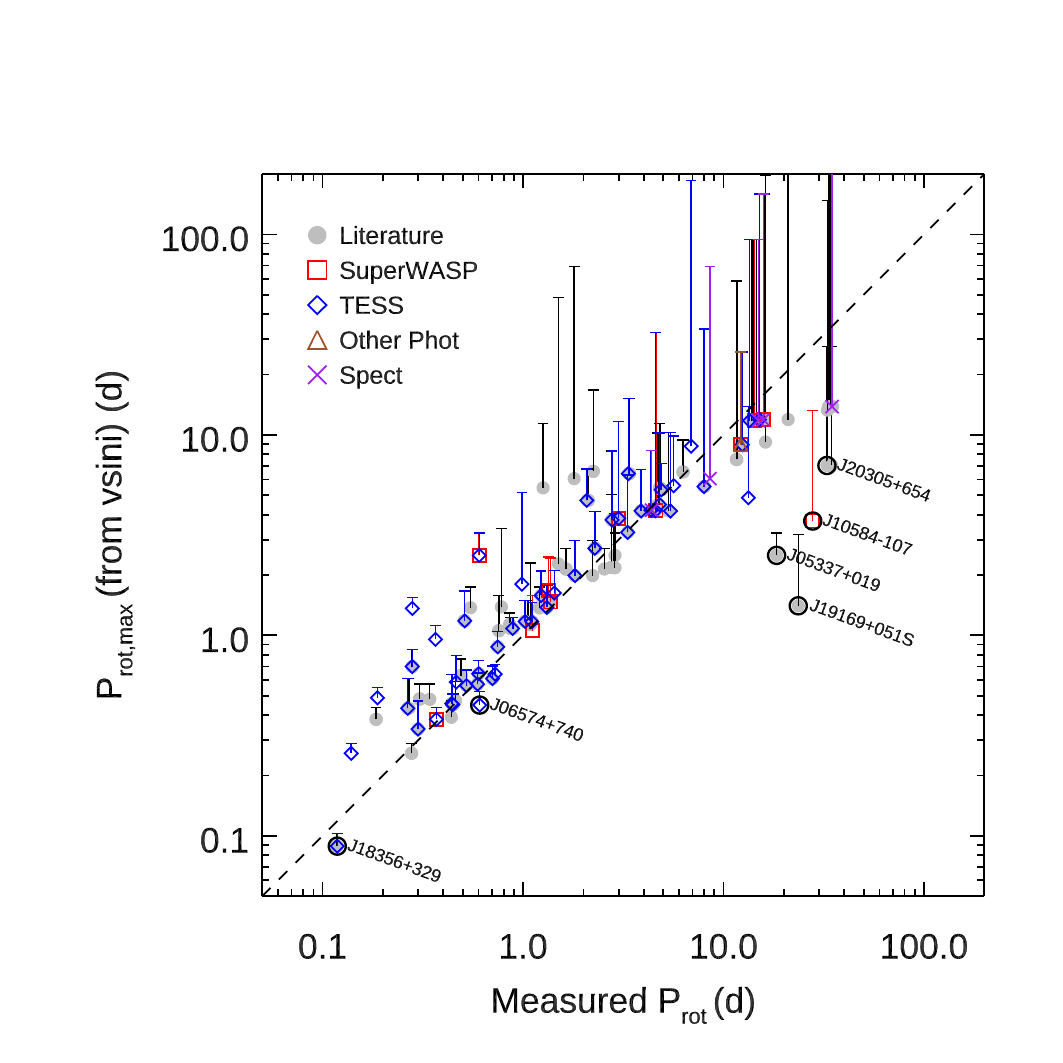}
\caption{Comparison between measured $P_{\rm rot}$ (from photometric or spectroscopic time series) and expected maximum $P_{\rm rot}$ from $v\sin i$ and $R_\star$. Error bars mark the $1\sigma$ tolerance on $P_{\rm rot, max}$ based on the uncertainties in $v\sin i$ and $R_\star$. Points falling in the region below the 1-to-1 line (dashed) have contradicting $P_{\rm rot}$, $v\sin i$ and $R_\star$ measurements. Those that fall significantly below (i.e. outside the $1\sigma$ error) are marked with a black circle and discussed in Sect. \ref{a:vsini_outliers}.  }
\label{fig:prot_vs_vsini}
\end{figure}

Figure \ref{fig:prot_vs_vsini} plots the measured $P_{\rm rot}$, either from this work or by other authors, against $P_{\rm rot, max}$. Also shown as upper error bars are the 1$\sigma$ `tolerance' on $P_{\rm rot, max}$, calculated from replacing $v\sin i$ and $R_\star$ in Equation \ref{eqn:prot_max} with $v\sin i - \delta v\sin i$ and $R_\star + \delta R_\star$, respectively. The vast majority of stars exhibit fully consistent measurements, meaning $P_{\rm rot} < P_{\rm rot, max, 1\sigma}$. The few outliers marked by thick black circles are discussed in Section \ref{a:vsini_outliers}, where we argue that all the significantly disagreeing periods are suspect.
%Below we make brief remarks on each object.  

\subsection{Correlation with $pEW'_{\rm H\alpha}$}\label{ss:pew_ha}

H$\alpha$ emission is a potent chromospheric activity indicator in M dwarfs. Using CARMENES spectra for this sample, \citet{Schoefer19} have uniformly characterized the amount of H$\alpha$ emission normalized by a quiescent reference star in the corresponding spectral type. The result is a metric defined as the H$\alpha$ pseudo-continuum width with spectral subtraction, $pEW'_{\rm H\alpha}$, which is straightforward to measure and should correlate with $P_{\rm rot}$. The $pEW'_{\rm H\alpha}$ values for our sample are given in Table \ref{t:bestprots}. 

\begin{figure}
\centering
%\vspace{-1cm}
\hspace{-1cm}
\includegraphics[width=1.1\hsize]{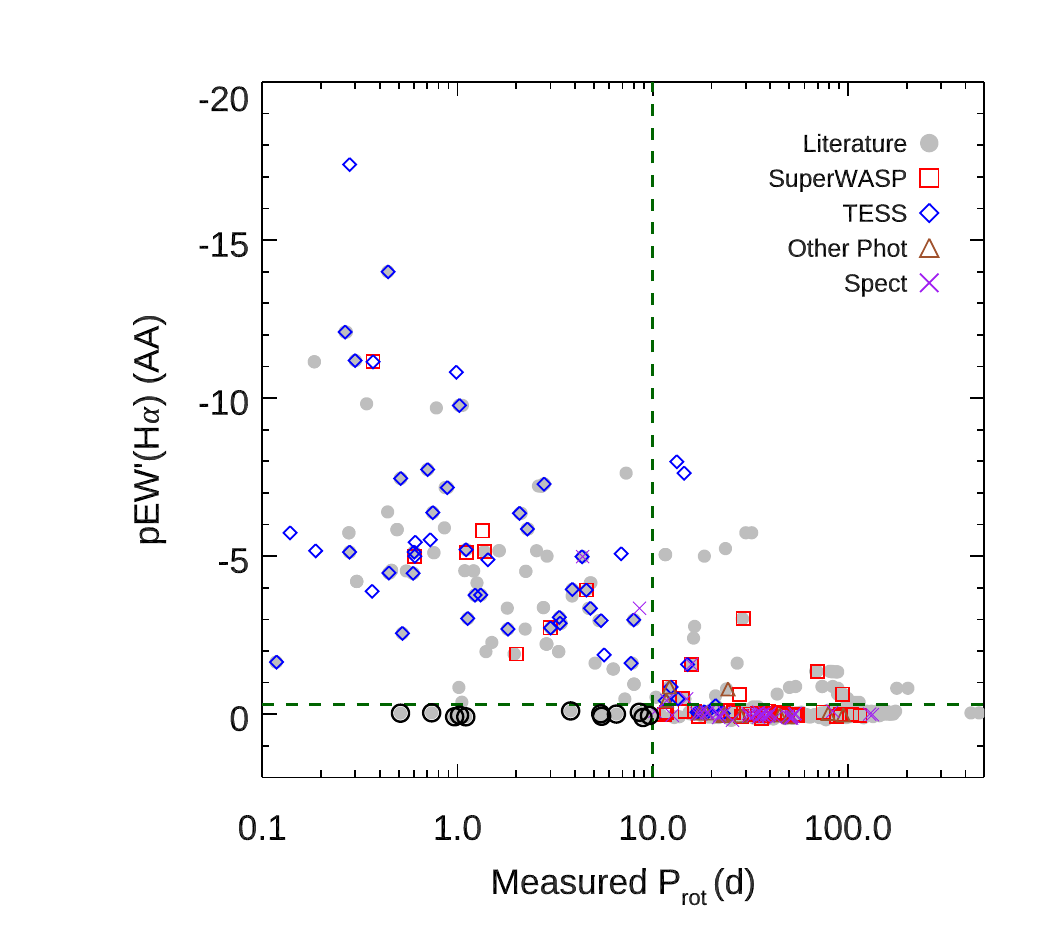}
\caption{H$\alpha$ pseudo-equivalent width from \citet{Schoefer19} versus measured $P_{\rm rot}$. Symbols as in Fig. \ref{fig:prot_vs_vsini}. The dashed green lines delineate $pEW'_{\rm{H}\alpha} = -0.3 \AA$ and $P_{\rm rot}$ = 10 d. Points to the lower-left of these bounds, marked with black circles, are considered to be outliers and discussed in Sect. \ref{a:ha_outliers}.}
\label{fig:prot_vs_halpha}
\end{figure}

Figure \ref{fig:prot_vs_halpha} compares the $P_{\rm rot}$ measurements available for this sample against $pEW'_{\rm H\alpha}$, showing the overall expected trend whereby stronger H$\alpha$ emission (i.e. larger negative value) is associated with faster rotators. Stars with $pEW'_{\rm H\alpha}\lesssim-0.3 \AA$ \citep[i.e. defined to be `H$\alpha$ active' by][]{Schoefer19} tend to have rotation periods shorter than 10 d, whereas stars with $pEW'_{\rm H\alpha} \sim 0$ typically have rotation periods longer than 10 d. While it appears not uncommon for stars with $P_{\rm rot} > 10$ d to exhibit H$\alpha$ in emission, very few stars with $pEW'_{\rm H\alpha} > -0.3 \AA$ are rapid rotators with $P_{\rm rot} < 10$ d. In fact, almost all of the points in the lower-left quadrant, representing supposed fast-rotators which appear to be H$\alpha$-inactive, are likely spurious. We discuss them under Section \ref{a:ha_outliers}.

\subsection{Best adopted rotation periods}\label{ss:prot_adopted}

The situation where multiple disagreeing periods are attributed to one given star is not uncommon. The fact that individual $P_{\rm rot}$ measurements do not always withstand scrutiny attests to persistent challenges in performing such measurements, as discussed in Sect. \ref{ss:lit_errors}. Our goal is to recommend one single $P_{\rm rot}$ to adopt for each star in the CARMENES sample, and provide an indication of its reliability. In assembling our catalog, we were sometimes confronted with the dilemma of having to select the most likely $P_{\rm rot}$ value out of several conflicting possibilities, or one that is at odds with activity signatures in the absence of alternative options. To make the most informed evaluation on the veracity of a given $P_{\rm rot}$, we developed a framework for leveraging multiple pieces of information.    

There are many possible approaches to assess period fidelity. Drawing from our investigations in Sects. \ref{ss:lit_compare}, \ref{ss:vsini}, and \ref{ss:pew_ha}, we choose to use three pieces of information, where available, in order of priority:  1) agreement with other measurements; 2) consistency with $v\sin i$; 3) consistency with $pEW'_{\rm H\alpha}$.

For each $P_{\rm rot}$ measurement, we assign a rating describing its likelihood to be correct on a coarse scale: `secure' (`S'), `provisional' (`P'), and `debated' (`D') based on whether there is independent evidence supporting or contradicting it. Each $P_{\rm rot}$ measurement is by default `provisional' unless one of the following applies: 

\begin{enumerate}
\item{If the $P_{\rm rot}$ agrees (or disagrees) with the majority of other independent $P_{\rm rot}$ measurements\footnote{The literature measurements are independent to varying degrees. While some works are based on entirely separate observations, others have performed independent analyses of the same data sets, for example DA19's use of the MEarth archival data, on which the periods from \citet{Irwin11}, \citet{Newton16}, and \citet{Newton18} are also based.} for the same star within (or outside) 15\%, then it is upgraded to `secure' (or downgraded to `debated'). Values that are obvious harmonics are treated as positive supporting evidence. If no other $P_{\rm rot}$ measurements are available, then}
%what about daily/yearly aliases? 
\item{Check plausibility with $v\sin i$. A strong agreement ($P_{\rm rot} \lesssim P_{\rm max, 1\sigma} \lesssim 30$\,d) leads to a `secure' designation. A strong disagreement ($P_{\rm rot} > P_{\rm max, 1\sigma}$) leads to a `debated' designation. If $v\sin i$ is not measured, then}
\item{For $P_{\rm rot} < 10$\,d, $pEW'_{\rm H\alpha} < -0.3 \AA$ leads to a `secure' designation, whereas $pEW'_{\rm H\alpha} > -0.3 \AA$ leads to a `debated' designation.}
\end{enumerate}

Based on these ratings, a single `best' known period ($P_{\rm rot, best}$) was assigned to each star where at least one period measurement is available. We take the period with the best rating (`secure' > `provisional' > `debated'). In cases where multiple periods are tied, 
we generally give precedence to photometric periods from this work.  
% TESS > SWASP > OSN > TESS-SAP
After this, we use the following hierarchy of preferred periods: periods from DA19 (which was the first comprehensive study for this sample), other photometric surveys from the literature, planet discovery papers, spectroscopic indicators from this work and the literature, and spectropolarimetry.  
The only periods we exclude from becoming $P_{\rm rot, best}$ are those of U-rotators from this work and from \citet{Newton16,Newton18}, the `*'-rotators of \citet{Fouque23}, the `marginal detections' of \citet{Donati23}, which are deemed by the authors themselves to be uncertain, as well as those from \citet{Oelkers18}, who presented variability periods (< 50 d) for a large sample ($\sim 5\times10^4$) of stars from the KELT survey, in which many are only `possible rotation periods' that did not undergo individual vetting. We noticed that \citet{Oelkers18} often reported the 1-day aliases of existing period measurements. These periods are valuable as supporting evidence for other period measurements, in the same way as the U-rotators, but not reliable enough to stand alone.  

In total, there are \HL{166} stars whose adopted periods have a high likelihood of being correct (`secure' or `S'), \HL{68} stars whose periods are tentative (`provisional' or `P'), and \HL{27} stars whose periods are disputed (`debated' or `D'). The `provisional' periods are most often long and difficult to refute with activity signatures, while also being difficult to confirm with multiple measurements. A typical `debated' period is a short one derived for a star that is not spectroscopically active.   

The final adopted periods, $P_{\rm rot, best}$, are given in Table \ref{t:bestprots}. These are used for the analyses in Sect. \ref{s:discussion}. For an estimate of the period errors, we refer to Sect. \ref{ss:lit_errors}. All literature periods that contributed to the assessment are given in Table \ref{t:allprots}.

\subsection{Empirical estimates of $P_{\rm rot}$ error}\label{ss:lit_errors}

\begin{figure}
\centering
\includegraphics[scale=0.5]{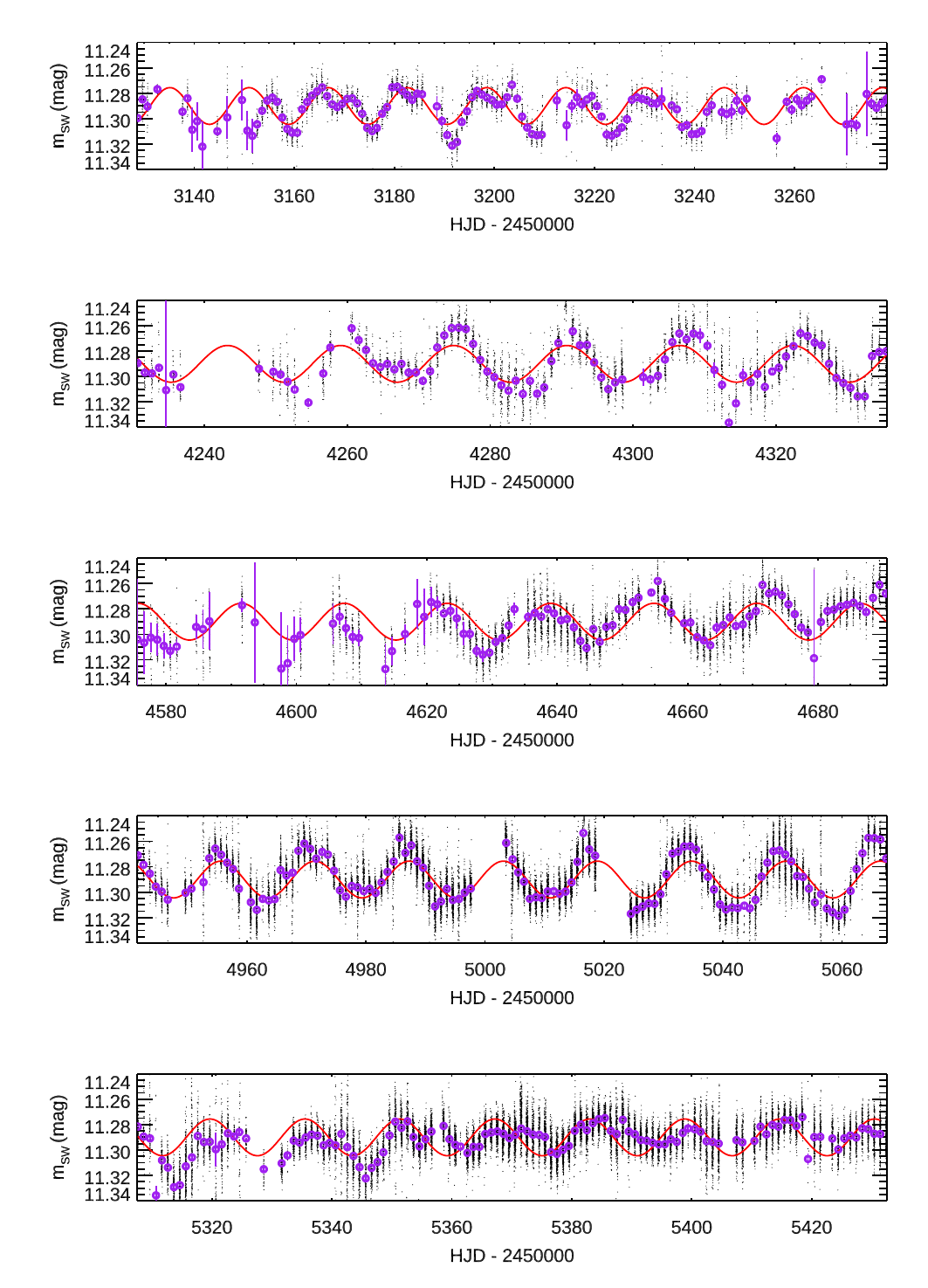}
\caption{SuperWASP light curve of J18174+483 in five seasons spanning six years. The best-fit sinusoidal signal to the global data set (15.83\,d) is plotted as the red curve. Purple circles are daily-binned data points. }
\label{fig:prot_evol_J18174}
\end{figure}

The existence of debated period measurements is a symptom of broader challenges with determining accurate stellar rotation periods. The detectability and reliability of a detected rotation signal are affected by many interconnected factors including: 
\begin{itemize}
\item{the characteristics and quality of the data sets (e.g. time and duration of observations, sampling strategy, data precision). For example, ground-based surveys are plagued by poorer precision, weather, and aliasing, while space-based surveys tend to have limited time baselines;}
\item{The choice of analysis methods and thresholds \citep[e.g.][]{Aigrain15}. For example, authors working independently with the same data set can still sometimes obtain discrepant results, such as \citet{Newton16} and DA19 on MEarth photometry, or \citet{Fouque23} and \citet{Donati23} on SPIRou spectropolarimetry;}
\item{the astrophysical behaviour of the star itself. Some example processes that can affect $P_{\rm rot}$ determination are differential rotation \citep[e.g.][]{Reinhold13}, spot configurations and lifetimes \citep[e.g.][]{Giles17}, and variability unrelated to rotation. The evolution of surface active regions over time can affect modulation amplitudes \citep[e.g.][]{Irwin11} and lead to intrinsically quasi-periodic variations \citep[e.g.][]{Angus18}. An example from our sample is shown in Fig. \ref{fig:prot_evol_J18174} for the star J18174+483, where five seasons of SuperWASP observations spanning six years reveal evolving morphology, amplitude and phase, visible against the backdrop of the global best-fit sinusoidal signal (15.8\,d). The best-fit periods differ slightly from season to season, varying between 15.6\,d and 17\,d; } 
\item{the nature and strength of the physical link between the modulating observable and stellar rotation \citep[see discussion in, e.g.][]{Lafarga21, Schoefer22}. For example, stellar activity can manifest differently in photospheric spots than in chromospheric lines, impacting their variation patterns and sensitivity to rotation. }
\end{itemize}

These considerations complicate the assignment of appropriate errors to a given period measurement \citep[e.g.][]{Irwin11,Newton16}. Authors who do report $P_{\rm rot}$ errors estimate them using a variety of methods, though it is unclear to what extent these formal errors adequately capture the complex and varied factors affecting the retrieved period values. For instance, one popular measure of $P_{\rm rot}$ error uses the breadth of the periodogram peak, which would be misleading if multiple peaks are present \citep{Newton16}. 

When multiple independent $P_{\rm rot}$ measurements are made for a given star, their level of disagreement provides an {\emph{empirical}} measure of the uncertainty. 
Using statistical dispersion as a proxy for level of disagreement, we investigate the behaviour of this empirical error ($\sigma_{\rm emp}$) for our sample.  
For each of our sample stars with more than one period measurement available (from the literature or from this work), we compute the standard deviation among this set of periods. 
To minimize the outsized effects of harmonics and aliases on our metric, we replaced periods that are obvious 2$\times$ or $1/2\times$ harmonics or daily aliases of the adopted period $P_{\rm rot, best}$ (i.e. within 15\%) with the harmonic or alias value closest to $P_{\rm rot, best}$. A few of the period discrepancies may also be attributed to yearly aliasing, although for this exercise we left them as they are.  

\begin{figure}
\centering
\hspace{-1cm}
\includegraphics[width=1.1\hsize]{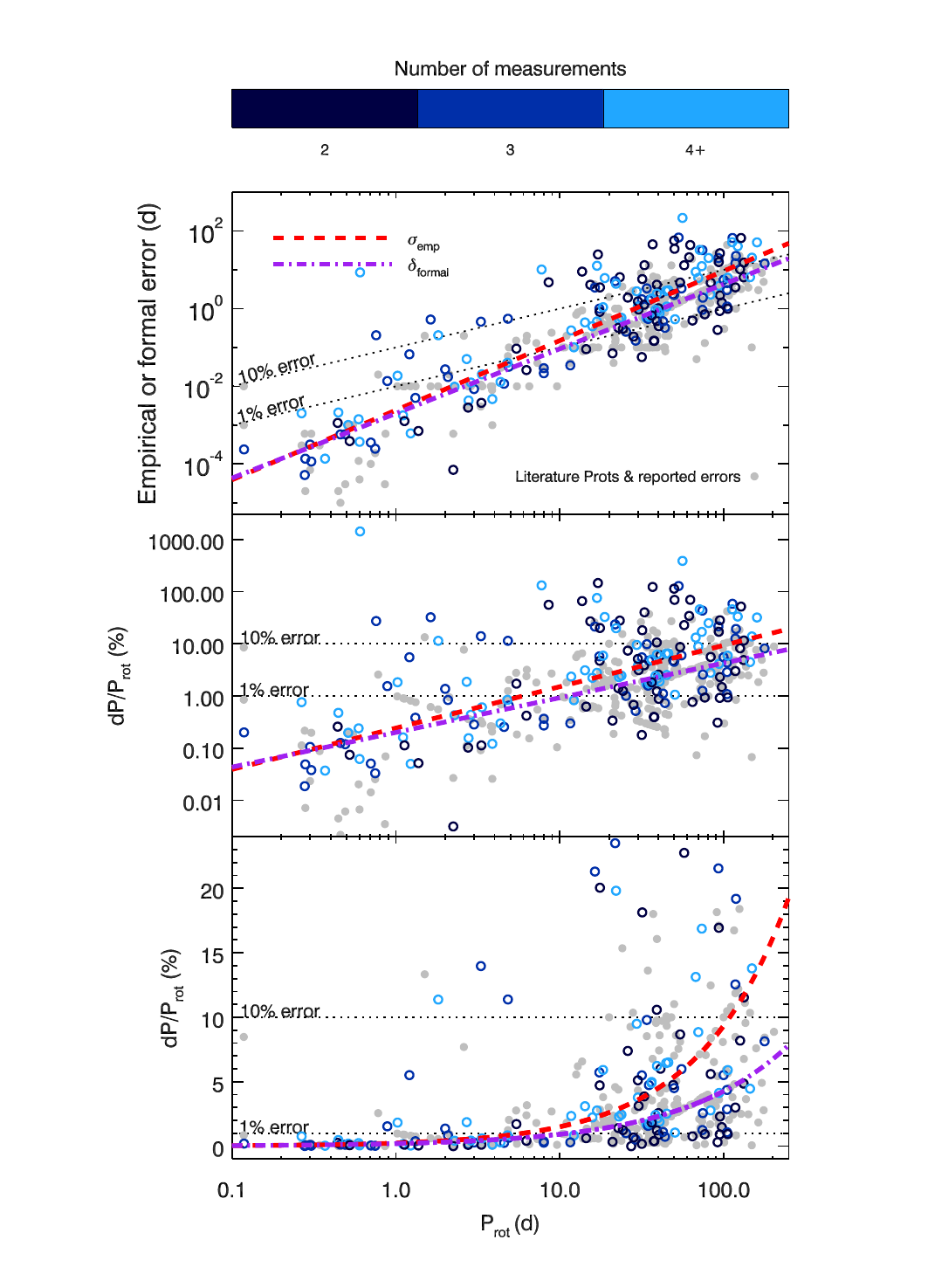}
\caption{Measures of error as a function of $P_{\rm rot}$. Blue unfilled circles plot the `empirical' dispersion $\sigma_{\rm emp}$, i.e. standard deviation of the measured periods, colour-coded by the number of measurements (see colour bar). Grey filled circles represent `formal' errors $\delta_{\rm formal}$, as reported in the literature. {\em Top}: $\sigma_{\rm emp}$ vs. $P_{\rm rot,best}$ (blue) and $\delta_{\rm formal}$ vs. reported $P_{\rm rot}$ (grey); {\em Middle}: fractional error vs. $P_{\rm rot}$, where fractional error is obtained by dividing $\sigma_{\rm emp}$ or $\delta_{\rm formal}$ (collectively termed `dP') by $P_{\rm rot}$. {\em Bottom}: same as the middle, but with the y-axis shown on a linear scale. The best-fit power laws are plotted for $\sigma_{\rm emp}$ (red dashed, as given in Eqn. \ref{eqn:stdev_prot}) and $\delta_{\rm formal}$ (purple dot-dashed, as given in Eqn. \ref{eqn:stdev_prot_lit}). The black dotted lines depict constant fractional errors of 10\% and 1\% for comparison.  Note that the bottom panel has restricted vertical plotting range in order to show the difference in the two power laws. Points beyond the vertical range are not shown.} 
\label{fig:prot_vs_spProt}
\end{figure}

\begin{figure}
\centering
\hspace{0cm}
\includegraphics[width=1\hsize]{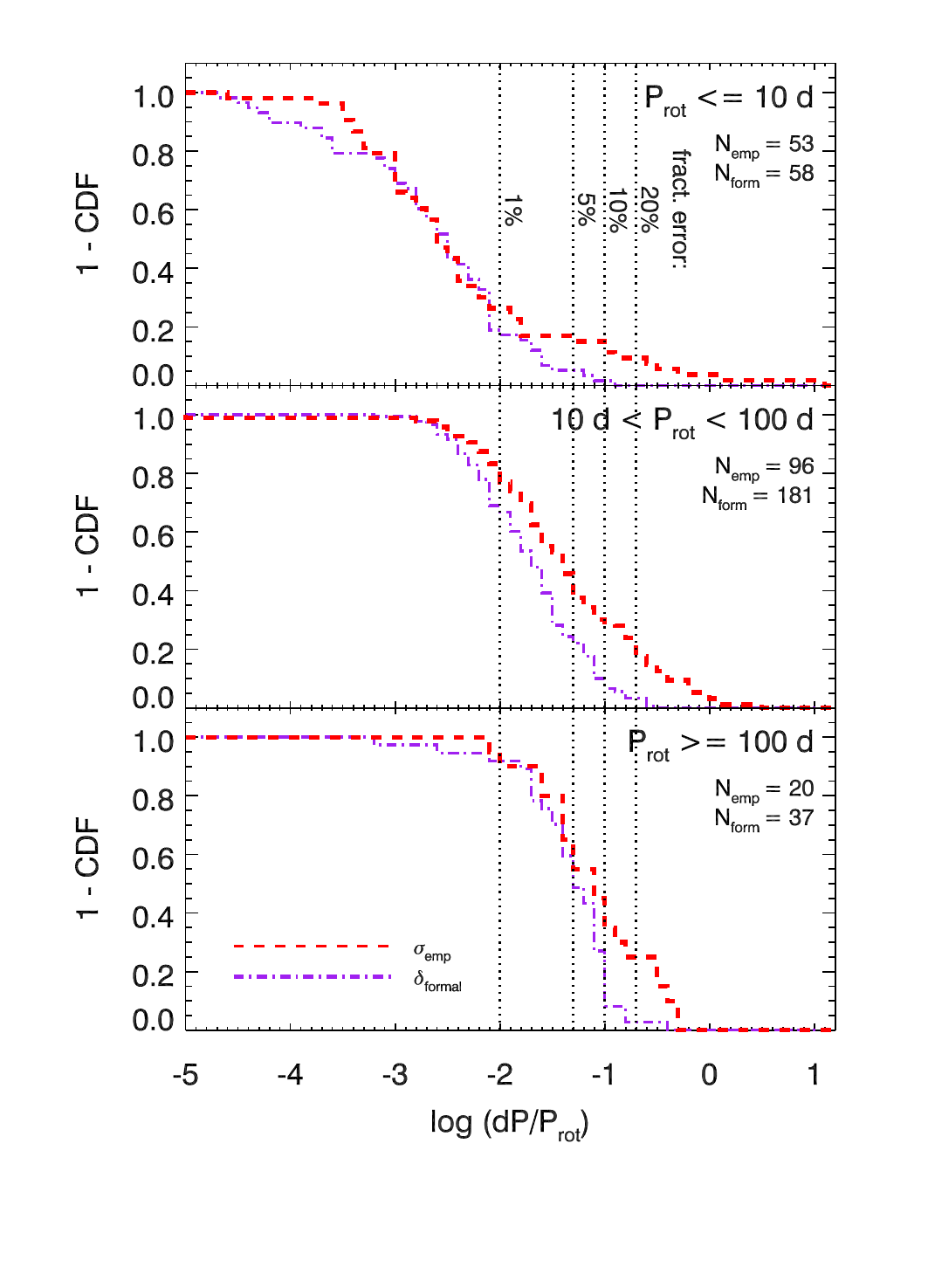}
\caption{Reverse cumulative distribution of fractional period errors: comparison between `empirical' ($\sigma_{\rm emp}$, red dashed) and `formal' errors ($\delta_{\rm formal}$, purple dot-dashed). {\em Top}: for $P_{\rm rot} \le 10$\,d; {\em middle}: for $10 < P_{\rm rot} < 100$\,d; {\em Bottom}: for $P_{\rm rot} \ge 100$\,d. The number of data points in each bin are given as $N_{\rm emp}$ for the empirical dispersion and $N_{\rm form}$ for the formal error distribution.}
\label{fig:eProt_cdf}
\end{figure}

In Fig. \ref{fig:prot_vs_spProt} we plot $\sigma_{\rm emp}$ against $P_{\rm rot, best}$ for 169 stars. A strong positive correlation is evident: larger $P_{\rm rot}$'s have larger dispersions in their measurements. %Fig. \ref{fig:eProt_cdf} depicts the fraction of stars 
In general, faster rotators (< 10\,d) display a high level of consistency among independent measurements, with typically $\lesssim 1$\% dispersion. Among the 116 slower rotators ($\gtrsim$ 10\,d), the median dispersion is modest at $\sim 5\%$. However, large outliers are also common: 38 stars have fractional dispersions $\gtrsim 10$\%, and 28 have dispersions $\gtrsim 20$\%. 
Figure \ref{fig:eProt_cdf} displays the cumulative fraction of stars exceeding a range of thresholds in fractional dispersion for < 10\,d, 10 -- 100\,d, and $\ge$100\,d rotators.   

The best-fit power law gives an empirical estimate of the typical error in an individual $P_{\rm rot}$ measurement as a function of $P_{\rm rot}$ value: 

\begin{equation}
{\sigma_{\rm emp}} (P_{\rm rot}) = 9.24 \left (\frac{P_{\rm rot}}{\rm 100 ~d}\right )^{1.79} \Rightarrow \frac{dP_{\rm rot,emp}}{P_{\rm rot}} = 9.24\% \left (\frac{P_{\rm rot}}{\rm 100 ~d}\right )^{0.79}, 
\label{eqn:stdev_prot}
\end{equation}   

\noindent where $dP_{\rm{rot, emp}}/P_{\rm rot}$ denotes the fractional period error. According to Eqn. \ref{eqn:stdev_prot}, the average fractional error reaches above 10\% for very slow rotators with $P_{\rm rot} > 109$\,d. However, since the dispersion on a number of $\gtrsim$10 \,d periods already far exceed the 10\% (see Fig. \ref{fig:prot_vs_spProt} and \ref{fig:eProt_cdf}), single $P_{\rm rot}$ measurements in the $10 - 100$ d range can also be more unreliable than Eqn. \ref{eqn:stdev_prot} would suggest. Additional measurements are important for verification, and are especially relevant for disentangling between harmonics and aliases, whose effects were filtered out in this exercise. Even in the absence of harmonics and aliases, Eqn. \ref{eqn:stdev_prot} likely gives a conservative representation of dispersion, because our assumption that the $P_{\rm rot}$ values from the literature are independent is not completely valid (e.g. authors may use the same data sets, and published results can be at least somewhat influenced by the knowledge of one another). Interestingly, the error-to-$P_{\rm rot}$ relation characterized by Eqn. \ref{eqn:stdev_prot} is very comparable to that found by \citet{Irwin11} using injection-retrieval simulations in MEarth data (see note `g' under Table 1 therein)

For comparison, in Fig. \ref{fig:prot_vs_spProt} we overplot 276 literature $P_{\rm rot}$ values with their reported formal errors $\delta_{\rm formal}$ collected for this sample. The best-fit power law,
\begin{equation}
\delta_{\rm formal}(P_{\rm rot}) = 4.27 \left (\frac{P_{\rm rot}}{\rm 100 ~d}\right )^{1.66} \Rightarrow \frac{dP_{\rm rot,formal}}{P_{\rm rot}} = 4.27\% \left (\frac{P_{\rm rot}}{\rm 100 ~d}\right )^{0.66},
\label{eqn:stdev_prot_lit}
\end{equation} 
characterizes the average formal error as a function of $P_{\rm rot}$, and is slightly shallower than our empirical relation (Eqn. \ref{eqn:stdev_prot}). While nearly indistinguishable at 1\,d, the discrepancy grows with $P_{\rm rot}$ such that, at 100\,d, $\sigma_{\rm emp}$ exceeds $\delta_{\rm formal}$ by more than a factor of two. Furthermore, as shown in Fig. \ref{fig:eProt_cdf}, the share of stars with large period errors appears to be also underestimated compared to reality. In this sample, only 3\% of periods above 10\,d have formal fractional errors above 20\%, compared to 24\% with fractional measurement dispersions greater than 20\%. In sum, we find that formal period errors tend to be somewhat more optimistic than the actual dispersion of measurements, especially for slower rotators with $P_{\rm rot} \gtrsim 10$\,d. 

Since Eqn. \ref{eqn:stdev_prot} is by construction derived using an ensemble of periods agnostic of the data, method, and stellar properties, it is not necessarily applicable to arbitrary data sets with very specific characteristics. Nevertheless, since we effectively `marginalize' over all the factors that could affect the measurement, Eqn. \ref{eqn:stdev_prot} has the advantage of being a very general and reasonable representation of uncertainty in the absence of a more principled error analysis. Therefore, we adopt $\sigma_{\rm emp}(P_{\rm rot})$ given in Eqn. \ref{eqn:stdev_prot} as our estimate of period error in this work, while stressing that periods $\gtrsim 10$\,d can be more erroneous than this equation would suggest. 

Equation \ref{eqn:stdev_prot}
also provides an instructive relation between the period uncertainty and period value. Intuitively, shorter periods should be much easier to measure because they undergo more cycles (for a fixed observing time baseline) and are associated with larger variability amplitudes and more stable spot configurations \citep[measured in number of rotations, see e.g.][]{Basri22}, which are favourable conditions to unambiguous and more precise detections. Moreover, shorter periods are more accessible by space-based missions, whose low noise, high cadence, and continuous monitoring capabilities facilitate the detailed characterization of stellar variability. 

However, the short-period advantage is difficult to scale up to the $\gtrsim 10$\,d-class rotators. Longer periods are challenging to measure because of the inherent difficulties involved in carrying out a long-term all-sky survey. 
With the exception of the original {\emph{Kepler}} mission (which monitored a small patch of the sky for four years),  
long time baseline surveys typically operate from the ground, where long-term stability in photometry cannot be guaranteed for months and years. In addition to day-to-night cycles and seasonal visibility, ground based observations are also affected by weather conditions and moon phases, all of which lead to non-uniform sampling, aliasing, and data points of varying quality, hindering the precise characterization of low-amplitude variability. 

The increased spread in measurements at larger periods could also be reflecting trends in differential rotation. It has been shown that slower rotators tend to have stronger relative differential rotation \citep[e.g.][]{Donahue96,Reiners03,Reinhold13}, which could further contribute to blurring their measured $P_{\rm rot}$. Under this interpretation, $P_{\rm rot}$ itself is not a unique quantity, and its `error' characterizes the dispersion in latitudinal rotation rates. This hypothesis should be tested in a data set with more uniform and widespread coverage and analysis techniques, as has been done with the {\emph{Kepler}} sample \citep[e.g.][]{Reinhold13}.

\section{Discussion}\label{s:discussion}

\subsection{Rotation-activity relations}\label{ss:rot-act}

Magnetic dynamos, powered by rotation, operate in stars and give rise to global magnetic fields that transform convective energy in the stellar interior to heat the surface, including the chromosphere and corona. While the exact process by which this happens is still not completely understood \citep[see, e.g.][]{Charbonneau05,Kochukhov21,Reiners22}, the theory may be constrained by a number of well-defined correlations between rotation rate of a star and energy output from the chromosphere and corona, which have been directly observed. The mass-dependence of such relations, in particular across the mass boundary at about $0.35 M_\sun$, is of especial interest due to the theoretical change in the dynamo mechanism for fully convective stars. 

Proxies for choromospheric activity include emission in the H$\alpha$ and Ca~{\sc ii} H\&K lines, while X-ray flux tracks coronal emission. In stars later than $\sim$ F-type, it has long been demonstrated that faster rotation maps into greater non-thermal flux \citep[e.g.][]{Pallavicini81,Noyes84}. Further investigation uncovered two distinct regimes for this correlation: an unsaturated regime where the luminosity of the activity indicator increases monotonically with rotation rate, and a saturated regime where the bolometrically normalized luminosity of the activity indicator stays roughly constant \citep[e.g.][]{Pizzolato03,Mamajek08,Wright11,Reiners14}. The saturation regime sets in below a certain threshold period, $P_{\rm sat}$. The correlation is characterized by the slope in each regime, the break point between the two regimes (i.e. $P_{\rm sat}$), and any systematic or random scatter around the relation. These relationships have also been studied extensively for M dwarfs for which rotation periods have been measured ($L_X$: \citealt{Stelzer16, GA19, Magaudda20}; $L_{\rm H\alpha}$: \citealt{Newton17}; $L_{\rm Ca}$: \citealt{AD17a,SM18,Boudreaux22}). 

Magnetic heating is thought to be the underlying driver of non-thermal emission. Recently, \citet{Reiners22} elucidated the mechanism responsible for activity-rotation relations. They measured the average surface magnetic field strengths $\langle B \rangle$ for a large subset of M dwarfs from the CARMENES survey, by using Zeeman broadening modelling of high-quality spectra. They demonstrated that $\langle B \rangle$ behaves identically to that of other activity indicators with respect to rotation. Saturation in magnetic activity is therefore a consequence of saturation in the magnetic field as it reaches the kinetic energy limit. Furthermore, non-thermal emission from the chromosphere and corona directly trace the magnetic flux, which is proportional to rotation rate.   

\begin{figure*}
%\centering
\begin{minipage}{0.5\textwidth}
    %\centering
    \hspace{-1cm}
\includegraphics[width=1.1\hsize]{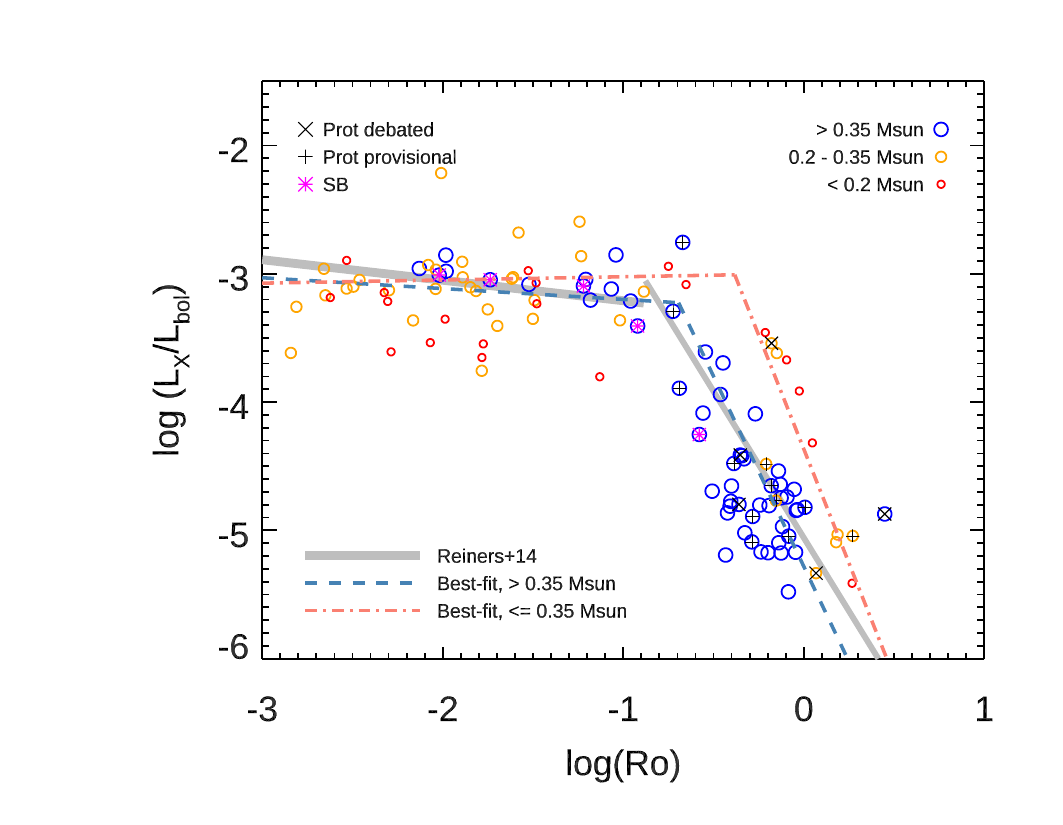}
\end{minipage}
\begin{minipage}{0.5\textwidth}
    %\centering
    \hspace{-1cm}
\includegraphics[width=1.1\hsize]{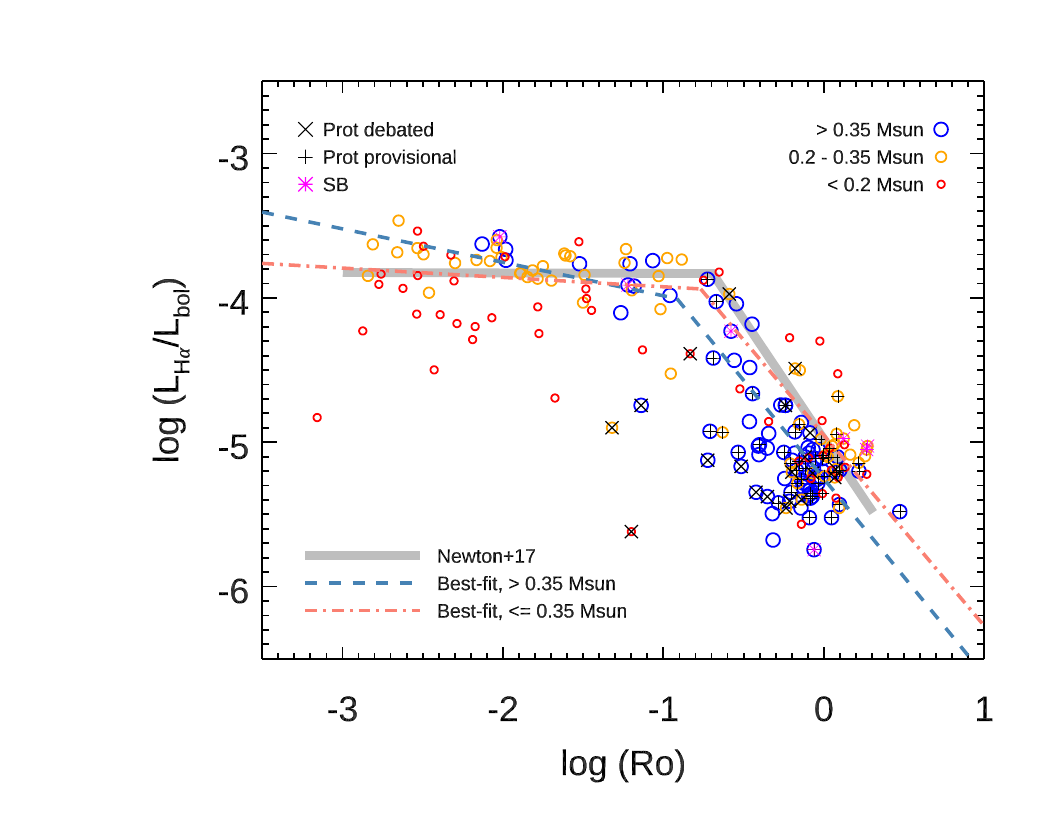}
\end{minipage}
\begin{minipage}{0.5\textwidth}
    %\centering
    \hspace{-1cm}
\includegraphics[width=1.1\hsize]{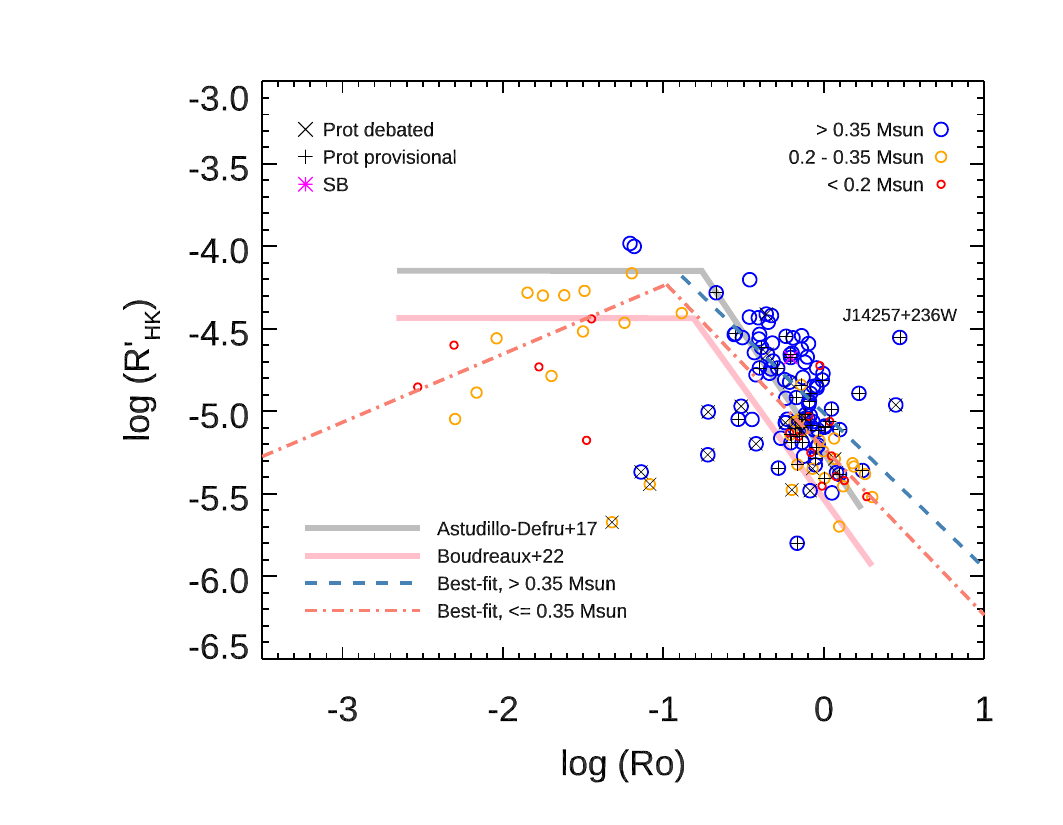}
\end{minipage}
\begin{minipage}{0.5\textwidth}
    %\centering
    \hspace{-1cm}
\includegraphics[width=1.1\hsize]{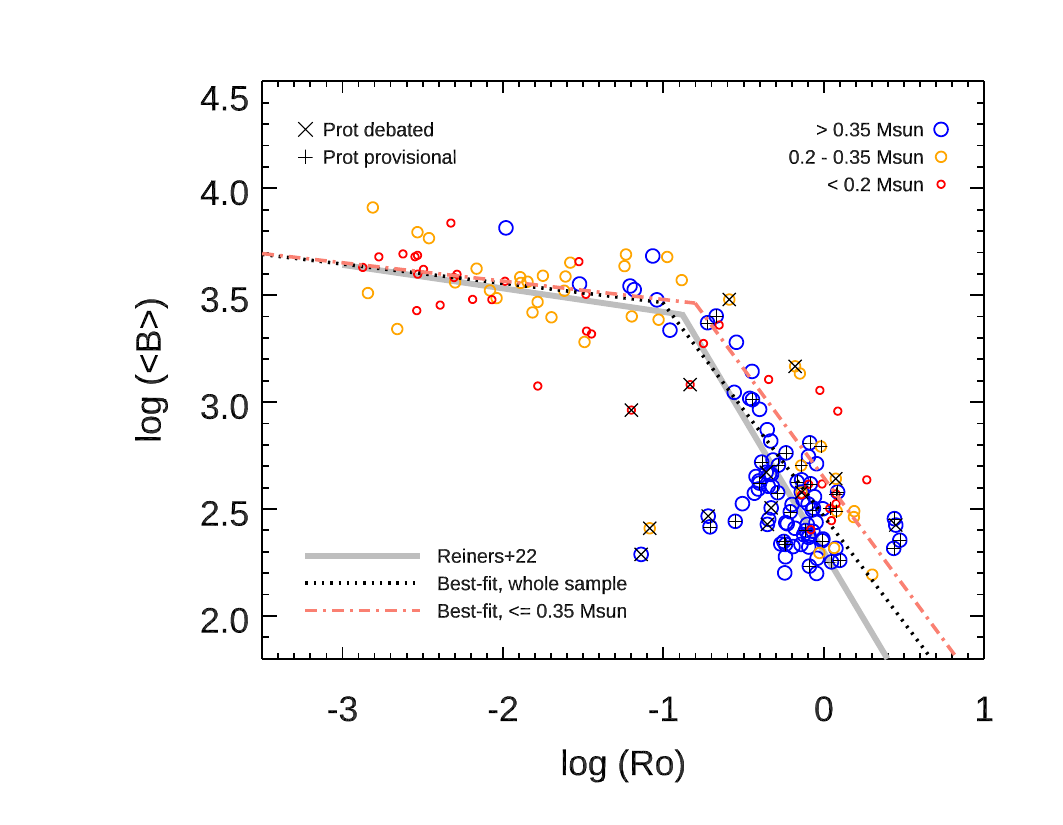}
\end{minipage}
\caption{
\HLr{Rotation-activity relations studied for our sample, as discussed in Sect. \ref{ss:rot-act}.} 
{\em Top left:} Normalized X-ray luminosity $L_X/L_{\rm bol}$ as function of the Rossby number $Ro$ for \HL{113} M dwarfs. The corresponding best-fit relation from \citet{Reiners14} is plotted in thick grey lines. 
{\em Top right:} Normalized H$\alpha$ luminosity $L_{\rm H\alpha}/L_{\rm bol}$ as function of $Ro$ for \HL{196} M dwarfs. The best-fit broken power-law relation from \citet{Newton17} is displayed in grey. 
{\em Bottom left:} 
Normalized Ca~{\sc ii} H\&K emission ($R'_{\rm HK} \equiv L_{\rm Ca}/L_{\rm bol}$) as function of $Ro$ for \HL{144} stars. The marked outlier J14257+236W was not included  in our fit. Best-fit relations from \citet{AD17a} and \citet{Boudreaux22} are overplotted.
{\em Bottom right:} Average magnetic field strength $\langle B \rangle$ versus $Ro$ for \HL{162} stars.
In all plots, earlier-, mid-, and later-type M dwarfs (i.e. $M_\star > 0.35\,M_\sun$,  $0.35\,M_\sun > M_\star > 0.2\,M_\sun$, $< 0.2\,M_\sun$) are denoted by large blue, medium-sized orange, and small red circles, respectively. Stars with non-debated period values are fitted with broken power laws, shown in broken dashed lines in black (whole sample), blue ($> 0.35 M_\sun$), and green ($\le 0.35 M_\sun$). Stars with debated periods are marked by black `X's, and with provisional periods they are marked by black `+'s. Known SBs are represented by magenta asterisks. Where applicable, best-fit rotation-activity relations for subsamples within defined mass ranges are overlaid (black dotted lines: whole sample, blue dashed lines: $> 0.35 M_\sun$, red dash-dotted lines: $\le 0.35 M_\sun$).}
\label{fig:rot-act}
\end{figure*}

In Figure \ref{fig:rot-act} we show several well-known rotation-activity relations for the CARMENES M dwarf sample in terms of the Rossby number, $Ro \equiv P_{\rm rot}/\tau_{\rm conv}$. We calculated the convective overturn time $\tau_{\rm conv}$ as a function of $(V - K_s)$ colour using the prescription of \citet{Wright18}, which is calibrated for the entire range of M dwarf spectral types. Johnson $V$ and Two-Micron All-Sky Survey $K_s$ magnitudes
were taken from \citet{Cifuentes20} and references therein (see also Section \ref{a:rossby}). We note that other calibrations for $\tau_{\rm conv}$ are available and systematically offset from one another (see Section \ref{a:rossby}), and that the exact prescription choice may affect the quantitative details of the activity-rotation relations \citep{Magaudda20}. The points in \HLr{Fig. \ref{fig:rot-act}} are colour-coded by stellar mass, divided into bins $>$ 0.35\,$M_\sun$, 0.20--0.35\,$M_\sun$, and $<$ 0.2\,$M_\sun$, the latter two corresponding to fully convective stars.  

All activity indicator measurements for this sample were compiled from the literature as described below. 
To these values we fitted classical rotation-activity relations in the following broken-power-law form (unless otherwise specified) where non-zero slopes are allowed in both the saturated and unsaturated regimes: 

\begin{equation}
{\rm Activity ~Indicator} = 
\begin{cases}
    A (Ro/Ro_{\rm sat})^{\beta_{\rm sat}} & {\rm for~} Ro < Ro_{\rm sat} \\ 
    A (Ro/Ro_{\rm sat})^{\beta_{\rm unsat}} & {\rm for~} Ro > Ro_{\rm sat}.
\end{cases}
\label{eqn:act_ro}
\end{equation}

\noindent Here $A$ is a normalization constant, $Ro_{\rm sat}$ is the saturation Rossby number, and $\beta_{\rm sat}$ and $\beta_{\rm unsat}$ are slopes in the saturated and unsaturated regimes, respectively. The fits were performed using the Markov Chain Monte Carlo fitting package {\tt{emcee}} \citep{ForemanMackey13}. We set uninformative priors on $\log(A)$, $\beta_{\rm sat}$, and $\beta_{\rm unsat}$, and imposed a flat prior on $\log(Ro_{\rm sat}) \in [-2, 0]$. In addition to the reported formal errors on the measured activity indicator, we added a parameter $\log(f)$ in constructing the likelihood function to account for underestimated variance as a fixed fraction $f$ of each activity indicator measurement, following standard procedures \citep[see, e.g.][]{Newton17, Nunez22}. We did not account for errors on $Ro$. The fits employed 32 walkers each taking 5000 steps, discarding the first 1000 steps. The best-fit parameters are given in Table \ref{t:rotact_bestpars}. In this work, all instances of $\log$ refer to the common logarithm in base 10.

\paragraph{X-rays.} 
The X-ray measurements are from ROSAT 1RXS \citep{Voges99}. We plot bolometric-normalized X-ray luminosity $L_{\rm X}/L_{\rm bol}$ versus $Ro$ relation for \HL{113} stars in our sample. Of these, \HL{108} stars have `S' and `P' period designations. For these stars we fitted Eqn. \ref{eqn:act_ro} to $L_{\rm X}/L_{\rm bol}$.  

We fitted to the whole sample as well as separately to partially convective ($> 0.35\,M_\sun$) and fully convective ($\le 0.35\,M_\sun$) M dwarfs. Figure \ref{fig:lx_cornerplot} in Appendix \ref{a:tables_figures} shows an example of the posterior distribution for the whole sample fit. Table \ref{t:rotact_bestpars} displays these best-fit parameters and their $1\sigma$ errors (i.e. 16th and 84th percentiles of the posterior distributions) for each mass range considered. Figure \ref{fig:rot-act} plots the corresponding relations for the partially convective ($> 0.35\,M_\sun$) and fully convective ($\le 0.35\,M_\sun$) M dwarfs. We overlay the relation shown in Fig. 3 of \citet{Reiners14}, which was based on a broader range of stellar masses, including predominantly sunlike stars. We note that \citet{Reiners14} computed the empirical $Ro$ based on \citet{Wright11}, which deviates from the \citet{Wright18} version by up to -0.2\,dex (see Fig. \ref{fig:rossby_compare}).

In the saturated regime, all fits are compatible with a null slope, although across the convective mass boundary the slopes are $\sim 1\sigma$ apart. The early-M dwarfs show a slightly negative slope ($\beta_{\rm sat} = -0.08 \pm 0.12$) consistent with that of \citet{Reiners14} of $-0.16$. However, the mid- and late-M dwarfs follow a slightly positive slope of $+0.02 \pm 0.10$, which is $2\sigma$ deviant from the slope of \citet{Reiners14}. A hint of this dichotomy is also visible in Fig. 4 of \citet{Magaudda20} and extensively discussed in \citet{Magaudda22}. 

The break points marking the end of saturation also appears to have a significant mass dependence. For early-M's, the decline in activity starts already at $Ro_{\rm sat} \sim 0.2$, whereas for later-M's it is close to $Ro_{\rm sat} \sim 0.4$. \citet{Magaudda20} found a very similar dependence for $P_{\rm sat}$ in the $L_{\rm X}-P_{\rm rot}$ relation for M dwarfs. These values are significantly larger than the $Ro_{\rm sat} = 0.13$ assumed by previous authors studying the $L_{\rm X}$-$Ro$ relation \citep{Wright11,Reiners14,Wright18}. Since $Ro_{\rm sat}$ plays a role in calibrating the empirical $\tau_{\rm conv}$, this difference could be related to the offsets between various $\tau_{\rm conv}$ parameterizations (see e.g. Appendix \ref{a:rossby}). The slope in the unsaturated regime is about $-3.0$ for the early-M's and $-3.5$ for the later-M's, the latter being poorly constrained due to the relatively few data points available.   

\begin{table*}[htb]
\begin{center}
\begin{small}
\caption{Best-fit activity-rotation parameters.}
\label{t:rotact_bestpars}
\begin{tabular}{lllllllllll}
\hline\hline
$M_\star$ range & $N_\star$ & $\log_{10}(A)$ & $\log_{10}(Ro_{\rm sat})$ & $\beta_{\rm sat}$ & $\beta_{\rm unsat}$ & $\log_{10}(f)$ & Lit. $Ro_{\rm sat}$ & Lit. $\beta_{\rm sat}$ & Lit. $\beta_{\rm unsat}$ & Ref\textsuperscript{b} \\
\hline
\noalign{\smallskip}
$L_{\rm X}/L_{\rm bol}$: \\
%\hline
\noalign{\smallskip}
        whole sample & 108 &$   -3.16^{+    0.11}_{-    0.10}$ &$   -0.78^{+    0.07}_{-    0.08}$ &$    -0.04_{-    0.09}^{+    0.09}$ &$    -1.90_{-    0.18}^{+    0.17}$ &$    +0.04^{+    0.06}_{-    0.06}$ & 
        0.13 (fixed) & $-0.16^{+0.03}_{-0.03}$ & & Rei14 \\
     $> 0.35 M_\sun$ &  53 &$   -3.23^{+    0.11}_{-    0.10}$ &$   -0.69^{+    0.03}_{-    0.05}$ &$    -0.08_{-    0.11}^{+    0.12}$ &$    -2.96_{-    0.17}^{+    0.16}$ &$   -0.11^{+    0.07}_{-    0.06}$\\
   $\le 0.35 M_\sun$ &  55 &$   -3.01^{+    0.16}_{-    0.15}$ &$   -0.38^{+    0.07}_{-    0.08}$ &$   +0.02_{-    0.10}^{+    0.10}$ &$    -3.55_{-    0.47}^{+    0.53}$ &$   -0.04^{+    0.08}_{-    0.07}$\\
\hline
\noalign{\smallskip}
$L_{\rm H\alpha}/L_{\rm bol}$: \\
%\hline
\noalign{\smallskip}
           whole sample & 176 &$   -3.96^{+    0.06}_{-    0.05}$ &$   -0.88^{+    0.05}_{-    0.06}$ &$   -0.08_{-    0.04}^{+    0.05}$ &$   -1.29_{-    0.06}^{+    0.06}$ &$   -0.21^{+    0.04}_{-    0.03}$ &
           $0.21^{+0.02}_{-0.02}$ & 0 (fixed) & $-1.7^{+0.1}_{-0.1}$ & New17 \\
     $> 0.35 M_\sun$ &  74 &$   -4.00^{+    0.15}_{-    0.14}$ &$   -0.92^{+    0.11}_{-    0.13}$ &$   -0.23_{-    0.15}^{+    0.16}$ &$   -1.36_{-    0.07}^{+    0.07}$ &$   -0.20^{+    0.06}_{-    0.05}$\\
 $0.2 - 0.35 M_\sun$ &  58 &$   -3.91^{+    0.06}_{-    0.06}$ &$   -0.91^{+    0.10}_{-    0.10}$ &$    -0.11_{-    0.05}^{+    0.05}$ &$    -1.12_{-    0.11}^{+    0.09}$ &$   -0.39^{+    0.06}_{-    0.05}$\\
    $\le 0.2 M_\sun$ &  44 &$   -3.98^{+    0.13}_{-    0.12}$ &$   -0.62^{+    0.14}_{-    0.14}$ &$   -0.02_{-    0.07}^{+    0.08}$ &$   -1.73_{-    0.40}^{+    0.28}$ &$   -0.17^{+    0.08}_{-    0.06}$\\
   $\le 0.35 M_\sun$ & 102 &$   -3.94^{+    0.06}_{-    0.06}$ &$   -0.77^{+    0.08}_{-    0.08}$ &$   -0.06_{-    0.04}^{+    0.04}$ &$   -1.32_{-    0.12}^{+    0.11}$ &$   -0.28^{+    0.05}_{-    0.04}$\\
\hline
\noalign{\smallskip}
$R'_{\rm HK}$: \\
%\hline
\noalign{\smallskip}
%        whole sample & 124 &$   -4.27^{+    0.06}_{-    0.06}$ &$   -1.22^{+    0.08}_{-    0.10}$ &$   -0.53^{+    0.14}_{-    0.15}$ &$    0.56^{+    0.06}_{-    0.05}$ &$   -0.20^{+    0.04}_{-    0.04}$\\
%    $> 0.35 M_\sun$ &  48 &$   -4.12^{+    0.12}_{-    0.12}$ &$\log_{10}(0.13)$  & ... &   $0.98^{+    0.17}_{-    0.17}$ &$   -0.31^{+    0.06}_{-    0.06}$\\
    $> 0.35 M_\sun$\textsuperscript{a} &  80 &$   -4.18^{+    0.08}_{-    0.08}$ &$\log_{10}(0.13)$  & ... &$    -0.94_{-    0.11}^{+    0.11}$ &$   -0.30^{+    0.05}_{-    0.04}$\\
   $\le 0.35 M_\sun$ &  43 &$   -4.23^{+    0.10}_{-    0.09}$ &$   -0.98^{+    0.20}_{-    0.18}$ &$   +0.41_{-    0.13}^{+    0.13}$ &$    -1.01_{-    0.26}^{+    0.20}$ &$   -0.36^{+    0.07}_{-    0.06}$\\
\hline
\noalign{\smallskip}
$\langle B \rangle$: \\
%\hline
\noalign{\smallskip}
        whole sample & 149 &$    +3.47^{+    0.05}_{-    0.04}$ &$   -1.01^{+    0.06}_{-    0.06}$ &$   -0.09_{-    0.04}^{+    0.04}$ &$   -1.00_{-    0.06}^{+    0.05}$ &$   -0.39^{+    0.04}_{-    0.03}$ & 
        0.13 (fixed) & $-0.11^{+0.03}_{-0.03}$ & $-1.26^{+0.10}_{-0.10}$ & Rei22 \\
   $\le 0.35 M_\sun$ &  69 &$    +3.46^{+    0.05}_{-    0.05}$ &$   -0.81^{+    0.16}_{-    0.11}$ &$   -0.09_{-    0.04}^{+    0.03}$ &$   -1.01_{-    0.18}^{+    0.12}$ &$   -0.49^{+    0.05}_{-    0.05}$\\
\hline
\end{tabular}
\end{small}
\end{center}
\textsuperscript{a}: This fit was done only in the unsaturated arm using a fixed $Ro_{\rm sat} = 0.13$ and without the severe outlier J14257+236W. \\
\textsuperscript{b}: Reference shorthands: Rei14: \citealt{Reiners14}; New17: \citealt{Newton17}; Rei22: \citealt{Reiners22} \\
\end{table*}

\paragraph{H$\alpha$.}
Bolometric-normalized H$\alpha$-luminosities were collected from \citet{Newton17} and supplemented by \citet{Schoefer19}. Preference was given to the \citet{Newton17} measurements for two reasons: 1) the \citet{Newton17} catalogue of $L_{\rm H\alpha}/L_{\rm bol}$ contains more stars in this sample than \citet{Schoefer19} (209 versus 96)\footnote{This difference is mainly due to the choice made by \citet{Schoefer19} to compute the H$\alpha$ luminosity only for stars considered to be active, i.e. $pEW'_{\rm H\alpha} < -0.3 \AA$, effectively setting a floor at $L_{H\alpha}/L_{\rm bol}\sim 10^{-5}$. In contrast, \citet{Newton17} computed $L_{\rm H\alpha}/L_{\rm bol}$ also for stars exhibiting H$\alpha$ fluxes below this threshold. Therefore, they reached lower values of luminosity ratios ($\sim 10^{-6}$) and apparently probed the unsaturated regime to a greater extent.}; 2) to facilitate comparison with the fitted activity-rotation relation of \citet{Newton17}. To put all measurements on the \citet{Newton17} scale, we fitted a linear relation to the $\log(L_{\rm H\alpha}/L_{\rm bol})$ of 56 shared stars to derive a small correction that was applied to all the \citet{Schoefer19} values.\footnote{The systematic deviation originates in differences in measurement methodologies. The two works used somewhat different zero-points: \citet{Newton17} measured the excess emission with respect to the strongest observed H$\alpha$ absorption as a function of colour, whereas \citet{Schoefer19} measured excess emission with respect to the slowest rotating star of the same spectral type. It is debated whether the H$\alpha$
line behaves in a monotonic way at low activity levels \citep[e.g.][]{Cram79}, and a good zero point may not even exist.}       

In Fig. \ref{fig:rot-act} we plot \HL{196} stars with both $L_{\rm H\alpha}$ and $P_{\rm rot}$ determinations. In the same way as for X-ray, we fitted the relation from Eqn. \ref{eqn:act_ro} to $L_{\rm H\alpha}/L_{\rm bol}$ for all stars except \HL{20} whose periods have `D' (debated) designations. We fit to the whole sample and the $> 0.35\,M_\sun$ and $\le 0.35\,M_\sun$ subsets, Fig. \ref{fig:rot-act} shows the functions corresponding to the \HLr{medians of the parameter posteriors} for the $> 0.35\,M_\sun$ and $\le 0.35\,M_\sun$ fits. In addition, the fully convective sample was itself large enough such that we could further subdivide it into the mid- and late-M dwarfs across $0.2\,M_\sun$ and still get convergent fits (not plotted). We used them to investigate any finer gradations in the mass-dependence of the fitted relations. The best-fit parameters for all five subsamples are given in Table \ref{t:rotact_bestpars}. Overplotted in Fig. \ref{fig:rot-act} is the relation given by \citet{Newton17}, whose sample spans a similar mass range to ours. Note that \citet{Newton17} computed $Ro$ using the equations from \citet{Wright11}, which were not properly calibrated for fully convective stars.   

The saturation regime presents a slight discrepancy from \citet{Newton17}. Their relation assumed a constant saturation emission level. In our overall sample, we find a slight downward slope of $-0.08 \pm 0.05$, formally incompatible with a null slope. The few early-M dwarfs on the saturation branch exhibit an even stronger downward trend ($-0.23 \pm 0.16$). Though the uncertainties are much larger, it seems to suggest that chromospheric emission as traced by H$\alpha$ could have a weak $Ro$-dependence even upon saturation, similar to that seen in X-ray and $B$-fields. On the other hand, the mid- and late-M dwarfs in the saturated regime have progressively milder slopes, mirroring the tentative trend observed in X-ray. 

The best-fit saturation breakpoint in the early-M subsample is somewhat smaller ($Ro_{\rm sat} \sim 0.12$) than that in the late-M subsample ($Ro_{\rm sat} \sim 0.17$), the latter being comparable to $Ro_{\rm sat} = 0.21 \pm 0.02$ from \citet{Newton17}. 
However, the slopes in the unsaturated arm are not as strong in our sample (about $-1.3$ to $-1.4$) as in \citet{Newton17} ($-1.7 \pm 0.1$ therein). In general, it may be more difficult to interpret comparisons in the unsaturated regime, because the measurement of H$\alpha$ luminosity is more problematic and uncertainties may be large. Figure 7 in \citet{Newton17} shows significant downward scatter in the tail end of the unsaturated branch, below $L_{\rm H\alpha}/L_{\rm bol} \sim 10^{-5}$, similar to here. Therefore, we do not assign too much meaning to the fit differences in the unsaturation regime.

\paragraph{Ca{~\sc ii} H\&K.}

Ca{~\sc ii} H\&K flux is commonly expressed with the metric $R'_{\rm HK}$. $R'_{\rm HK}$ time series were measured for 200 stars in the CARMENES sample using assorted public archival spectra covering the blue wavelength ranges \citep{Perdelwitz21}.  
For each star with multiple $R'_{\rm HK}$ measurements, we took the mean of its time series weighted by the errors as its $R'_{\rm HK}$ value. We plot $R'_{\rm HK}$ against $Ro$ for \HL{144} stars, of which \HL{124} are deemed to have non-debated periods. 

Since there are very few early-M dwarfs with $R'_{\rm HK}$ on the saturation branch, we fitted the two-arm rotation-activity relation Eqn. \ref{eqn:act_ro} for $R'_{\rm HK}$ only to \HL{43} fully convective M dwarfs. For the early-M dwarfs, we fitted the relation to the unsaturated arm\footnote{This fit was done without the severe outlier J14257+236W, whose long period is provisional and whose inclusion would have drastically skewed the fit.} using a fixed $Ro_{\rm sat} \equiv 0.13$. Intriguingly, for the fully convective M dwarfs, we find a very significant upward trend with slope $0.41 \pm 0.13$ in the saturation regime, implying receding emission with faster rotation for the rapid rotators. If robust, this could be evidence of chromospheric supersaturation \citep{Christian11}. However, the fact that this trend is not matched in H$\alpha$ or coronal emission suggests possible systematics in the measurement of $R'_{\rm HK}$ for these late-type fast rotators \citep[see below as well as discussion in][]{Perdelwitz21}.

Several previous works have explored the $R'_{\rm HK}$-rotation relation for low-mass stars. \citet{AD17a} studied the $R'_{\rm HK}-P_{\rm rot}$ relation in 38 predominantly early-type M dwarfs. On the other hand, \citet{Boudreaux22} measured $R'_{\rm HK}$ for 53 fully convective M dwarfs with known rotation periods. In Fig. \ref{fig:rot-act}, we overplot the two relations, given by \citet{Boudreaux22} in terms of $R'_{\rm HK}$-$Ro$, based on \citet{Wright18}'s $\tau_{\rm conv}$-$(V-K_s)$ relation. 

For rapid rotators with $P_{\rm rot} < 10$\,d  (corresponding to $Ro \lesssim 0.14$), \citet{AD17a} found a constant $R'_{\rm HK} = -4.045$. \citet{Boudreaux22} showed that later-type M dwarfs have a notably lower Ca~{\sc ii} H\&K saturation level at about $R'_{\rm HK} = -4.4$. However, neither found evidence of a dropoff in activity as $Ro$ decreases. In the unsaturated regime, the previous works find a log-linear correlation that is also traced out by our sample. The \citet{AD17a} and \citet{Boudreaux22} relations are offset from one another, suggesting a mass-dependent level of normalized Ca~{\sc ii} H\&K emission in the unsaturated regime. Qualitatively, such a mass-dependence manifests similarly in our sample.    

\begin{figure}
\centering
\includegraphics[width=\hsize]{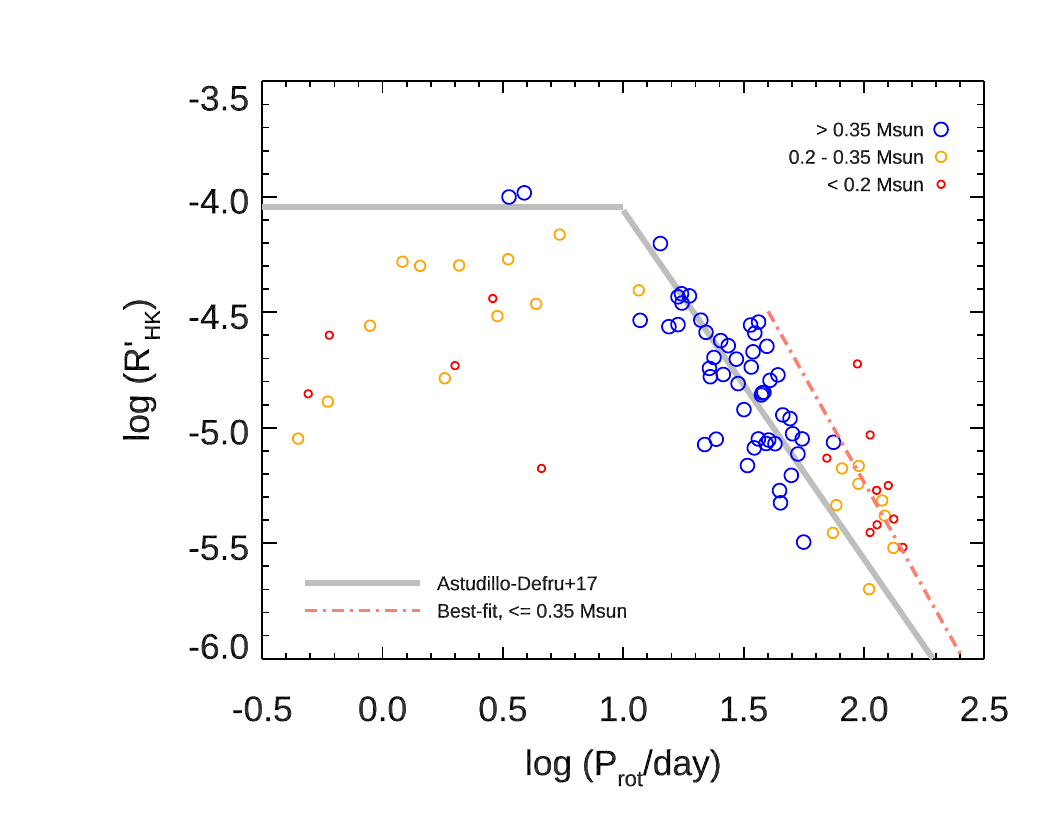}
\caption{Normalized Ca~{\sc ii} H\&K emission ($R'_{\rm HK}$) as a function of $P_{\rm rot}$. `D' and `P' rotators are omitted from the plot. The relation from \citet{AD17a} is overplotted in grey. To illustrate the offset with that relation for our fully convective stars in the unsaturated regime, we also show the best-fit power law for 18 `S' rotators with $M_\star \le 0.35 M_\sun$ and $P_{\rm rot} > 40$\,d in dashed red lines. }
\label{fig:rhk-prot}
\end{figure}

Following \citet[][Fig. 6 therein]{AD17a}, we plot our $R'_{\rm HK}$ against $P_{\rm rot}$ for stars with `secure' periods in Fig. \ref{fig:rhk-prot}. The correlation quantified by \citet{AD17a} is overplotted. While this empirical relation has limited physical meaning in itself, it has been widely applied in crudely estimating rotation periods of inactive stars in the unsaturated regime, whose $R'_{\rm HK}$ can be readily obtained. 

The agreement with our measurements is decent overall, with a root-mean-square scatter of $\sim0.17$\,dex in $P_{\rm rot}$. While the early-M dwarfs appears to follow the relation of \citet{AD17a}, our smallest stars ($< 0.2$\,$M_\sun$) appear to have systematically higher $R'_{\rm HK}$ for a given $P_{\rm rot}$. For illustration, we fitted a power law to the 18 fully convective stars with `secure' $P_{\rm rot} > 40$\,d to obtain the following:  
\begin{equation}
\log R'_{\rm HK} = -1.86 \log (P_{\rm rot}/\rm d) -1.52. 
\label{eqn:rhk_prot_conv}
\end{equation} 

Evaluated for $\log (R'_{\rm HK}) = -5.0$, \citet{AD17a} gives $P_{\rm rot} = 42$\,d whereas Eqn. \ref{eqn:rhk_prot_conv} would give $P_{\rm rot} = 74$\,d. Considering that the sample of \citet{AD17a} contained only 7 fully convective M dwarfs with $P_{\rm rot} > 40$\,d, it may be tempting to conclude that their relation is simply less accurate for mid- and late-M dwarfs. On the other hand, such a large discrepancy could also be explained by methodological differences in the way $R'_{\rm HK}$ is derived between the works. \citet{Perdelwitz21} pointed out that their $R'_{\rm HK}$ measurements are systematically offset by $\sim$\,0.2\,dex from that of \citet{AD17a} towards lower activity values and later-type stars, especially for $\log (R'_{\rm HK}) < -5.0$. Such an offset roughly corresponds to the size of the observed discrepancy. Therefore, when using an $R'_{\rm HK}$-$P_{\rm rot}$ relation to infer $P_{\rm rot}$, it is important to use the version correctly calibrated to the method-specific $R'_{\rm HK}$ measurements.

\paragraph{$B$-field.}
\citet{Reiners22} measured average surface magnetic field strengths (or upper limits) for \HL{283} M dwarfs in our sample by modelling high-resolution CARMENES spectra. They showed that the rotation-$B$ field relation takes an analogous form to that of other well-known rotation-activity relations. This has significant implications for the magnetic origin of stellar activity phenomena, as discussed extensively by \citet{Reiners22}. 

We show $\langle B \rangle$ versus $Ro$ for the \HL{162} stars in our sample with both $\langle B \rangle$ (excluding upper limits) and $P_{\rm rot}$ measurements in Fig. \ref{fig:rot-act}. Of these, 149 have `S' (secure) and `P' (provisional) period designations. As before, we fitted the canonical rotation-activity relation Eqn. \ref{eqn:act_ro} to $\langle B \rangle$. Due to the dearth of early-M dwarfs with B-field measurements on the saturated arm, these fits were not carried out on the early-M subset. The best-fit parameters are printed in Table \ref{t:rotact_bestpars}.  

The whole-sample fits agree well with the original relation given in \citet{Reiners22} (despite slight difference in the computation of $\tau_{\rm conv}$, see Section \ref{a:rossby}). The saturated regime exhibit a downward slope of $\sim -0.09$ significant to $>2\sigma$, comparable to the $-0.11\pm 0.03$ slope found by \citet{Reiners22}, indicating that surface magnetic field strengthens with decreasing Rossby number even in the nominal saturation regime \citep[see also][]{Kochukhov21}. The trend appears to extend down to $Ro=10^{-3}$.  

The free-fitted saturation breakpoint settled on $Ro_{\rm sat}\sim0.1$ and $\sim 0.16$ for the whole sample and the fully convective subset, respectively, which are fairly similar to the 0.13 assumed by \citet{Reiners22}. However, the slope of $-1.0$ in the unsaturated arm is somewhat in tension with that of \citet[][$\beta_{\rm unsat} = -1.26 \pm 0.10$]{Reiners22}. This discrepancy could be partially attributed to updates to the sample, as well as in differences in the fitting methods (i.e. in \citealt{Reiners22}, the ordinary least squares bisector method was used, which accounts for error in both X and Y variables).

\subsection{Period-mass, period-kinematics relations}\label{ss:mass_kinematics}

Several previous works have investigated the distribution of rotation periods as a function of stellar mass and age, in order to shed light on the  mechanisms of rotational evolution \citep[e.g.][]{Irwin09,McQuillan14,Newton16,Rebull18,Popinchalk21}. A striking finding relates to the shape of the upper and lower envelopes in $P_{\rm rot}$. The maximum $P_{\rm rot}$ has been shown to increase from $\sim 50$\,d in early-M dwarfs to $\sim 150$\,d in the latest-type M dwarfs, while the minimum $P_{\rm rot}$ decreases from $\sim 0.3$ to $\sim 0.1$\,d over the same mass range. Another key insight from these studies is the bimodality of the distribution among the mid- to late-type M dwarfs, clustering around $P_{\rm rot} < 5$ d and $ > 50$\,d. Relatively few stars have a rotation period intermediate between the two extremes, with the gap opening wider towards lower masses. On the other hand, earlier-type stars are much less likely to be found as fast-rotators. Since stars across the mass range are born with statistically indistinguishable primordial rotation period distributions \citep{Venuti17,Popinchalk21}, any differences in their final rotation distributions must mean they have experienced different angular momentum shedding pathways over time. 

\begin{figure*}
\centering
%\vspace{-1cm}
\hspace{-1cm}
\includegraphics[width=\hsize]{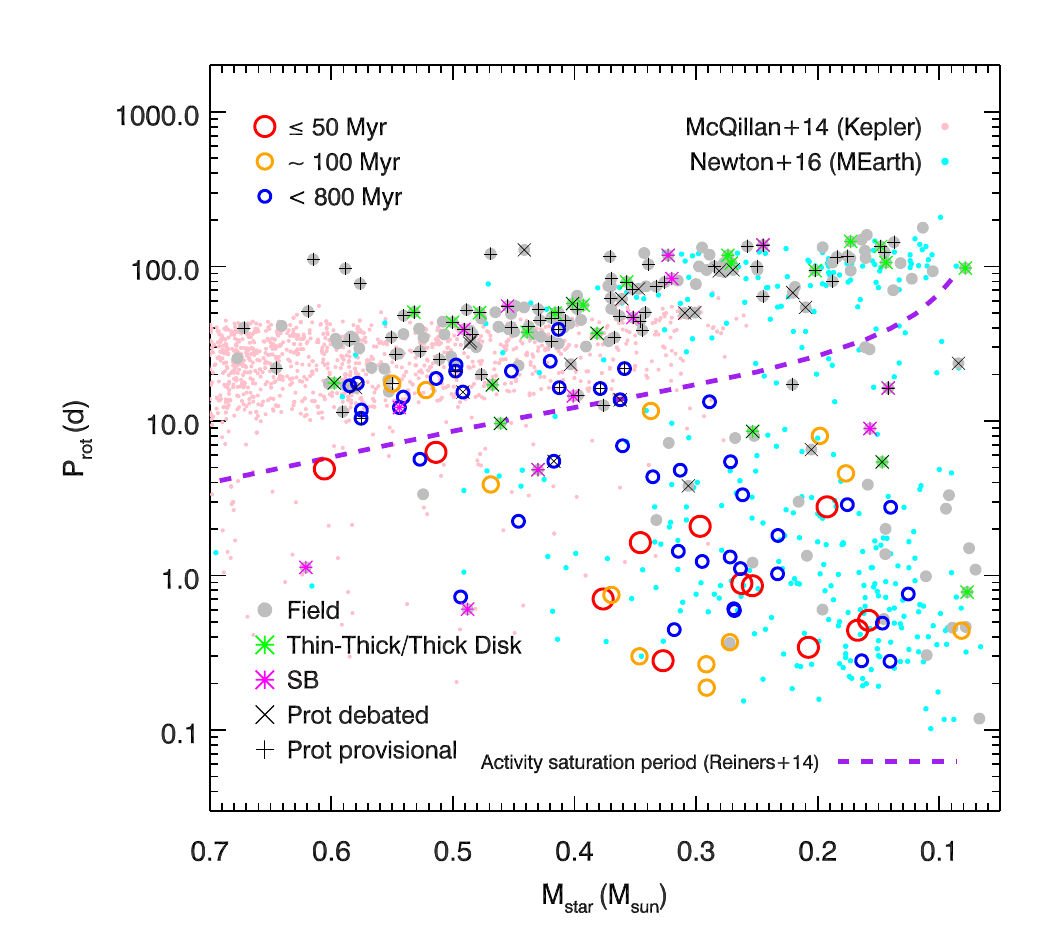}
%\vspace{-0.5cm}
\caption{Measured $P_{\rm rot}$ versus $M_{\rm star}$. Pink and cyan points represent stars from \citet{McQuillan14} and \citet{Newton16} (`A' and `B' rotators only), respectively. The purple dashed line represents the saturation period as a function of stellar mass, based on Eqn 10 of \citet{Reiners14} and $L_{\rm bol}$ from 5-Gyr PARSEC isochrones \citep{Bressan12}. The empty circles denote young stars in the CARMENES sample assigned to stellar kinematic groups by \citet{Cardona23}. They are colour- and size-coded by age, as indicated in the legend. Filled grey points represent the rest of the stars from the CARMENES sample \HLr{(collectively termed `field')}. Additional symbols overlaid on the CARMENES stars: magenta asterisks mark known SBs or spectroscopic triples; green asterisks highlight stars belonging to the thin-thick or thick disk based on Galactic kinematics; black `X's and `+'s denote stars with debated or provisional periods, respectively.  %similar to Fig 9 in Jeffers+18
}
\label{fig:prot_vs_mstar_vs_age}
\end{figure*}

How the CARMENES survey sample fits into the context of the period-mass relationship for M dwarfs is displayed in Fig. \ref{fig:prot_vs_mstar_vs_age}. We plot the $P_{\rm rot}$ values against the stellar masses for our sample. The background shows measurements by \citet{McQuillan14} and \citet[][`A' and `B' rotators only]{Newton16} from {\em Kepler} and MEarth light curves, respectively. Unlike previous studies, which often used optical colours as proxies for mass, the stellar masses for the CARMENES sample are homogeneously determined as detailed in Sect. \ref{ss:vsini}.

The CARMENES sample traces out the major features that have been noted by previous authors. Though our sample is not as large as some of its predecessors, it generously covers both sides of the partial-to-fully convective boundary, thereby better resolving the upper envelope in the transition region between the early and mid-M dwarfs at $\sim 0.4\,M_\sun$ \citep[see also Fig. 12 in][]{Popinchalk21}. The large quantity of measurements from Kepler light curves shows a sharp boundary at 40--50\,d, which seemingly represents the maximum rotation period attained for early M dwarfs \citep{McQuillan14}. Our sample appears to challenge this notion, exhibiting a fuzzy upper envelope sprinkled with several stars with $P_{\rm rot}\sim100$\,d. However, none of these long periods have the status `secure'. To understand the end state of early-M dwarf spin evolution, it would be useful to clarify these periods with additional measurements.     

An underlying variable in the period-mass distribution is stellar age, which governs the stage of a star's spin-down. While precise ages for M dwarfs are difficult to determine directly, Galactic kinematic information can provide constraints. Since CARMENES is an RV survey, all stars have their $v_{\rm rad}$ measured from high-resolution spectra. Combining these with Gaia proper motions gives a full picture of the 3D Galactic kinematics of these stars. \citet{Cardona23} matched Galactic kinematics of the CARMENES sample to known young moving groups in the solar neighbourhood and deduced ages for a young subsample up to $\sim800$\,Myr. The age-resolved period-mass diagram is also shown in Fig. \ref{fig:prot_vs_mstar_vs_age}. Despite its coarseness, the age dimension gives structure to the period-mass scatter plot. The earliest M dwarfs ($\gtrsim 0.5\,M_\sun$) already begin to settle onto the slow-rotating sequence at $\sim100$\,Myr. By 800\,Myr the sequence includes most stars with $\gtrsim 0.4\,M_\sun$. SBs identified by \citet{Baroch18,Baroch21} are marked and can account for most of the few early-M field stars that fall significantly below the slow-rotator sequence. The fact that the slow rotator sequence is not yet in place by 800 Myr across the entire M dwarf mass range has also been established by other authors using K2 data in open clusters \citep[e.g.][]{Douglas17, Popinchalk21}. 

The sample exhibits a bimodality of periods below $\sim 0.4 \,M_\sun$, with a gap of intermediate periods separating the slow- and fast-rotating sequence. This is a consequence of the much longer overall timescales of spin-down in fully convective stars. Assuming that the star formation history in the solar neighbourhood has not been punctuated, \citet{Newton16} proposed that the gap feature among fully convective stars can be explained by first stalled, then rapid spin-down past a given $P_{\rm rot}$. A similar phenomenon has been reported in sunlike stars and early M dwarfs \citep[e.g.][]{Curtis20}. Here we show that a large fraction of the fast-rotators can be associated with a young age. Within this range of ages below $\sim 1$ Gyr, the collection of rapid rotators among the late-M dwarfs lacks clear stratification. This region in the $P_{\rm rot}$ - $M_\star$ space likely comprises a mixture of the natal distribution of rotation rates, spin-up due to size contraction during the long pre-main sequence lifetimes (a few hundred Myr), and subsequent spin-down \citep[see also][]{Popinchalk21}.

Since there are few rotators in the intermediate-period transition region ($\sim 30$\,d) and none have an inferred age, it is difficult to directly study the hypothesis of \citet{Newton16}. Interestingly, the position and shape of the gap appear to be correlated with the phenomenon of activity-saturation, which could provide a clue to its origin. As in \citet{Jeffers18}, we plot in Fig. \ref{fig:prot_vs_mstar_vs_age} the X-ray saturation period as a function of stellar mass based on Eqn. 10 from \citet{Reiners14}. To apply this equation, we mapped $M_\star$ into $L_{\rm bol}$ using a 5-Gyr PARSEC isochrone at solar metallicity \citep{Bressan12}. The saturation period curve divides the clusters of slow and fast rotators and traces the shape of the lower envelope of the slow period sequence. There is a marked deficit of stars in the neighbourhood of this curve, which coincides with the gap. This seems to illustrate a physical link between angular momentum evolution and activity saturation: a star experiences efficient spin-down until it has exited the saturation phase \citep[e.g.][]{ReinersMohanty12}. Many mid- to late-type M dwarfs appear to remain in the saturation regime into field age (i.e. $\gtrsim 1$\,Gyr), which is also likely related to their observed activity lifetimes of several Gyr \citep{West08}.  

To observationally resolve the spin evolution sequence between $\sim 1$\,Gyr and the field requires ideally a comprehensive study of rotation periods for M dwarfs in benchmark open clusters of intermediate age. However, few such clusters are nearby, and none close enough to probe down to M dwarf brightness \citep[e.g.][]{Curtis20}, at least not without resource-intensive campaigns on large ground-based telescopes \citep[e.g. M67,][]{Dungee22}. 
Another crude age indicator is membership in the Galaxy's structural components. In Figure \ref{fig:prot_vs_mstar_vs_age} we indicate the stars kinematically classified to be in the thin-thick or thick disk component of the Galaxy \citep{MCC17}, which correspond to stars $\gtrsim 8$ Gyr \citep[e.g.][]{Fuhrmann98}. Most of them seem to fall on the upper envelope of the slow rotator sequence, consistent with the spin-down framework. 
%\HLr{The rest of the stars are part of the disk or young disk.}   

\begin{figure}
\centering
\includegraphics[width=\hsize]{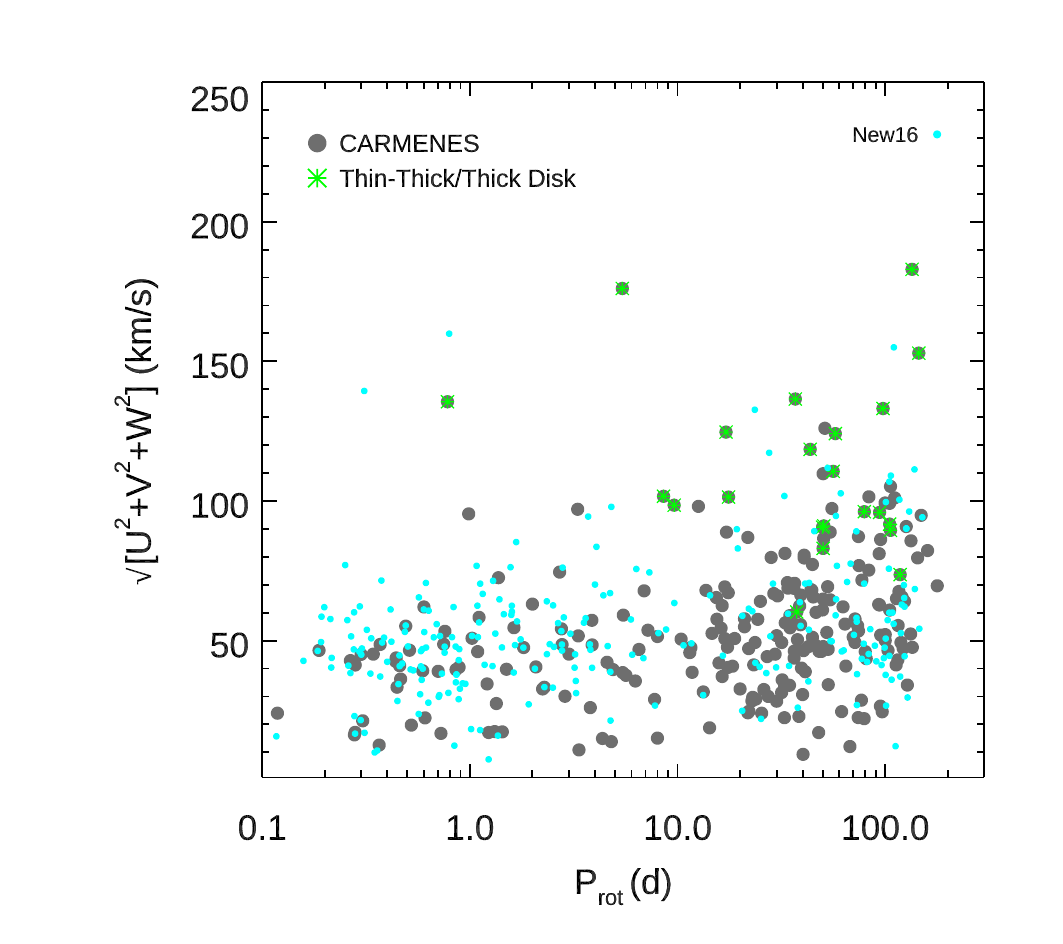}
\caption{Measured $P_{\rm rot}$ vs. galactic kinematics. Grey filled points represent the CARMENES sample presented in this work. Cyan points denote stars from \citet{Newton16} (`A' and `B' rotators only). The green asterisks are classified to be thin-thick or thick disk stars. }
\label{fig:prot_vs_uvw}
\end{figure}

%To probe the transition towards the slow rotator sequence, one needs to look to older M dwarfs in the field. 
Relatedly, Galactic velocity dispersion can also serve as a direct (if crude) age proxy, whereby larger velocity dispersion is associated with older age \citep[e.g.][]{Newton16,Kiman19,Angus20,Lu21}. Figure \ref{fig:prot_vs_uvw} compares the Galactic kinematics (as measured by Gaia and CARMENES RV) with the measured $P_{\rm rot}$ for the CARMENES GTO sample, as well as for the `A' and `B' rotators in \citet{Newton16}. As in \citet{Newton16}, we use $\sqrt{U^2+V^2+W^2}$ from Gaia EDR3 to quantify the total space motion of each star, where $U$, $V$ and $W$ are the radial, azimuthal, and vertical velocity components with respect to the local standard of rest ($U_\odot, V_\odot, W_\odot = [11, 12, 7]$ km s$^{-1}$). Stars with larger rotation periods tend to have a larger spread in space velocities and larger average space motions, which is expected given slower rotators tend to be older stars. This is an average trend and does not apply in the one-on-one sense to individual stars, since the process of dynamically heating the Galactic disk is stochastic.

Recently, \citet{Pass22} showed that yet another promising avenue to investigate the rotation history of M dwarfs is to study those with wide binary companions (FGK or white dwarfs) whose ages can be constrained. They narrowed the window for rapid spin-down to around 2--3\,Gyr. Using gyro-kinematic ages on a large sample of stars, \citet{Lu23} identified evidence of fundamental changes to the spindown mechanism across the fully-convective boundary, with fully convective stars experiencing significantly higher angular momentum loss rates. 
%\HLr{If true, perhaps it could explain the transition in the upper envelope of periods in early- and late-M dwarfs.} 
The advent of new gyrochronological relations calibrated for M dwarfs \citep{Dungee22,Gaidos23,Engle23,Lu23b} will shed further light on the spin evolution of low-mass stars and cement $P_{\rm rot}$ as a critical input in constraining ages.  

\subsection{Planet hosts}\label{ss:planethosts}

\begin{figure}
\centering
\hspace{-1cm}
\includegraphics[width=1\hsize]{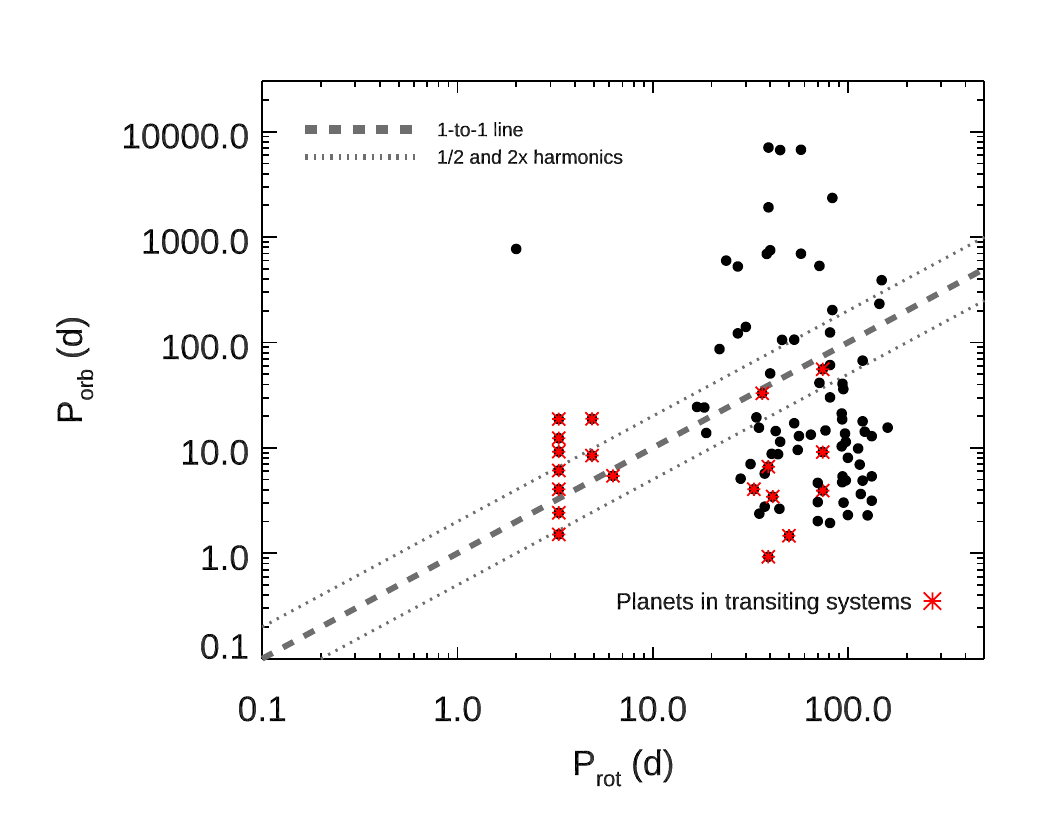}
\caption{Orbital period of planets versus spin period of host stars. Planets in transiting systems are denoted by red asterisks. The 1-to-1 relation is in grey dashed lines, while the 1/2$\times$ and 2$\times$ harmonics are in grey dotted lines. }
\label{fig:porb_vs_prot}
\end{figure}

\begin{figure}
\centering
\vspace{-1cm}
\hspace{-0.5cm}
\includegraphics[width=1.1\hsize]{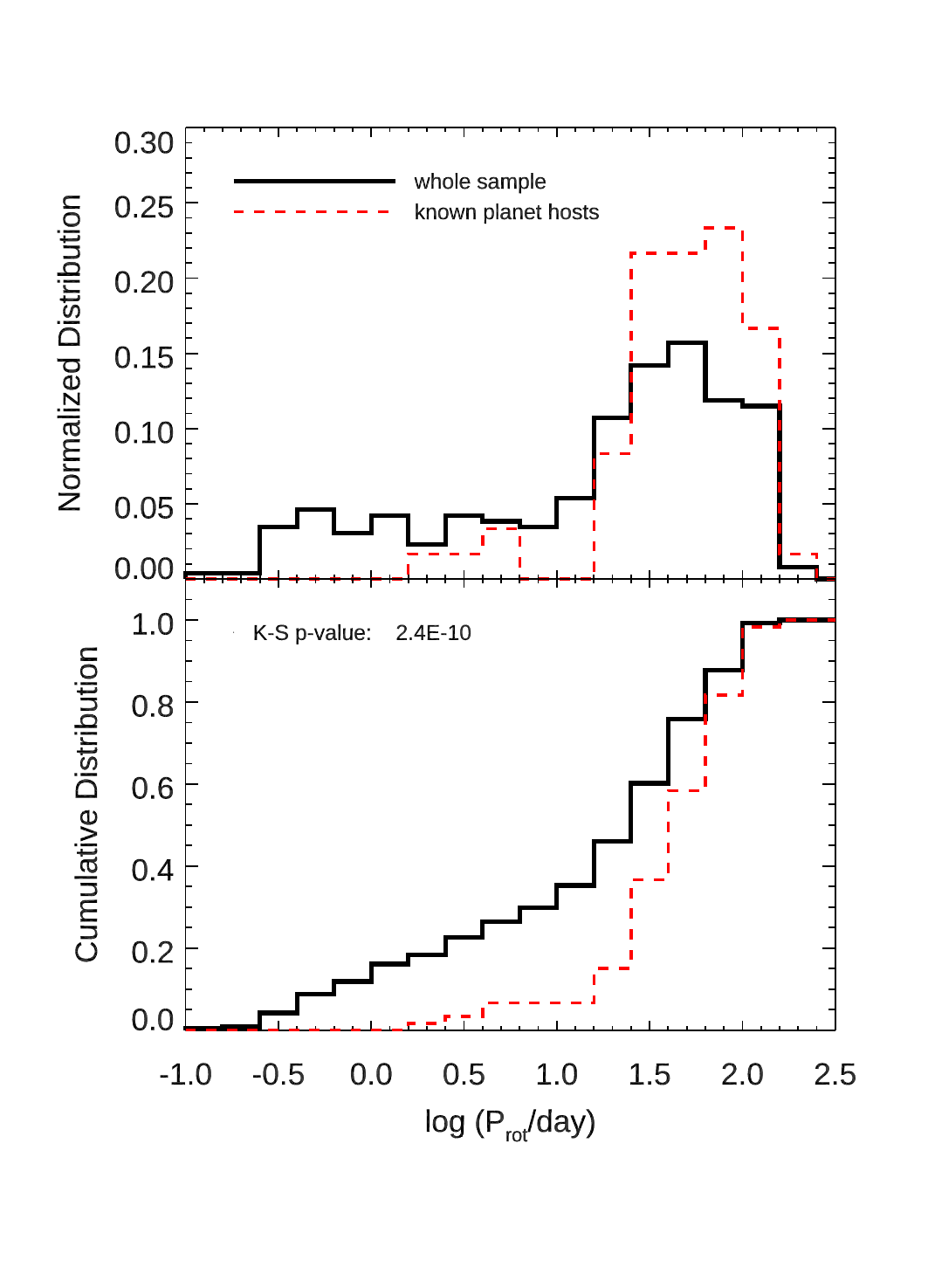}
\vspace{-2cm}
\caption{Rotation period distribution of all known planet hosts (red dashed) compared to the entire sample (black solid). {\em Top:} differential distribution. {\em Bottom:} cumulative distribution. The p-value of the 2-sample K-S test is $2.4 \times 10^{-10}$, i.e. the two distributions are very likely to be distinct. }
\label{fig:planet_hosts}
\end{figure}

There are 97 known planets across 63 systems in our 348-star sample (see Table \ref{t:planets}). Figure \ref{fig:porb_vs_prot} shows the planet orbital vs. stellar rotation periods, and Figure \ref{fig:planet_hosts} plots the raw rotation period distribution of known planet hosts compared to the overall sample. Relative to the ambient stellar background, planet hosts appear to have a significant preference for slower rotation. According to a two-sample Komolgorov-Smirnov (K-S) test, the distributions are distinct at high significance ($p$-value = $2.4\times 10^{-10}$). 

The discrepancy could be rooted in astrophysics or due to observational bias. RV surveys have more complex selection effects, in part because their sampling strategies are often concerned with maximizing planet yields. A clue that favours observational bias as the explanation for the relative dearth of fast-rotating planet hosts in our sample is that, while the majority of planets in this sample were found by RV, three out of the four planet systems around the stars with $P_{\rm rot} < 15$\,d were discovered by transit (K2-33, AU Mic, TRAPPIST-1), the exception being TZ Ari. Since the RV method is less sensitive to planets around active stars, the monitoring of certain targets, many of which are fast rotators, were discontinued early on \citep{Sabotta21,Ribas23}. 

To control for the effect of RV survey selection bias against active stars, we cross-matched our sample of stars with $P_{\rm rot}$ measurements with that used in the planet occurrence rate analysis of \citet{Ribas23}, which is also based on the M dwarfs observed in the CARMENES GTO survey. The sample was debiased by discarding RV-loud stars \citep{Talor18}, spectroscopic binaries and triples (mostly from \citealt{Baroch18} and \citealt{Baroch21}), as well as excluding systems with transiting planets not in the original input catalogue (especially TESS planets), or with fewer than ten RV measurements. Of the 170 stars in common between the two samples, 51 have known planets. In addition, there are 68 stars in the sample of \citet{Ribas23} without a $P_{\rm rot}$ determination, four of which have planets. On the other hand, there are another 91 stars, nine of which have known planets, with a $P_{\rm rot}$ determination but discarded from the planet occurrence rate analysis of \citet{Ribas23}. 

\begin{figure}
\centering
\hspace{-1cm}
\includegraphics[width=1.1\hsize]{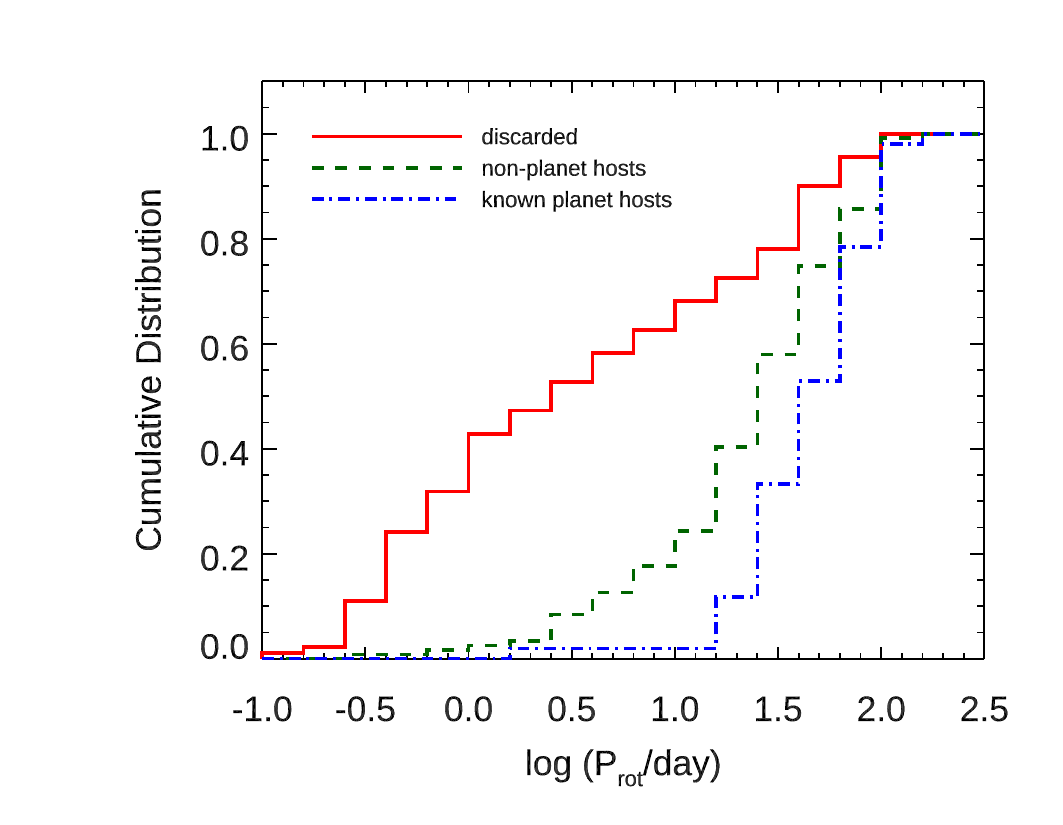}
\caption{Cumulative distribution of rotation periods in the context of RV selection biases. Red solid: rotation period distribution of stars discarded in the planet occurrence analysis of \citet{Ribas23}; green dashed: stars in common with \citet{Ribas23} that are not known to be planet hosts; blue dot-dashed: known planet hosts in \citet{Ribas23}. }
\label{fig:planet_hosts_ribas23}
\end{figure}

We compare the $P_{\rm rot}$ distribution for the discarded and de-biased sample in Fig. \ref{fig:planet_hosts_ribas23}. About 75\% of the discarded stars have $P_{\rm rot} < 15$\,d, with a broad maximum at about 0.5--2\,d. More than 40\% of these stars are RV-loud stars, among \HLr{them} young transiting planet systems such as K2-33 \citep{David16} and AU~Mic \citep{Plavchan20}. In the de-biased sample of \citet{Ribas23}, the distributions of planet hosts and non-hosts are identical for $P_{\rm rot} \gtrsim 15$\,d. Previous works mostly focused on sunlike stars have found that there is no relation between the $P_{\rm rot}$ of moderate and low rotators and the presence of planets \citep[e.g.][]{Walkowicz13,Ceillier16}. However, to our knowledge, our work represents one of the first times that this is established for M dwarfs.  

At $P_{\rm rot} \lesssim 15$\,d, a significant \HLr{deficit of stars with known planets compared to those without} persists even in the de-biased sample. This could still be attributed to detection bias. That is, everything else being equal, it is more difficult to detect a low-RV amplitude planet of a given mass and orbital period around an M dwarf with a shorter $P_{\rm rot}$, owing to stronger magnetic field and greater RV jitter \citep{Reiners18,Reiners22}. Such a bias would not be completely eliminated by the target exclusion criteria of \citet{Ribas23}. This explanation is most plausible because fast-rotation is an evolutionary phase in the star's lifetime, and the slow-rotating stars, virtually all of which have on average at least one planet \citep{Ribas23}, started out as fast-rotators. It is also possible that planets in M dwarf systems tend to start out on larger orbits and migrate secularly inwards over time, making younger planets more difficult to detect in RV surveys than their mature counterparts. 

The fact that astrophysics may play a role in sculpting the dearth of planets discovered around fast-rotators has been proposed before. Using three years of data from {\em Kepler}, a transiting planet survey largely immune to RV biases, \citet{McQuillan13b} discovered a notable lack of close-in planets ($P_{\rm orb} \lesssim 3$\,d) around fast-rotating stars ($P_{\rm rot} \lesssim 10$\,d). They discussed aspects of physical coupling in the angular momentum evolution of the star-planet system, including processes such as stellar magnetic braking, planet migration, and tidal interactions, which could affect the demographics of planets as a function of stellar rotation state. However, the exact mechanism leading to their specific observed pattern was unclear. Recently, \citet{Sibony22} found evidence that planet-hosting stars rotate 1.6 d slower than non planet-hosts, a small but significant difference. However, they could not identify the root cause responsible for this difference. Both works dealt chiefly with sunlike stars.    
%For example, perhaps stars that rotate faster kept the angular momentum that would have otherwise gone into the formed planets. 

To disentangle the two scenarios would require a detailed completeness analysis on the data as well as star and planet formation and evolution models beyond the scope of this manuscript. Nevertheless, our work demonstrates that, even with a state-of-the-art observing programme such as CARMENES, there are still significant challenges in studying planets around young and active M dwarfs using the RV technique, restricting our ability to probe nascent planet populations in early stages of evolution.

%It is plausible that star-planet connections, such as formation and co-evolution processes that couple their angular momentum evolution, could affect the demographics of planets as a function of stellar rotation state. 

\begin{figure}
\centering
\hspace{-1cm}
\includegraphics[width=1.1\hsize]{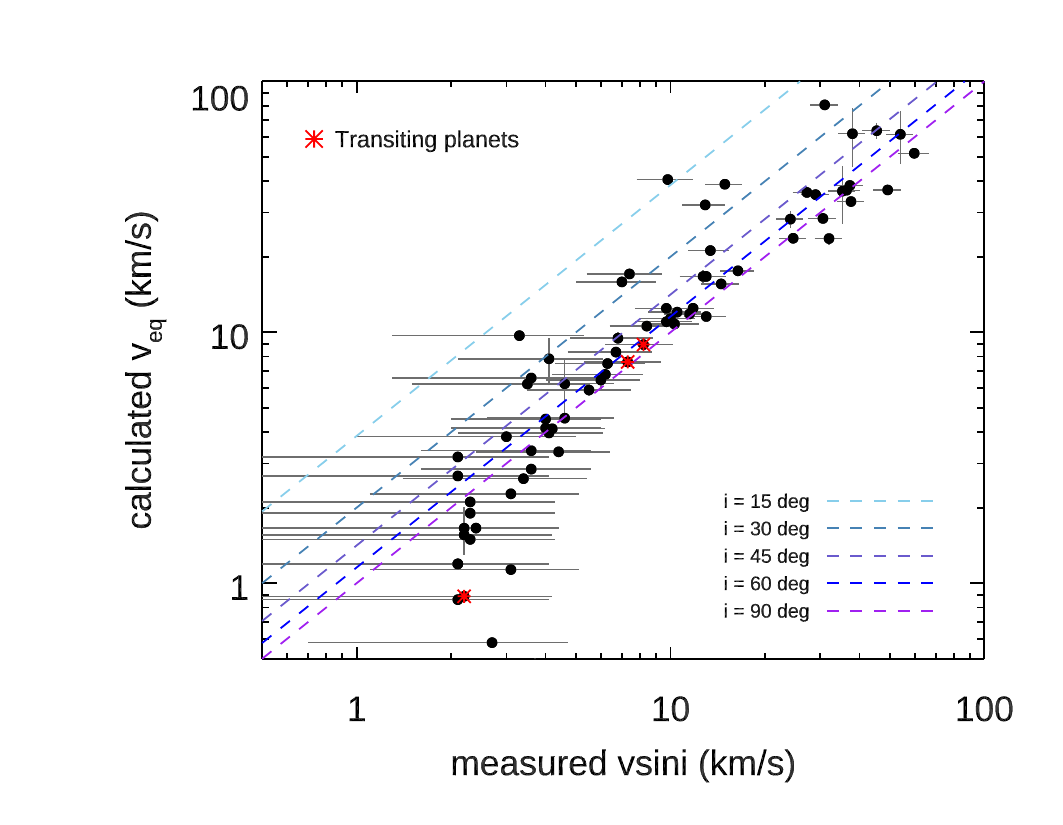}
\caption{Projected spin inclination angles for stars with measured $v\sin i$. Transiting planet systems are highlighted in red asterisks and consistent with a spin-orbit-aligned configuration. The rest are not known to be planet hosts. }
\label{fig:spin_inclinations}
\end{figure}

For stars with known $v\sin i$, $R_\star$, and $P_{\rm rot}$, the line-of-sight inclination of the stellar rotation axis can be constrained (see Eqn. \ref{eqn:prot_max}). Figure \ref{fig:spin_inclinations} presents these stars overplotted on iso-inclination contours. Those stars falling beyond the $90$\,deg contour are outliers discussed in Sect. \ref{ss:vsini} and Appendix \ref{a:vsini_outliers}. The only known planet hosts among these stars are transiting planet systems, and they are consistent with spin-orbit alignment.         

\section{Summary}\label{s:sum}

In this work, we further characterize the CARMENES survey input sample of 348 nearby bright M dwarfs by presenting new rotation period determinations from photometric and spectroscopic time series data. Our data come from ground-based photometric observations (e.g. the SuperWASP survey), from the TESS satellite, and from the CARMENES spectroscopic survey itself. Standard periodogram analyses followed by careful vetting yielded \HL{158} periods for \HL{129} stars (some stars have measurements in multiple data sets), ranging between 0.12\,d and 140\,d. While the majority of short periods (< 5\,d) were measured by single-sector TESS-PDCSAP data and longer periods (> 10\,d) by ground-based surveys spanning year-long baselines, we found multi-sector TESS SAP light curves to be effective for detecting or verifying a number of intermediate periods ($\sim$5-- 20\,d). Together with measurements already available in the literature, the total number of stars in this sample with a measured rotation period is now \HL{261} or \HL{75\%}.       

Leveraging the comprehensive and uniform catalogueof stellar properties for this sample that has been recently made available, we study $P_{\rm rot}$ as a function of various stellar parameters. We find the periods to have very good consistency with $v\sin i$ and $R_\star$ measurements in the subsample for which all three properties have been determined. $P_{\rm rot}$ exhibits the expected overall negative correlation with the $pEW'_{\rm H\alpha}$ metric, a proxy for H$\alpha$ emission, though with considerable scatter. Combining information from $v\sin i$, $pEW'_{\rm H\alpha}$, and agreement with literature measurements, we recommend one $P_{\rm rot}$ value for each star and assign a confidence rating of `secure', `provisional', or `debated'.

Since many of the stars in our sample have multiple period measurements available in the literature, we assess their typical degree of consistency as a function of period itself. 
Fitting a power law, we find the dispersion of independent period measurements for a star to be strongly positively correlated with period value. We adopt this power law function as an empirical estimate of $P_{\rm rot}$ error. Short periods (<10 \,d) are most likely to be very precise ($\sim 1$\%), whereas longer periods (> 100\,d) can be determined to no better than $\sim 10$\%. Furthermore, 25\% of periods longwards of 10\,d have a dispersion $> 20$\%. In general, published formal errors fall short of the empirical errors, by 1.5$\times$ for $P_{\rm rot} \gtrsim 10$\,d and 2$\times$ for $P_{\rm rot} \gtrsim 100$\,d. The trend of growing error with period value chiefly reflects the fact that longer $P_{\rm rot}$'s are in practice more difficult to measure owing to a variety of factors, and that the difficulty scales more than proportionally with the length of the period. We recommend using independent measurements to verify $P_{\rm rot}$'s longwards of $\sim 10$ d. Since the habitable zones around all but the latest M dwarfs lie beyond the orbital period of $\sim 10$\,d \citep{Newton16b,Kopparapu13}, careful determination of stellar rotation periods in this range is critical to rule out false positives and confirm the most interesting RV planet detections.

We show the sample in the context of rotation-activity relations for chromospheric (H$\alpha$, Ca~{\sc ii} H\&K) and coronal (X-ray) emission, as well as surface B-field strength as a function of the Rossby number. Our fits are broadly consistent with those obtained by previous authors. We find small but suggestive evidence for mass-dependence in the saturation behaviour of X-ray and H$\alpha$ emission across the fully-convective boundary, as well as a possible non-constant level of H$\alpha$ emission in the saturation regime. A larger sample building on the current one is needed to test the robustness of these results. We show that choice of methodologies in measuring Ca{~\sc ii} H\&K emission can have a significant quantitative impact on the $R'_{\rm HK}-P_{\rm rot}$ relation.  

We also examine $P_{\rm rot}$ versus stellar mass, galactic kinematics, and age, confirming the previous findings of an upper envelope \HLr{in $P_{\rm rot}$} that increases with stellar mass across the M dwarf range, and a relative dearth of rotators with intermediate periods ($\sim$5--50\,d) among fully convective stars. Early-M dwarfs reach the bottom of the slow-rotator sequence by 800\,Myr and continue to spin down. Fully convective M dwarfs remain as fast rotators in the activity-saturated regime until beyond 1\,Gyr, but eventually reach $> 100$\,d periods. The deficit of intermediate-period rotators among mid- to late-M dwarfs appears to coincide with their activity-saturation periods, suggesting a physical link. The slowest rotators are associated with the largest space motion, consistent with belonging to older kinematic components of the Galaxy. Several early-M dwarfs appear to have rotation periods in the $\sim 100$\,d range, much slower than the upper limit suggested by Kepler data \citep{McQuillan14}, which warrants further investigation with independent measurements.

Finally, we point out that the known planet hosts in this sample are overwhelmingly preferentially slow rotators ($P_{\rm rot} \gtrsim$15\,d). Since the majority of planets in this sample were found by RV, such an outcome predominantly reflects selection and detection biases in blind RV surveys. Considering most M dwarfs with ages $\lesssim 1$\,Gyr have $P_{\rm rot} \lesssim $ 15\,d, there are evidently important challenges in using the RV technique to discover and study young planet systems. 

%\HL{what about metallicity? Metallicity: no trend visible for this sample in Prot-Mstar space} 

%Importance of having a catalog of well-characterized M dwarfs 

%Importance of vetting the periods, esp. longer periods (more scatter in the literature? aliases)  

%==================================================================
\begin{acknowledgements}
We thank the anonymous reviewer for insightful and constructive suggestions that significantly improved the quality of this work. CARMENES is an instrument for the Centro Astron\'omico Hispano-Alem\'an de Calar Alto (CAHA, Almer\'{\i}a, Spain).   CARMENES is funded by the German Max-Planck-Gesellschaft (MPG), the Spanish Consejo Superior de Investigaciones Cient\'{\i}ficas (CSIC), the European Union through FEDER/ERF FICTS-2011-02 funds, and the members of the CARMENES Consortium 
  (Max-Planck-Institut f\"ur Astronomie,
  Instituto de Astrof\'isica de Andaluc\'ia,
  Landessternwarte K\"onigstuhl,
  Institut de Ci\`encies de l'Espai,
  Institut f\"ur Astrophysik G\"ottingen,
  Universidad Complutense de Madrid,
  Th\"uringer Landessternwarte Tautenburg,
  Instituto de Astrof\'{\i}sica de Canarias,
  Hamburger Sternwarte,
  Centro de Astrobiolog\'{\i}a and
  Centro Astron\'omico Hispano-Alem\'an), 
  with additional contributions by the Spanish Ministry of Economy, 
  the German Science Foundation through the Major Research Instrumentation 
    Programme and DFG Research Unit FOR2544 ``Blue Planets around Red Stars'', 
  the Klaus Tschira Stiftung, 
  the states of Baden-W\"urttemberg and Niedersachsen, 
  and by the Junta de Andaluc\'{\i}a.
The Joan Or\'o Telescope (TJO) of the Montsec Observatory (OdM) is owned by the Generalitat de Catalunya and operated by the Institute for Space Studies of Catalonia (IEEC). This work makes use of observations from the Las Cumbres Observatory global telescope network.
We acknowledge financial support from
the Agencia Estatal de Investigación of the Ministerio de Ciencia e Innovación 
through projects PID2019-109522GB-C5[1:4],
%We acknowledge financial support from the Agencia Estatal de Investigación of the Ministerio de Ciencia, Innovación y Universidades through projects PID2019-109522GB-C52, 
PID2019-107061GB-C64, and PID2019-110689RB-100 and the Centre of Excellence `Severo Ochoa' Instituto de Astrof\'\i sica de Andaluc\'\i a (SEV-2017-0709). 
Data were partly collected with the 90-cm and 150-cm telescopes at Sierra Nevada Observatory (OSN), operated by the Instituto de Astrof\'\i fica de Andaluc\'\i a (IAA, CSIC). This work was also funded by the Generalitat de Catalunya/CERCA programme, and the DFG through the priority programme SPP 1992 ``Exploring the Diversity of Extrasolar Planets'' (JE 701/5-1) and the Research Unit FOR2544 ``Blue Planets around Red Stars''.
\end{acknowledgements}

\bibliographystyle{aa}
\bibliography{refs}

\begin{appendix}

\section{Outliers}\label{a:outliers}

\subsection{Literature comparison outliers}\label{a:lit_outliers}

Below, we list and discuss the discrepant periods between this work and the literature described in Section \ref{ss:lit_compare} and shown in Fig. \ref{fig:lit_compare}. We distinguish between cases where the periods are likely harmonically related (Sect. \ref{aa:harmonics}) and those lacking a straightforward explanation (Sect. \ref{aa:other_outliers}). We note again that this is not an exhaustive list of all instances where measured periods disagree. Instead, these examples illustrate the type and extent of deviations confronted, as well as the thought processes and evidence used to decide on the best period we adopt. For each case, we also present final period assessments, following the reasoning described here and mirroring the guidelines given in Sect. \ref{ss:prot_adopted}.

\subsubsection{Harmonics}\label{aa:harmonics}

When two contending periods are harmonics of each other, it is in general expected that one of them, typically the larger one, is correct. This is because the first harmonic of the true period can manifest itself as the prevailing period of modulation when, for example, the star's spot distribution is bimodal across its hemispheres, resulting in a `double-dipping' feature throughout each full rotation \citep[e.g.][]{Basri20, Jeffers22, Schoefer22}. On the other hand, there is no simple mechanism that may explain the production of one dip over two rotations. While the double-dip phenomenon is usually temporary and minute differences in the dip morphologies may be resolved with high-precision, high-cadence photometric data, it is more difficult to distinguish nearly equal-sized dips in ground-based data or data that cover only a short time baseline. We evaluate the following cases with this principle in mind.

\paragraph{J01019+541.}
The best TESS period of 0.139\,d is exactly half that from MEarth data 0.278\,d (DA19, \citealt{Newton16}). We give preference to the longer period from the literature. 

\paragraph{J04173+088.}
The period of this fast rotator was measured to be 0.185\,d using MEarth data \citep{West15}. However, both the TESS light curve and SuperWASP data spanning 6 seasons unambiguously indicate a period of 0.3697\,d, exactly twice as long \citep[see discussion in][]{Johnson21}. We strongly favour the longer period from our work.  

\paragraph{J10182$-$204.} 
\citet{Baroch18} obtained a period of 7.3\,d from archival SuperWASP data. TESS SAP light curves resolve a double-dipping modulation pattern with periodicity twice that at 14.5 d, which we deem to be convincing. 

\paragraph{J16581+257.}
The period measured from one season of SuperWASP observations is 11.73\,d, about half that reported from ASAS data in the literature (23.8\,d, DA19, \citealt{Oelkers18}). We consider the longer periods to be correct.

\paragraph{J20450+444.} 
DA19 measured a 19.9\,d period from archival SuperWASP data. However, a period of $\sim 39$ 
d, about twice the literature period, is implicated in spectroscopic indicators (the two Ca-IRT lines, H$\alpha$, and TiO). In fact, TESS SAP light curves also rule the shorter period for J20450+444 as implausible. Therefore, we update this period to the longer one from our work.

In a couple of cases, there is additional evidence in favour of the shorter period:  

\paragraph{J06371+175.} 
\HLr{A 37\,d period} is strongly implicated by APT photometry presented in the planet discovery paper of \citet{Luque22} and also significant in both seasons of SuperWASP data. \HLr{Although the second season of SuperWASP data slightly prefers a $\sim 74$\,d period, we still adopted the shorter period.}

\paragraph{J20525$-$169.} 
The literature period is $\sim$68\,d (ASAS, DA19, \citealt{SM16}) or 85\,d (MEarth, \citealt{Newton18}), which are approximately yearly aliases of one another, and 134 d from our work in H$\alpha$ and Ca-IRTb, nearly exactly twice the ASAS period. We adopted the 68\,d period but note that it is `debated'.

\subsubsection{Other discrepant cases}\label{aa:other_outliers}
 
\paragraph{J03213+799.}
The literature measurement of 32.4\,d is based on photometric data from AstroLAB that were available at the time. In this work, we perform a new analysis using a more complete light curve from AstroLAB. To be conservative, we indicate that the new period is still `debated' and would benefit from further confirmation.

\paragraph{J03463+262.} 
The literature value of $\sim 16$\,d comes from KELT photometry \citep{Oelkers18}, and is supported by SuperWASP data as well as Ca-IRTa. However, multiple spectroscopic indicators H$\alpha$, Ca-IRTb, and TiO show a $\sim 10$\,d period. We prefer the value derived from photometry, but indicate that it is `debated'.      

\paragraph{J04153$-$076.}
This active mid-M dwarf ($pEW'_{\rm H\alpha} = -3.3 \AA$) is expected to have a relatively short period. Using CARMENES data, \citet{Lafarga21} detected a significant signal at 1.8\,d in three spectroscopic indicators: RV, CRX, and BIS. With the latest spectroscopic time series, we detect a 8.56\,d period in the two Ca-IRTab lines. The measured $v \sin i = 2.2$ km/s indicates that the $P_{\rm rot} \lesssim 6$\,d, slightly preferring the value from \citet{Lafarga21}. However, the limited precision in $v \sin i$ means the 8\,d period can also be easily accommodated. We take the period in this work as the nominal `best' period, but designate it as `debated'.    

\paragraph{J05337+019.} 
SB1 analyzed in detail by \citet{Baroch21}. The rotation period of 0.6\,d robustly determined from TESS (see Figure \ref{fig:J05337+019}) and unbinned SuperWASP data agree with each other and also with the orbital period of the binary, as measured from RV, which is consistent with tidal synchronization. One of the reported periods in the literature is 2.8\,d, which is an alias of the 0.6\,d period. Therefore, we are confident in the 0.6\,d period derived in this work and assign it a `secure' rating. 

\paragraph{J09140+196.} 
The literature period of 89.9\,d comes from DA19's analysis of archival ASAS data, whereas the SuperWASP data exhibit an unequivocal sinusoidal signal at 39\,d over a 127\,d time baseline. The same SuperWASP data were analyzed in \citet{Baroch21}, reaching a similar value. It is possible that the two periods are harmonically related. However, a reanalysis of the data used by DA19 shows that the 89.9\,d peak, while formally significant, competes with many other peaks of similar or greater height, therefore cannot be unambiguously interpreted. Therefore, we prefer the $\sim 39$ d period, but indicate it as `provisional'.   

\paragraph{J11000+228.} The literature period is 171\,d from longitudinal B-fields \citep{Donati23}, which is rather long for an early-M dwarf. We see a much shorter, 53\,d period in both TiO and $R'_{\rm HK}$ time series, corroborated at least tentatively by SuperWASP light curves. We deem our period to be more reasonable, and brand it as `secure' owing to multiple lines of support.

\paragraph{J11509+483.}
This star has many periods in the literature. Earlier works measured a period of $\sim$125\,d \citep{Irwin11,Newton16,DA19}, all based on MEarth data. However, a new analysis of the latest MEarth data presented a longer period at 149\,d \citep{Medina22b}, and subsequent measurements using spectropolarimetric time series support this longer period of \citet{Medina22b} \citep[158\,d and 176\,d, ][respectively]{Fouque23, Donati23}. Three of our spectroscopic indicator time series (H$\alpha$, Ca-IRTb, TiO) show an $\sim$83\,d period, which could represent the first harmonic of the longer periods. We list the $P_{\rm rot}$ of \citep{Medina22b}, the middle ground within $\sim15$\% of all the available measurements, as the best period with a `secure' rating.  

\paragraph{J12230+640.} 
The literature period of 32.9\,d originates from the analysis of archival NSVS photometry data by DA19. The Ca-IRT time series contains 151 epochs and reveals a 51.5\,d signal, which incidentally coincides with a smaller peak in the NSVS data. While we deem the photometric period more likely to be correct, the veracity of the other signal cannot be ruled out. Therefore, we list the DA19 period as the best period with a `debated' designation.       

\paragraph{J14257+236E.} 
The period from DA19 is 17.6\,d and is derived from unbinned public SuperWASP data. A reanalysis of the same data with daily binning shows that this peak, while present, is not formally significant. Multiple spectroscopic indicators (H$\alpha$, Ca-IRT) reflect a period of $\sim 12.8$ d, casting further doubt into the photometric period but not enough to flat-out refute it. Therefore, we list the DA19 period as the best period but flag it as `debated'.  

\paragraph{J14321+081.} 
The 0.757\,d period comes from MEarth data recently presented by \citet{Pass23} and previously by \citet{Newton16} as a `U'-rotator. This is a very faint ($V >$ 15\,mag) but very active late-M star whose $v\sin i$ bounds the period to $\lesssim 1.7$\,d (see also Section \ref{ss:vsini}). While our unbinned SuperWASP data exhibit a highly significant period at 1.115\,d and 1.025\,d (and their aliases) in the first season, the weight of the two literature values in agreement tips the value in favour of the shorter period, which we indicate as `secure'.

\paragraph{J19346+045.}
The literature value of 12.9\,d (DA19) is from ASAS photometry with FAP = 0.33\%. We find a 21.8\,d period in H$\alpha$ and Ca-IRTab. An identical period in dLW \citep{Lafarga21} strengthens our preference for the longer period measured in this work, which we assign as the best period with a rating of `secure'.

\paragraph{J23381$-$162.} 
The literature value of 61.66\,d comes from \citet{Watson06}, which refers to the AAVSO database of variable stars. The SuperWASP data show conflicting periods (therefore a U-designation). While there is a large-amplitude sinusoidal signal of $\sim 91$\,d in season 6, a similarly clear signal at $\sim 54$\,d in season 7 is only slightly less formally significant. In summary, we consider the evidence insufficient to reject the period in the literature, though flag it as `debated'.

\subsection{$v\sin i$ outliers}\label{a:vsini_outliers}

Here we list and describe the $v\sin i$ outliers as shown in Figure \ref{fig:prot_vs_vsini} in Section \ref{ss:vsini}.

\paragraph{J05337+019.} 
This is an SB1 analyzed in detail in \citet{Baroch21}, with the companion suspected to be a white dwarf (See also Section \ref{a:lit_outliers}). The $v\sin i$ measurement implicates a maximum rotation period of 2.5\,d (1$\sigma$: 3.2\,d), which is completely consistent with the rotational modulation of 0.6 d well-resolved in TESS and SuperWASP data. \citet{Oelkers18} found a period of 2.8\,d, which is an alias of 0.6\,d. The clearly contradicting period of 18.4 d is from \citet{Howard20}. 

\paragraph{J10584$-$107.}
The $v\sin i$ measurement implicates a maximum rotation period of 3.7\,d (1$\sigma$: 13 d). The SuperWASP data give a 28\,d period which, as discussed in Section \ref{sss:swasp_analysis}, is suspected to be moon-related and likely to be unreliable.

\paragraph{J18356+329.} With $P_{\rm rot} \sim 0.1$\,d, this is the fastest rotator in the sample. $P_{\rm max} = 0.089$\,d and the $1\sigma$ upper bound is 0.103 d, which is only marginally exceeded by all the photometric measurements of 0.118\,d (e.g. DA19, TESS). The discrepancy could be easily accounted for if the error on either $v\sin i$ or $R_\star$ is slightly underestimated.

\paragraph{J19169+051S.} This very late-type (M8V) and faint star is in a wide binary with J19169+051N (73\farcs8). It exhibits very large H$\alpha$ activity, although at its radius ($\sim 0.1 R_\star$) the saturation period is $\sim 150$\,d (see Section \ref{ss:rot-act}), which implies that even extremely slow rotators can be expected to display considerable level of activity. The $v\sin i$ measurement constrains the maximum period to be $\sim 1.4$\,d ($1\sigma$ upper limit is 3.2\,d), which severely contradicts the photometric rotation period of 23.6\,d reported by DA19. However, it is worth noting that the DA19 period comes from MEarth, the same data that led \citet{Newton16} to conclude that no reliable period can be derived.

\paragraph{J06574+740, J20305+654.} The measured $P_{\rm rot}$ for these stars exceed their expected $P_{\rm rot, max}$ by 16\%, just over the threshold of 15\%. They could be easily explained if errors in the stellar radius or rotational velocity are slightly underestimated. 
J06574+740 is interesting because its light curve from every sector shows a beat pattern of two closely-spaced frequencies at 0.61 and 0.63\,d (see Fig.~\ref{fig:J06574+740}). This star is catalogued in {\em Gaia} as a duplicate source and has a high astrometric excess noise. Furthermore, it has been classified as RV-loud after the criteria of \citet{Talor18}, exhibiting peak-to-peak RV variability of > 600 m/s over 11 epochs of monitoring (spread over 450\,d). Both clues indicate that this object could be an unresolved binary. However, in this interpretation, the apparent spin-synchronization of the binary components at $\sim 0.6$\,d is more difficult to explain. The RV amplitude of an equal-mass binary consistent with the observed spectral type and spin-orbit synchronized at 0.6\,d is expected to be of order $\sim 100$ km/s. On the other hand, the fact that J06574+740 does not show overluminosity in a colour-magnitude diagram argues against the binary hypothesis. The beating pattern could also be due to two latitudinally displaced spots exhibiting differential rotation \citep[see, e.g.][]{Nagel16}. In both possible scenarios, the modulation signal is attributed to rotation and has relatively good consistency with the measured H$\alpha$ activity and $v\sin i$. %despite having an apparently more complex configuration that could complicate the measurement of the rotation velocity and stellar parameters. 
We consider the most significant of the two peaks at 0.61\,d to be the `securely' determined $P_{\rm rot}$ for this target, and recommend follow-up observations to resolve the nature of this system.

%==========================================

\subsection{$pEW'_{\rm H\alpha}$ outliers}\label{a:ha_outliers}

\HLr{In Fig. \ref{fig:prot_vs_halpha} of Sect. \ref{ss:pew_ha},} the clump of extreme outliers with $P_{\rm rot} \lesssim 1$\,d correspond to measurements from \citet[][J00184+440, J09439+269, J17115+384, J22115+184]{Oelkers18} and a `U' rotator from \citet[][J01026+623]{Newton16}. All five stars have $v \sin i < 2$\,km/s which, given their radii, rules out $P_{\rm rot} < 1$\,d in all but a narrow range of inclination angles. All except J00184+440 have alternative $P_{\rm rot}$ measurements in the literature longwards of 10\,d.

The rest of the points in this quadrant have modestly rapid rotations with 3\,d $\lesssim P_{\rm rot} < 10$\,d. Though closer to the overall trend, there are reasons to suspect these particular periods may be unreliable. The periods for J04225+105, J08119+087 and J14155+046 are measured from MEarth light curves by DA19 and \citet{Baroch18}, but have been deemed `N' rotators by \citet{Newton16} who analysed the same data. The period of J09561+627 is 8.54\,d according to \citet{Oelkers18}, which is about half of the $\sim18$ d measured by \citet{Lafarga21} and 16.9\,d from TESS SAP in this work, as discussed in Section \ref{ss:lit_compare}. The sub-10 d periods of J16554$-$083N, J19072+208, and J20567$-$104 were measured by DA19 from rather noisy ASAS data, the latter two with non-trivial FAPs (0.91 \% and 0.28\%). %note: they were also period outliers according to Ansgar's analysis 

%==========================================

\section{Rossby number computation}\label{a:rossby}

Since the dynamo mechanism relies on the motion of charged fluids in the stellar interior, stellar rotation and convection are thought to be relevant. Leading theories posit that if the timescale of rotation is short compared with that of convection, then the dynamo would operate efficiently, inducing more vigorous magnetic field generation \citep[e.g.][]{Parker55,Durney78,Noyes84}. Therefore, a commonly used parameter in physically connecting stellar rotation and activity phenomena is the Rossby number ($Ro$), a dimensionless quantity that characterizes the ratio between the two timescales: $Ro \equiv P_{\rm rot}/\tau_{\rm conv}$. $\tau_{\rm conv}$ is the convective overturn time and depends on the stellar type, but cannot be directly measured. While theoretical calculations of $\tau_{\rm conv}$ are becoming available \citep[e.g.][]{Kim96}, they are subject to stellar model assumptions and, importantly, do not work for fully convective stars. In practice, $\tau_{\rm conv}$ is often estimated empirically from minimizing the scatter in observed activity-rotation relations \citep[e.g.][]{Pizzolato03,KS07,Wright11}. A recent calibration as a function of $(V-K_s)$ colour comes from \citet{Wright18}, which is applicable across the entire M dwarf range:
\begin{equation}
\centering
    \log \tau_{\rm conv} = 0.64^{+0.1}_{-0.12} + 0.25^{+0.08}_{-0.07} ~(V-K_s). 
    \label{eqn:tau_w18}
\end{equation}

\noindent The relation is also given in terms of $M_\star$ as follows \citep{Wright18}:
\begin{equation}
\centering
    \log \tau_{\rm conv} = 2.33 ^{+0.06}_{-0.05} - 1.50^{+0.21}_{-0.20} ~(M_\star/M_\sun) + 0.31^{+0.16}_{-0.17}~(M_\star/M_\sun)^2. 
    \label{eqn:tau_w18m}
\end{equation}

\noindent These relations supersedes the previous widely-used versions from \citet{Wright11}. The earlier version is less accurate for late-type stars, deviating from the updated one by up to $\sim +0.2$ dex. 

\begin{figure}
\centering
\includegraphics[width=\hsize]{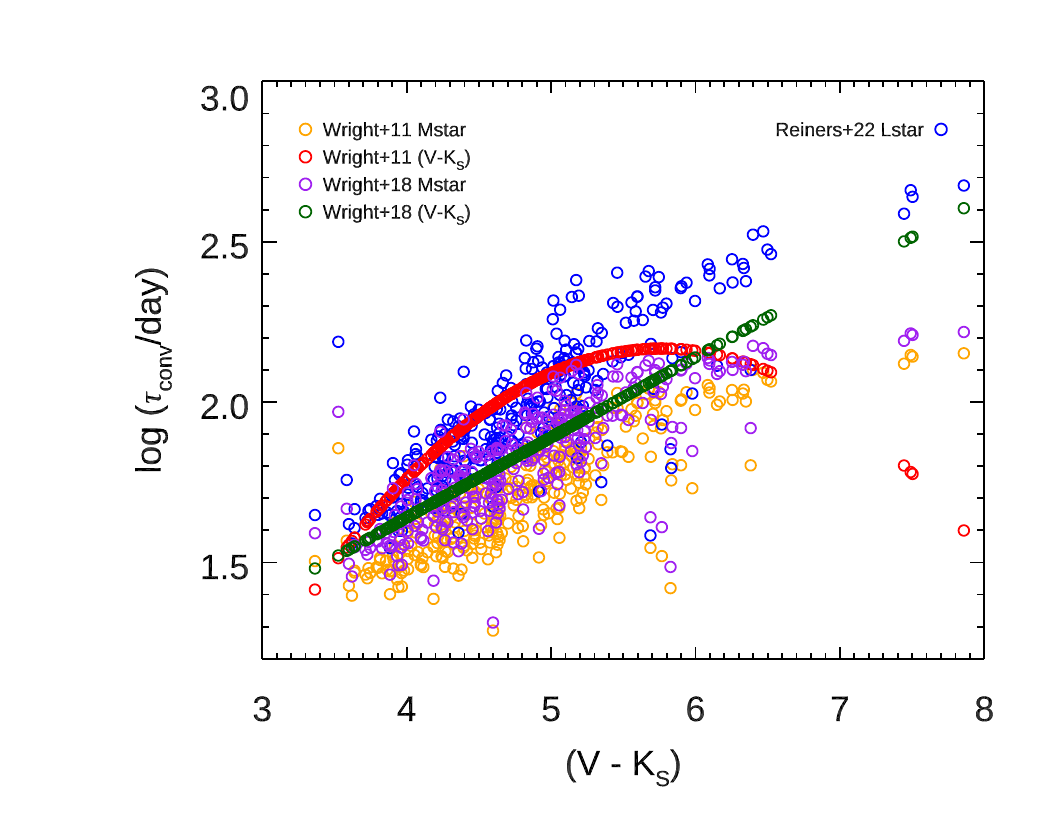}
\caption{Systemtic differences in calculated $\tau_{\rm conv}$ from popular formulae for our sample, which can result in differences in $Ro$.}
\label{fig:rossby_compare}
\end{figure}

An alternative prescription, used by \citet{Reiners22}, is given in terms of $L_\star$: $\tau_{\rm conv} = 12.3 {\rm d} \times (L_{\rm bol}/L_\sun)^{-1/2}$. Figure \ref{fig:rossby_compare} shows the $\tau_{\rm conv}$ versus $(V-Ks)$ resulting from using different version of the relations for our sample.   

A number of authors have also remarked that activity measures, when not luminosity normalized, are tightly correlated with rotation period itself, and that $\tau_{\rm conv}$ in fact has a similar scaling with colour as $L_{\rm bol}$. Therefore, the significance of the $Ro$ as the fundamental determinant of magnetic activity over the simple and more readily measured $P_{\rm rot}$ has been questioned \citep[e.g.][]{Pizzolato03, Reiners14}. Furthermore, \citet{Reiners14} advocated for the utility of the generalized Rossby number $P_{\rm rot}^{-2}R_\star^{-4}$, demonstrating that it provides a slightly tighter relation with normalized X-ray luminosity over a wide range of dwarf spectral types (F to M) than the traditional $Ro$. Nevertheless, many activity-rotation studies continue to use $Ro$. In order to facilitate direct comparison with results from the literature, we present most of our activity-rotation relations in terms of $Ro$ in addition to $P_{\rm rot}$. We use $(V-K_s)$ colours from \citet{Cifuentes20} and Eqn. \ref{eqn:tau_w18} combined with the measured $P_{\rm rot}$, where available, to compute $Ro$ for our sample. For the stars without a $V$-mag measurement in our sample, we used the $M_\star-Ro$ relation given in Equation \ref{eqn:tau_w18m}.

%==========================================

\section{Long tables and additional figures}\label{a:tables_figures}

\HLr{Tables \ref{t:bestprots}, \ref{t:allprots}, and \ref{t:planets} are available in their entirety in electronic form at the CDS via anonymous ftp to cdsarc.u-strasbg.fr (130.79.128.5) or via http://cdsweb.u-strasbg.fr/cgi-bin/qcat?J/A+A/. Below we show representative snapshots of each table to provide guidance on their content.}

\begin{figure*}
\begin{minipage}{\textwidth}
    \centering
    \includegraphics{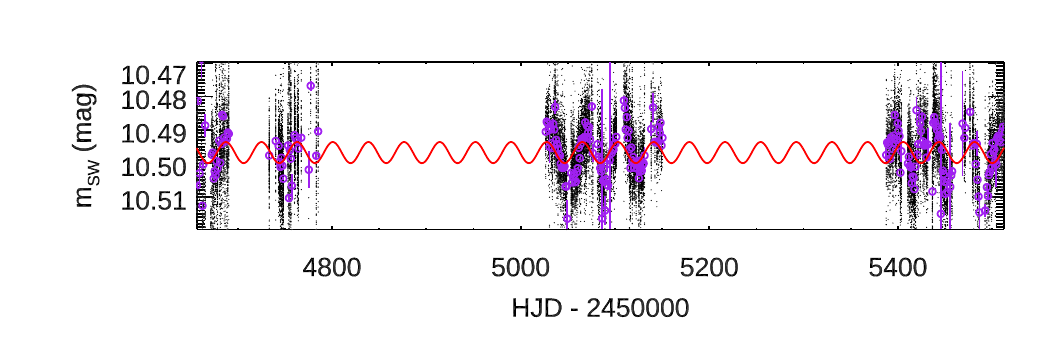}
\end{minipage}    
\begin{minipage}{\textwidth}
    \centering
    \includegraphics{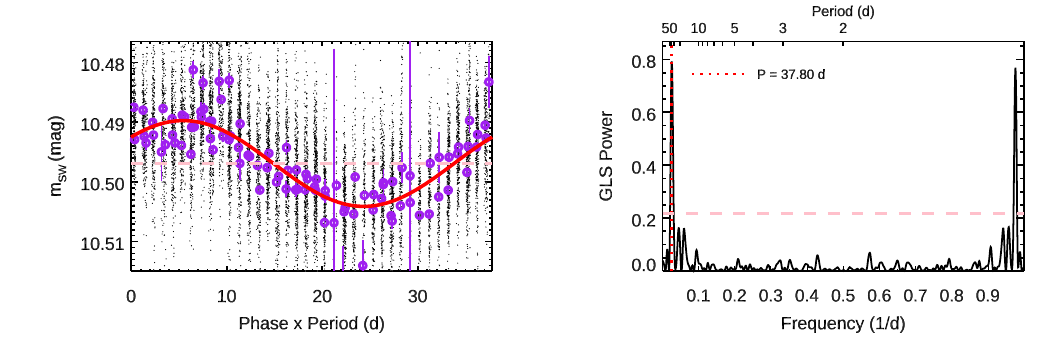}
\end{minipage}
\begin{minipage}{\textwidth}
    \centering
    \includegraphics{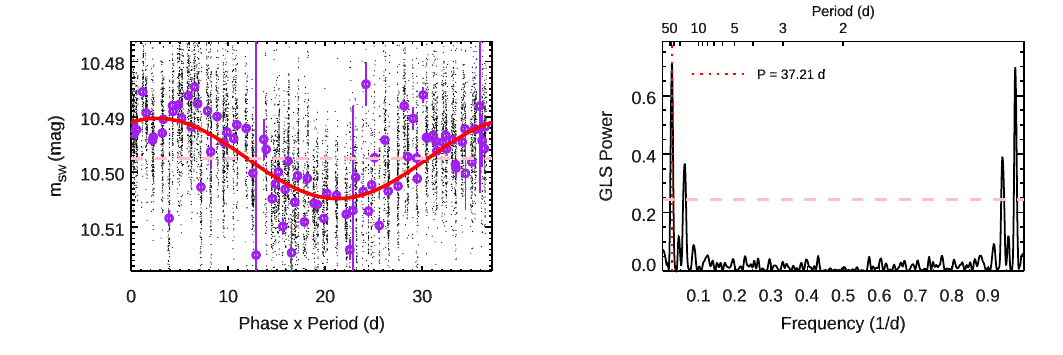}
\end{minipage}
    \caption{Example SuperWASP light curve for J22330+093. Top: all three seasons with black dots being the data points at every epoch. The purple circles and corresponding error bars represent the daily-binned points, computed as described in the text. In red we overplot the sinusoid parametrized by the best period from the second season. The best period is at 37.8 d in the second season (HJD $\sim 5100$, middle panels) and agrees well with that in season 3 (37.2 d, HJD $\sim 5400$, bottom panels). The pink dashed line in the light curve denotes the best-fit mean magnitudes, whereas in the periodograms it corresponds to FAP = 0.1\%. }
    \label{fig:J22330+093_SWASP}
\end{figure*}

\begin{figure*}
\centering
\includegraphics[width=\hsize]{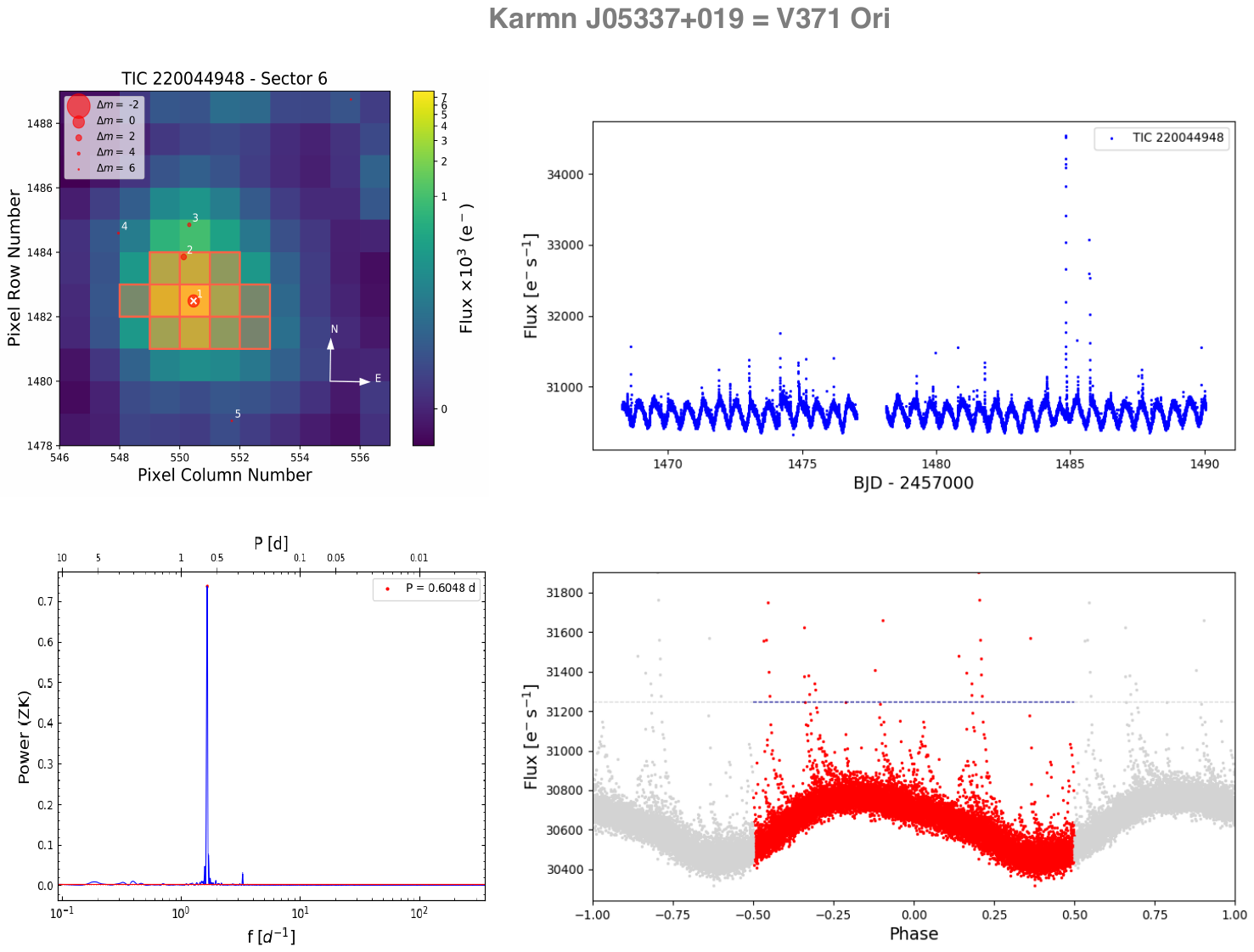}
\caption{Summary page for the analysis of PDCSAP light curves from J05337+019, TESS sector 6. Top left: target pixel map; top right: PDCSAP light curve; bottom left: GLS periodogram; bottom right: light curve phase-folded on the peak period (0.6048 d). }
\label{fig:J05337+019}
\end{figure*}

\begin{figure*} 
\centering
\begin{minipage}{\textwidth}
\includegraphics[width=0.5\hsize]{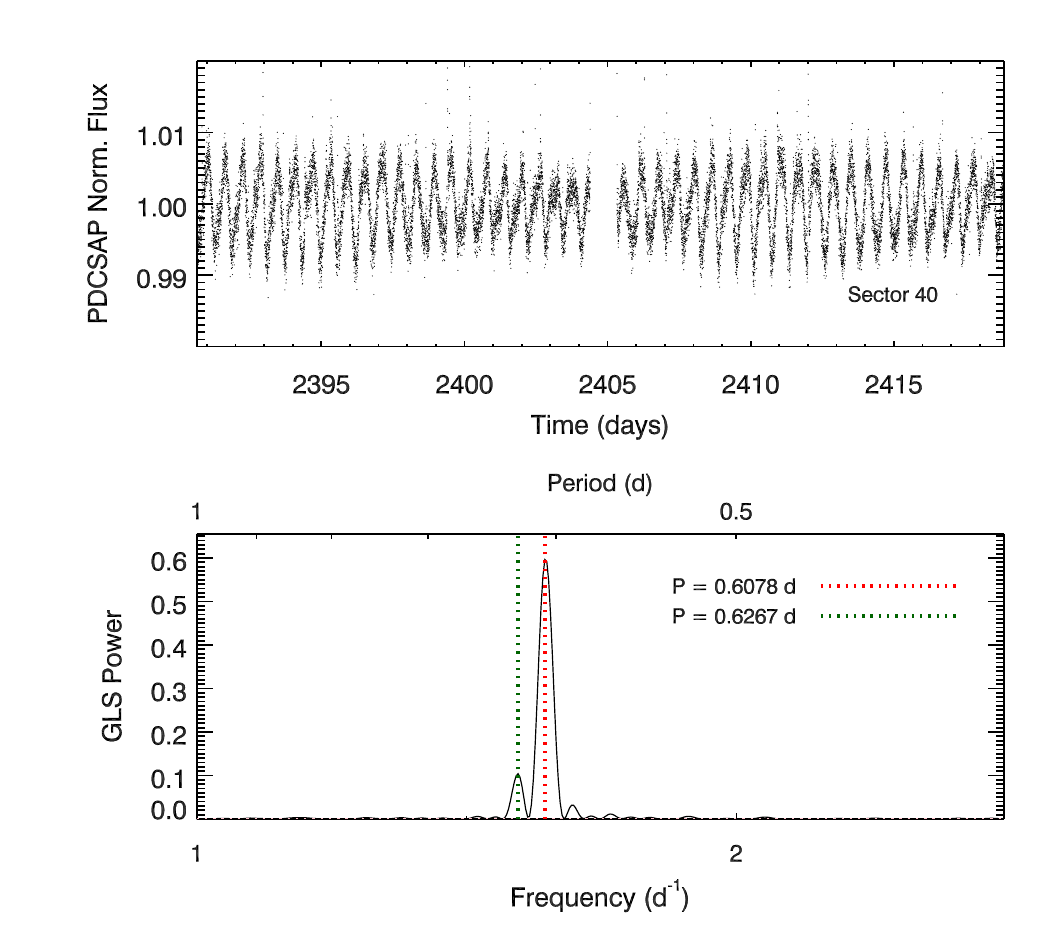}
%\end{minipage}
%\begin{minipage}{\textwidth}
\includegraphics[width=0.5\hsize]{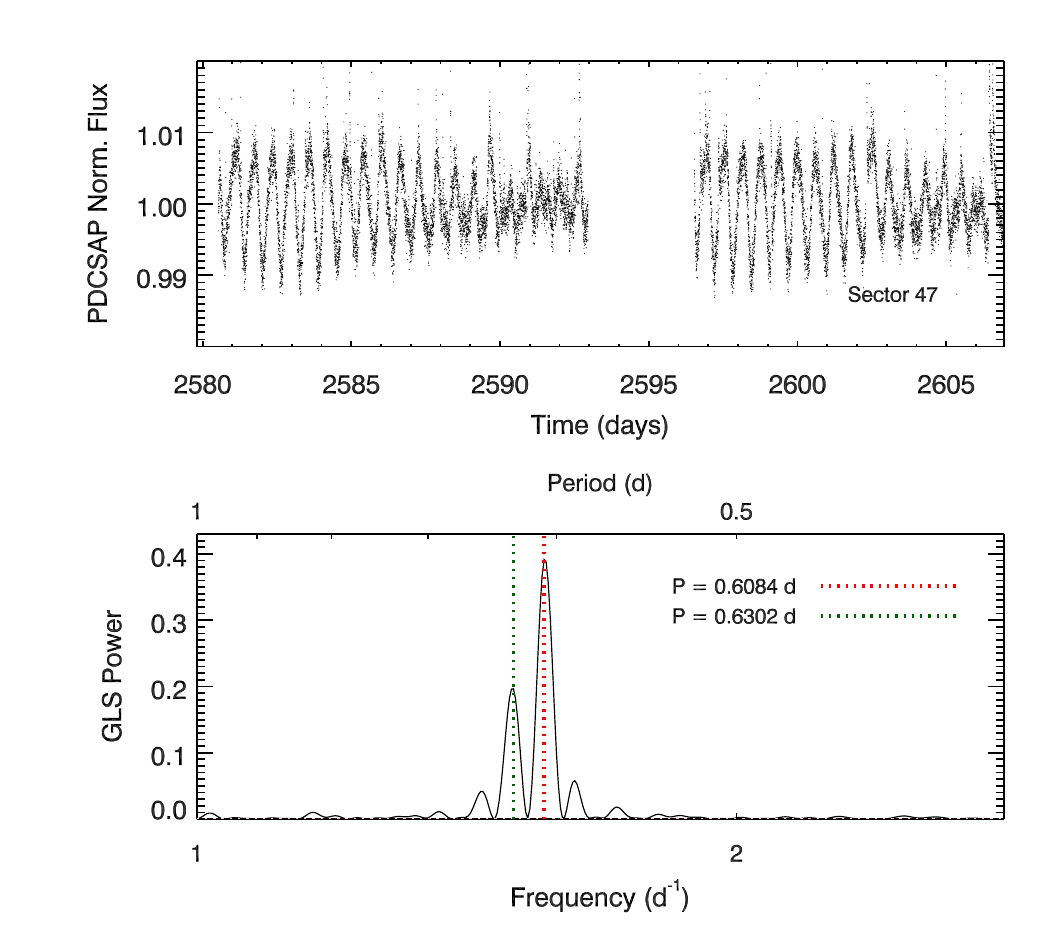}
\end{minipage}
\caption{\HLr{TESS light curve analysis for J06574+740.} Top: PDCSAP light curves for J06574+740 from sector 40 (\HLr{left}) and 47 (\HLr{right}). The light curves exhibit a `beating' pattern, suggestive of there being two stars with very similar rotation periods. Bottom: periodograms showing a main peak at 0.608\,d and a smaller but still highly significant peak at $\sim$0.63\,d, which could be the rotational signal from a binary companion or from differential rotation. }
\label{fig:J06574+740}
\end{figure*}

%\section{Analysis of spectroscopic indicators}

%\iffalse
\begin{figure*}
\begin{minipage}{1.05\textwidth}
    \centering
    \hspace{-1cm}
\includegraphics[width=0.9\hsize]{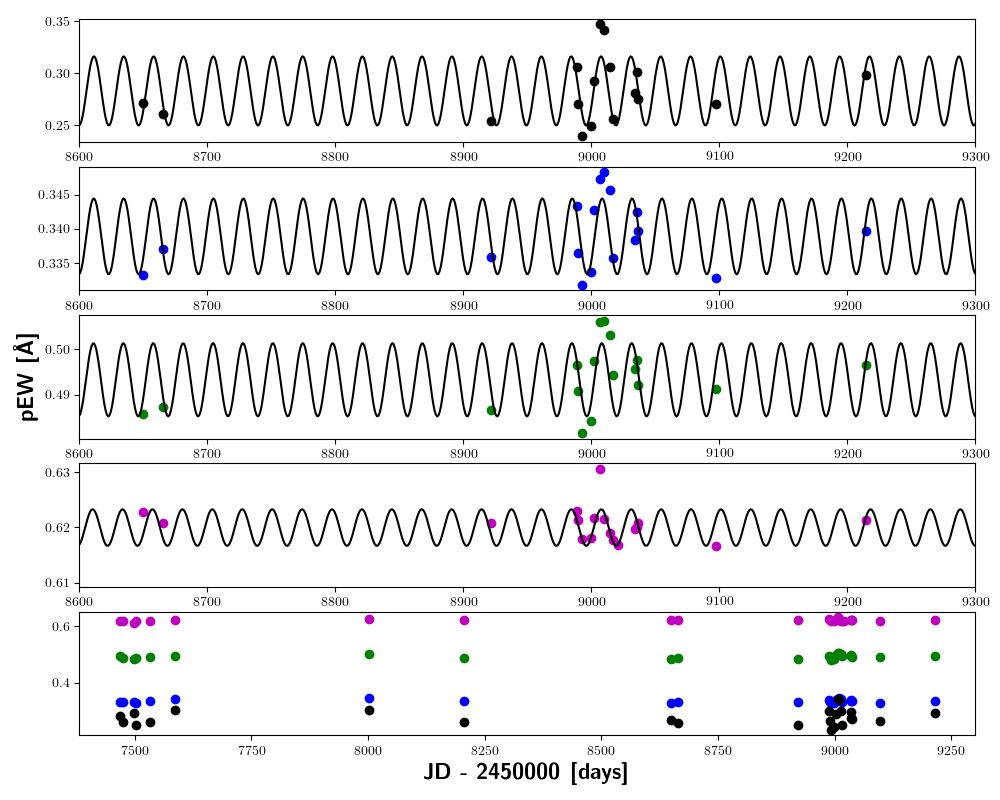}
\end{minipage}
\begin{minipage}{0.4\textwidth}
    \centering
\includegraphics[width=\hsize]{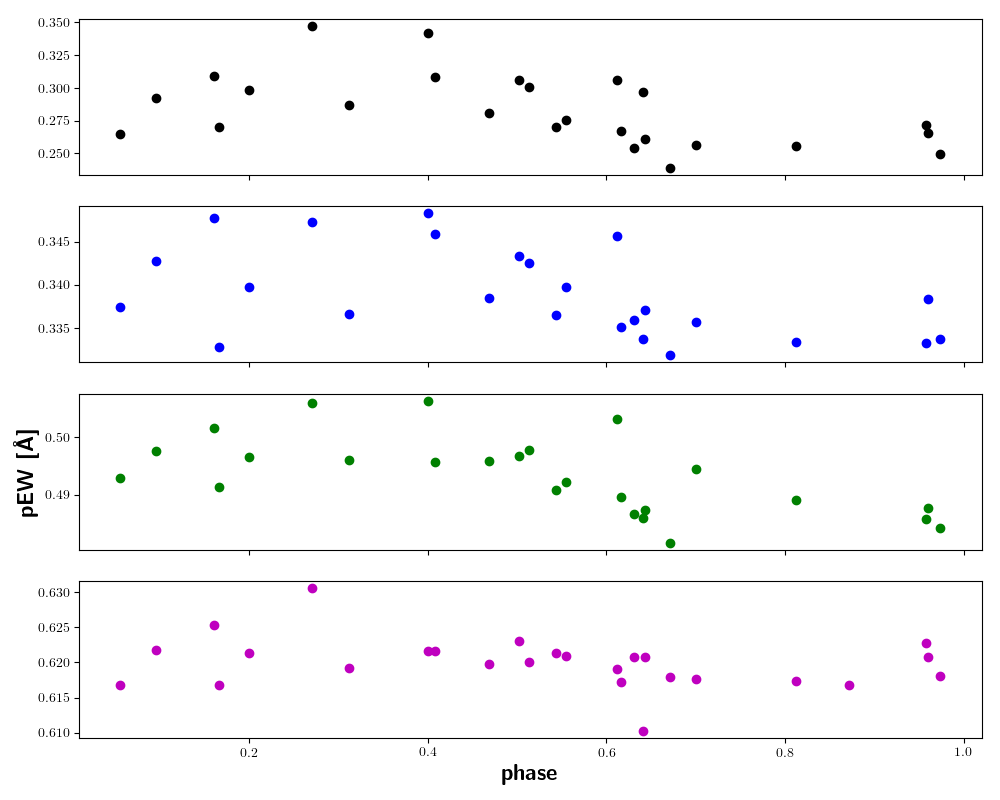}
\end{minipage}
\begin{minipage}{0.6\textwidth}
    \centering
\includegraphics[width=\hsize]{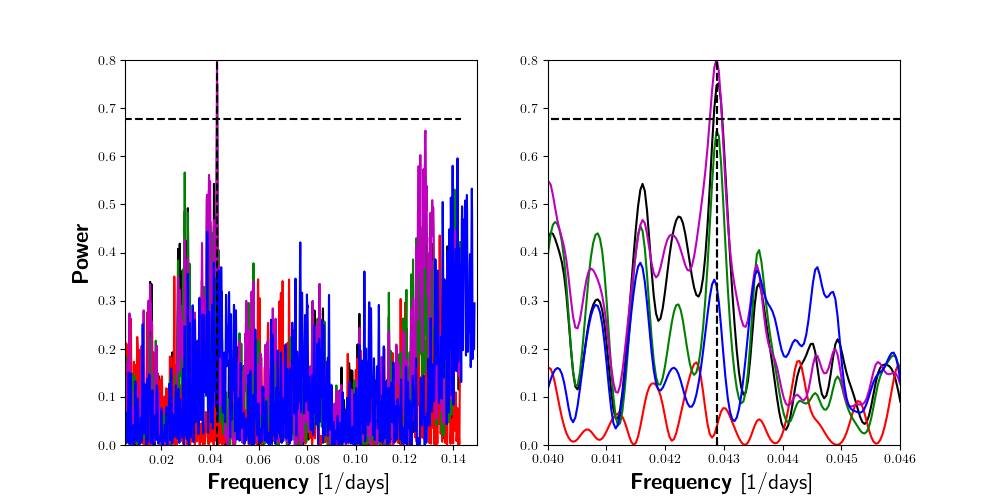}
\end{minipage}
\caption{Example of spectroscopic indicator time series analysis for J14524+123. Top: time series data with sinusoids corresponding to the best-fitting period (23.3 d) overplotted; Middle: time series of the individual indicators over the entire baseline; Bottom left: phase-folded on the strongest period; Bottom right: GLS periodogram (middle) and zoom-in on the strongest period (right) of the 4 indicators, with horizontal dashed line denoting FAP = 0.1\% and the peak highlighted by the vertical line. Black = H$\alpha$, blue = Ca-IRTa, green = Ca-IRTb, magenta = TiO. Red denotes the window function. }
\label{fig:J14524_spectro_inds}
\end{figure*}
%\fi

\begin{figure*}
\centering
\includegraphics[width=\hsize]{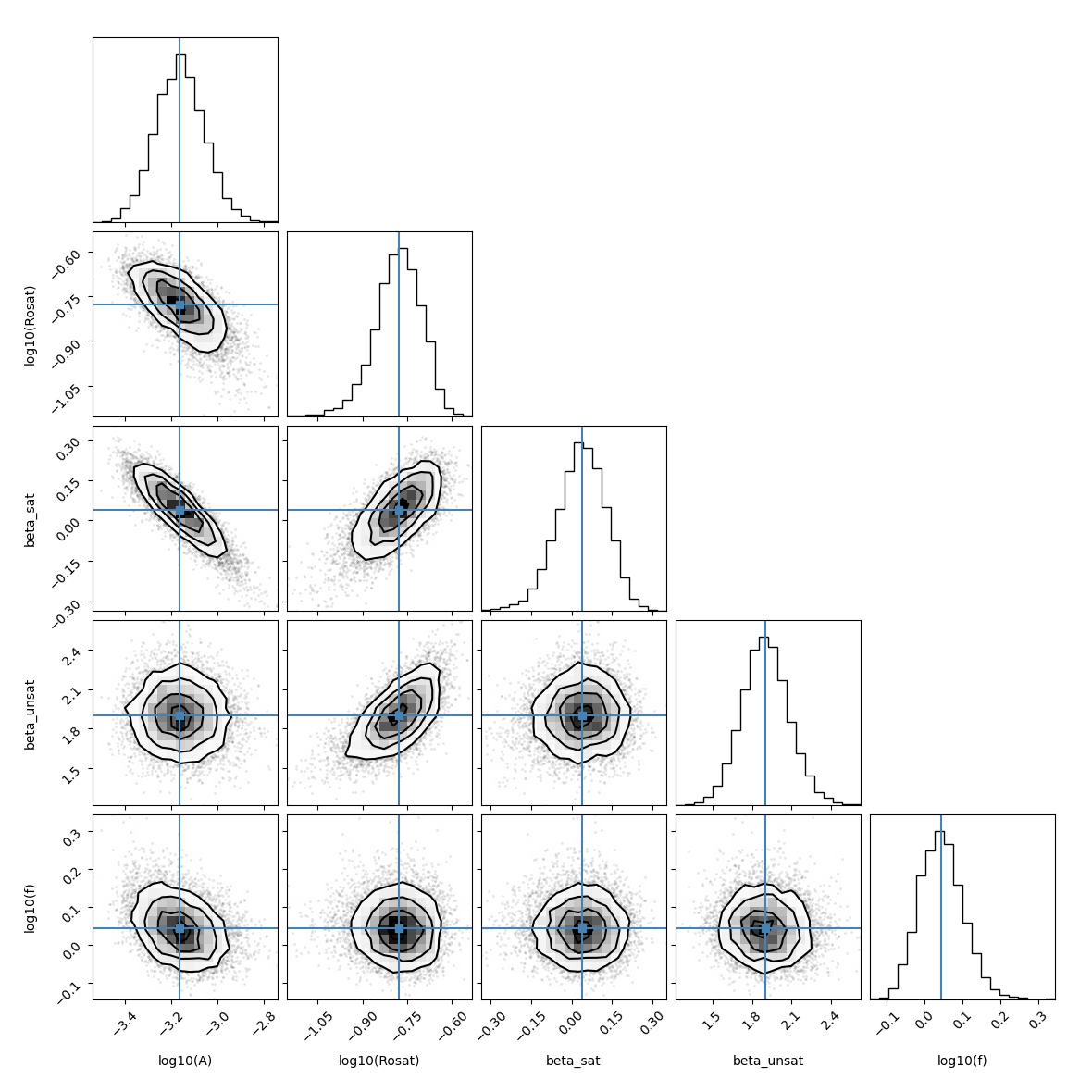}
\caption{Posterior distributions for each parameter fit in the rotation-activity relation $L_{\rm X}/L_{\rm bol}$ versus $Ro$, as described in Sect. \ref{ss:rot-act} and Eqn. \ref{eqn:act_ro}. This MCMC fit is performed over the whole stellar mass range. The contours shown are $0.5\sigma$, $1\sigma$, $1.5\sigma$ and $2\sigma$. The median value for each parameter is marked in blue. 
This corner plot is generated using {\tt{corner.py}} {\tt{(https://corner.readthedocs.io/en/latest/)}}.
}
\label{fig:lx_cornerplot}
\end{figure*}

\onecolumn
\begin{landscape}
\begin{tiny}
\begin{longtable}{llrccccccccccccccccc}
\caption{Stellar parameters of the investigated stars.}
\label{t:bestprots}
\\
\hline\hline
Karmn & Spectral & Best $P_{\rm rot}$ & Source \textsuperscript{b} & Secure & $M_\star$ & $\log \frac{\tau}{{\rm d}}$ & $\log Ro$ & $v\sin i$ & $R_\star$ & $P_{\rm rot, max}$ & $pEW'_{\rm H\alpha}$ & $L_{\rm bol}$ & $\log \frac{L_X}{L_{\rm bol}}$ & $\log \frac{L_{\rm H\alpha}}{L_{\rm bol}}$ & $\log R'_{HK}$ & $\log\frac{\langle B \rangle}{G}$ & SB & Age\textsuperscript{d} & Kin. \\
 & Type \textsuperscript{a} & (d) &  & flag \textsuperscript{c} & ($M_\sun$) & & & (km/s) & ($R_\sun$) & (d) & ($\AA$) & ($10^{-2} L_\sun$) &  & & &  & flag & (Myr) & Member \\
\hline
\endhead
\hline
\endfoot
  J00051+457 & M1.0 V &   15.37 &        DA19 &  D & 0.49 &  1.70 & -0.52 &  <   2.0 &  0.49 &... &  -0.06& 4.36&    ... &  -5.17 &  -4.97 &    ...  &      &  800 &    YD\\
  J00067-075 & M5.5 V &   93.00 &       Fou23 &  S & 0.11 &  2.24 & -0.27 &  <   2.0 &  0.12 &... &   0.00& 0.14&    ... &    ... &    ... &    ...  &      &  ... &     D\\
 J00162+198E & M4.0 V &  105.00 &        DA19 &  S & 0.27 &  1.93 &  0.10 &  <   2.0 &  0.28 &... &   0.05& 0.86&    ... &  -5.46 &  -5.70 &    ...  &      &  ... &  TD-D\\
 J00162+198W &  M4.0 V &    4.83 &        DA19 &  S & 0.43 &  1.90 & -1.22 &      4.0$\pm$   2.0 &  0.43 &    5.4 &  -4.16& 2.18&  -3.09 &  -3.91 &    ... &    ...  &  SB2 &  ... &  TD-D\\
  J00183+440 & M1.0 V &   45.00 &        SM18 &  S & 0.39 &  1.71 & -0.05 &  <   2.0 &  0.40 &... &  -0.07& 2.39&  -4.68 &    ... &  -5.33 &    ...  &      &  ... &     D\\
  J00184+440 & M3.5 V &  113.30 &       Don23 &  S & 0.16 &  1.93 &  0.13 &  <   2.0 &  0.17 &... &  -0.03& 0.33&    ... &  -5.02 &  -5.42 &    ...  &      &  ... &     D\\
  J00286-066 & M4.0 V &     ... &             &    & 0.34 &  1.87 &   ... &  <   2.0 &  0.35 &... &   0.12& 1.50&    ... &    ... &    ... &    ...  &      &  ... &     D\\
  J00389+306 & M2.5 V &   50.20 &        DA19 &  P & 0.41 &  1.74 & -0.04 &  <   2.0 &  0.41 &... &   0.04& 2.46&    ... &  -5.28 &  -5.11 &    ...  &      &  ... &     D\\
\hline\hline
\end{longtable}
\end{tiny}
Note: This table is available in its entirety in ASCII format online.\\
{\textsuperscript{a}: For known SBs, we tabulate the combined spectral type or that of the primary.} \\
{\textsuperscript{b}: T.W. refers to This Work.} \\
{\textsuperscript{c}: Secure flags: `S' = secure; `P' = provisional; `D' = debated.} \\
{\textsuperscript{d}: Ages of 800 Myr are upper limits.} \\
\end{landscape}

\onecolumn
\begin{landscape}
\begin{tiny}
\begin{longtable}{lcccccccccccccccc}
\caption{Adopted, literature, and all measured $P_{\rm rot}$ of our investigated stars.}
\label{t:allprots}
\\
\hline\hline
Karmn & Best $P_{\rm rot}$ & Source & Secure & $P$1 \textsuperscript{b} & $\delta P$1 & Ref1 & $P$ (Planet) & Ref & $P$ (Laf21) & $P$ (Other & Other phot & $P$ (TESS) & TESS & $P$ (SWASP) & SWASP & $P$ (Spect) \\
 & (d) &  & flag\textsuperscript{a} & (d) & (d) & & (d) & (Planet) & (d) & phot) (d) & facility & (d) & LC type & (d) & flag & (d) \\
\hline
\endhead
\hline
\endfoot
 J00051+457&  15.37&     DA19&  D&  15.37&   0.09&      DA19&    ...&            &    ...&    ...&             &    ...&      &    ...&  &    ...\\
 J00067-075&  93.00&    Fou23&  S&  93.00&   1.60&     Fou23& 126.00&        SM23&    ...&    ...&             &    ...&      &    ...&  &    ...\\
J00162+198E& 105.00&     DA19&  S& 105.00&  44.00&      DA19&    ...&            &    ...&    ...&             &    ...&      &    ...&  &    ...\\
J00162+198W&   4.83&     DA19&  S&   4.83&   0.06&      DA19&    ...&            &    ...&    ...&             &    ...&      &    ...&  &    ...\\
 J00183+440&  45.00&     SM18&  S&  45.00&   4.40&      SM18&  43.82&       How14&  46.19&    ...&             &    ...&      &    ...&  &    ...\\
 J00184+440& 113.30&    Don23&  S& 113.30&   4.30&     Don23&    ...&            &    ...&    ...&             &    ...&      &    ...&  &    ...\\
 J00286-066&    ...&         &   &    ...&    ...&          &    ...&            &    ...&    ...&             &    ...&      &    ...&  &    ...\\
 J00389+306&  50.20&     DA19&  P&  50.20&   1.30&      DA19&    ...&            &    ...&    ...&             &    ...&      &    ...&  &    ...\\
\hline\hline
\end{longtable}
\end{tiny}
Note: This table is available in its entirety in ASCII format online.\\
{\textsuperscript{a}: Secure flags: `S' = secure; `P' = provisional; `D' = debated.} \\
{\textsuperscript{b}: Other literature periods can be found in online version of table.} \\
\end{landscape}

\begin{table}[htb]
\begin{center}
\begin{small}
\twocolumn
\caption{List of reference abbreviations to Table \ref{t:allprots}. \label{t:reflist_prots1}}
\begin{tabular}{ll} 
\hline\hline
\noalign{\smallskip}
Abbreviation & Reference \\
\hline
\noalign{{\smallskip}}
AD17b& \citealt{AD17b} \\
AD17c& \citealt{AD17c} \\
Aff19& \citealt{Affer19} \\
Ama21& \citealt{Amado21} \\
Bar18& \citealt{Baroch18} \\
Bla23& \citealt{BlancoPozo23} \\
Blu20& \citealt{Bluhm20} \\
Bon18& \citealt{Bonfils18} \\
Bou18& \citealt{Bourrier18} \\
Cab22& \citealt{Caballero22}\\
Clo17& \citealt{Cloutier17} \\
Clo20& \citealt{Cloutier20} \\
DA19& \citealt{DA19} \\
Don08& \citealt{Donati08} \\
Dav16& \citealt{David16} \\
Ded21& \citealt{Dedrick21} \\
Dia19& \citealt{Diaz19} \\
Don23& \citealt{Donati23} \\
Dre20& \citealt{Dreizler20} \\
Fen20& \citealt{Feng20} \\
Fek00& \citealt{Fek00} \\
Fou23& \citealt{Fouque23} \\
GA20& \citealt{GA20} \\
GA21& \citealt{GA21} \\
GA23& \citealt{GA23} \\
Gil16& \citealt{Gillon16} \\
Gor23& \citealt{Gorrini23} \\
GR98& \citealt{GR98} \\
Hal08& \citealt{Hallinan08} \\
Hag10& \citealt{Haghighipour10} \\
Har11& \citealt{Hartman11} \\
Heb16& \citealt{Hebrard16} \\
Hob18& \citealt{Hobson18} \\
Hob19& \citealt{Hobson19} \\
How14& \citealt{Howard14} \\ 
How20& \citealt{Howard20} \\
Irv23& \citealt{Irving23} \\
Irw11& \citealt{Irwin11} \\
Joh07& \citealt{Johnson07} \\
Joh10& \citealt{Johnson10} \\
%Kem20& \citealt{Kemmer20} \\
Kir12& \citealt{Kiraga12} \\
Kos23& \citealt{Kossakowski23} \\
KS07& \citealt{KS07} \\
KS13& \citealt{KS13} \\
Laf21& \citealt{Lafarga21} \\
Lal19& \citealt{Lalitha19} \\
\noalign{{\smallskip}}
\hline
\end{tabular}
\end{small}
\end{center}
\end{table}

\addtocounter{table}{-1}
\begin{table}[htb]
\begin{center}
\begin{small}
\twocolumn
\caption{Continued.}
\begin{tabular}{ll} 
\hline\hline
\noalign{\smallskip}
Abbreviation & Reference \\
\hline
\noalign{{\smallskip}}
Lot18& \citealt{Lothringer18} \\
Luq18& \citealt{Luque18} \\
Luq19& \citealt{Luque19} \\ 
Luq22& \citealt{Luque22} \\
Med20& \citealt{Medina20} \\
Med22& \citealt{Medina22b} \\
Mes11& \citealt{Messina11} \\
%Mon14& \citealt{Montet14} \\
Mor08& \citealt{Morin08} \\
Mor10& \citealt{Morin10} \\
Mor19& \citealt{Morales19} \\
Mou17& \citealt{Moutou17} \\
Nag19& \citealt{Nagel19} \\
New16& \citealt{Newton16} \\
New18& \citealt{Newton18} \\
Nor07& \citealt{Norton07} \\
Oel18& \citealt{Oelkers18} \\
Ola16& \citealt{Olah16} \\
Pas22& \citealt{Pass22} \\
Pas23& \citealt{Pass23} \\
Per19& \citealt{Perger19} \\
Pin19& \citealt{Pinamonti19} \\
Poj05& \citealt{Pojmanski05} \\
Qui22& \citealt{Quirrenbach22} \\
Rae20& \citealt{Raetz20} \\
Rei20& \citealt{Reinhold20} \\
Rib18& \citealt{Ribas18} \\
Riv05& \citealt{Rivera05} \\
Sar18& \citealt{Sarkis18} \\
SM15& \citealt{SM15} \\
SM16& \citealt{SM16} \\
SM17& \citealt{SM17a} \\
SM17b& \citealt{SM17b} \\
SM18& \citealt{SM18} \\
SM23& \citealt{SM23} \\
Ste16& \citealt{Stelzer16} \\
Sto20a& \citealt{Stock20a} \\
Sto20b& \citealt{Stock20b} \\
Ter22& \citealt{Terrien22} \\
Tes04& \citealt{Testa04} \\
TP19& \citealt{TP19} \\
TP21& \citealt{TP21} \\
%Tri18& \citealt{Trifonov18} \\
%Tuo14& \citealt{Tuomi14} \\
Tri21& \citealt{Trifonov21} \\ 
Vid17& \citealt{Vida17} \\
Wat06& \citealt{Watson06} \\
Wes15& \citealt{West15} \\
\hline

\end{tabular}
\end{small}
\end{center}
\end{table}
%\end{center}

\onecolumn
\begin{landscape}
\begin{small}
\begin{longtable}{lccccccccccccc}
\caption{Planets table.}
\label{t:planets}
\\
\hline\hline
Karmn & Best $P_{\rm rot}$ & Secure & Transit & $P_{\rm orb}$1 & $m_p\sin i1$ & Ref1 & $P_{\rm orb}$2 & $m_p\sin i2$ & Ref2 & $P_{\rm orb}$3 & $m_p\sin i3$ & Ref3 & Add.  \\
 & (d) & flag & flag & (d) & ($m_\oplus$) & & (d) & ($m_\oplus$) & & (d) & ($m_\oplus$) & & planets  \\
\hline
\endfirsthead
\caption{continued}\\
\hline
Karmn & Best $P_{\rm rot}$ & Secure & Transit & $P_{\rm orb}$1 & $m_p\sin i1$ & Ref1 & $P_{\rm orb}$2 & $m_p\sin i2$ & Ref2 & $P_{\rm orb}$3 & $m_p\sin i3$ & Ref3 & Add.  \\
 & (d) & flag & flag & (d) & ($m_\oplus$) & & (d) & ($m_\oplus$) & & (d) & ($m_\oplus$) & & planets  \\
\hline
\endhead
\hline
\endfoot
\hline
\endlastfoot
%\endhead
%\hline
%\endfoot
 J00067-075&  93.00& S&  &  10.35&   1.08&        SM23&  21.20&   1.36&        SM23&    ...&    ...&            &  \\
 J00183+440&  45.00& S&  &  11.44&   3.98& Tri22 Rib23&6694.00&  50.40& Tri22 Rib23&    ...&    ...&            &  \\
 J01026+623&  18.82& S&  &  13.85&   5.63&      Per19 &    ...&    ...&            &    ...&    ...&            &  \\
 J01125-169&  70.10& S&  &   2.02&   0.70&     Sto20a &   3.06&   1.14&     Sto20a &   4.66&   1.09&     Sto20a &  \\
 J02002+130&   2.00& S&  & 771.40&  67.00&      Qui22 &    ...&    ...&            &    ...&    ...&            &  \\
 J02530+168&  97.56& S&  &   4.91&   1.05&      Zec19 &  11.41&   1.11&      Zec19 &    ...&    ...&            &  \\
 J03133+047& 126.20& S&  &   2.29&   3.95&      Bau20 &    ...&    ...&            &    ...&    ...&            &  \\
 J04167-120&  33.00& P& Y&   4.05&  30.80&      Dre20 &    ...&    ...&            &    ...&    ...&            &  \\
\hline\hline
\end{longtable}
\end{small}
Note: This table is available in its enterity (including additional planets) in ASCII format online.\\
\end{landscape}

\begin{table*}[htb]
\begin{center}
\begin{small}
\caption{List of reference abbreviations to Table \ref{t:planets}.}
\label{t:reflist_planets}
\begin{tabular}{ll}
\hline\hline
Abbreviation & Reference \\
\hline
Ago21& \citealt{Agol21} \\
AD17c& \citealt{AD17c} \\
Ama21& \citealt{Amado21} \\
Bau20& \citealt{Bauer20} \\
Bla23& \citealt{BlancoPozo23} \\
Blu20& \citealt{Bluhm20} \\
Bon18& \citealt{Bonfils18} \\
Cal21& \citealt{Cale21} \\
Dam22& \citealt{Damasso22} \\
Dav16& \citealt{David16} \\
Ded21& \citealt{Dedrick21} \\
Dre20& \citealt{Dreizler20} \\
Fen20& \citealt{Feng20} \\
GA20& \citealt{GA20} \\
GA21& \citealt{GA21} \\
GA23& \citealt{GA23} \\
Gor23& \citealt{Gorrini23} \\
Hob19& \citealt{Hobson19} \\
How10& \citealt{Howard10} \\
Joh10& \citealt{Johnson10} \\
Kam18& \citealt{Kaminski18} \\
Kem20& \citealt{Kemmer20} \\
Lal19& \citealt{Lalitha19} \\
Luq18& \citealt{Luque18} \\
Luq19& \citealt{Luque19} \\ 
Luq22& \citealt{Luque22} \\
Mon14& \citealt{Montet14} \\
Mor19& \citealt{Morales19} \\
Nag19& \citealt{Nagel19} \\
Pal23& \citealt{Palle23} \\
Per19& \citealt{Perger19} \\
Pin19& \citealt{Pinamonti19} \\
Qui22& \citealt{Quirrenbach22} \\
Rei18b& \citealt{Reiners18b} \\
Rib18& \citealt{Ribas18} \\
Rib23& \citealt{Ribas23} \\
Sar18& \citealt{Sarkis18} \\
SM17b& \citealt{SM17b} \\
SM23& \citealt{SM23} \\
Sto20a& \citealt{Stock20a} \\
Sto20b& \citealt{Stock20b} \\
TP21& \citealt{TP21} \\
Tri18& \citealt{Trifonov18} \\
Tri20& \citealt{Trifonov20} \\
Tri21& \citealt{Trifonov21} \\ 
Tri22& Trifonov et al., in prep \\
Wri16& \citealt{Wright16} \\
Zec19& \citealt{Zechmeister19} \\
\hline

\end{tabular}
\end{small}
\end{center}
\end{table*}

\end{appendix}

\end{document}